\pdfoutput=1 
\documentclass[useAMS,fleqn,usenatbib]{mnras}

\usepackage[T1]{fontenc}
\usepackage{ae,aecompl}

\usepackage{etoolbox}
\makeatletter
\patchcmd\@combinedblfloats{\box\@outputbox}{\unvbox\@outputbox}{}{\errmessage{\noexpand patch failed}}
\makeatother

\usepackage{amsmath}
\usepackage{amsopn}
\usepackage[british]{babel}
\usepackage[varg]{txfonts}
\usepackage{biblio}
\usepackage{natbib}
\usepackage{color}

\usepackage{graphicx}
\usepackage{xpatch}
\makeatletter
\xpatchcmd\Gin@setfile{\expandafter\strip@prefix\meaning\@tempa}{\makebox[\Gin@req@width]{Image place holder}}{}{}
\makeatother

\usepackage{marginnote} 

\definecolor{orange}{rgb}{1.0,0.5,0.}

\def\MDM{\ifmmode{\>M_{\textnormal{\sc dm}}}\else{$$M_{\textnormal{\sc dm}}}\fi}

\def\XH{\ifmmode{\>X_{\textnormal{\sc h}}} \else{$X_{\textnormal{\sc h}}$}\fi}
\def\nH{\ifmmode{\>n_{\textnormal{\sc h}}} \else{$n_{\textnormal{\sc h}}$}\fi}

\def\maspyr{\ifmmode{\>\textnormal{mas~yr}^{-1}}\else{mas~yr$^{-1}$}\fi}

\def\mG{\ifmmode{\>\mu\mathrm{G}}\else{$\mu$G}\fi}
\def\erg{\ifmmode{\> {\rm erg}}\else{erg}\fi}
\def\keV{\ifmmode{\> {\rm keV}}\else{keV}\fi}

\def\deg{\ifmmode{\>^{\circ}}\else{$^{\circ}$}\fi}
\def\onedeg{\ifmmode{\>1^{\circ}}\else{$1^{\circ}$}\fi}

\def\xvir{\ifmmode{\>\!x_{vir}}\else{$x_{vir}$}\fi}
\def\Mvir{\ifmmode{\>\!M_{vir} }\else{$M_{vir} $}\fi}
\def\rvir{\ifmmode{\>\!r_{vir}}\else{$r_{vir}$}\fi}
\def\vvir{\ifmmode{\>\!v_{vir}}\else{$v_{vir}$}\fi}
\def\Vvir{\ifmmode{\>\!V_{vir} }\else{$V_{vir} $}\fi}

\def\tratio{\ifmmode{\>\tau}\else{$\tau$}\fi}

\def\rms{\ifmmode{\>r_{\textnormal{\sc ms}}}\else{$r_{\textnormal{\sc ms}}$}\fi}

\def\Mpc{\ifmmode{\>\!{\rm Mpc}} \else{Mpc}\fi}
\def\kpc{\ifmmode{\>\!{\rm kpc}} \else{kpc}\fi}
\def\pc{\ifmmode{\>\!{\rm pc}} \else{pc}\fi}

\def\Gyr{\ifmmode{\>\!{\rm Gyr}} \else{Gyr}\fi}
\def\Myr{\ifmmode{\>\!{\rm Myr}} \else{Myr}\fi}
\def\yr{\ifmmode{\>\!{\rm yr}} \else{yr}\fi}
\def\pyr{\ifmmode{\>\!{\rm yr}^{-1}}\else{yr $^{-1}$} \fi}
\def\s{\ifmmode{\>\!{\rm s}}\else{s}\fi}
\def\ps{\ifmmode{\>\!{\rm s}^{-1}}\else{s$^{-1}$}\fi}
\def\Hz{\ifmmode{\>\!{\rm Hz}}\else{Hz}\fi}

\def\kms{\ifmmode{\>\!{\rm km\,s}^{-1}}\else{km~s$^{-1}$}\fi}

\def\K{\ifmmode{\>\!{\rm K}}\else{K}\fi}

\def\sr{\ifmmode{\>\!{\rm sr}}\else{sr}\fi}
\def\psr{\ifmmode{\>\!{\rm sr}^{-1}}\else{sr$^{-1}$}\fi}
\def\arcs{\ifmmode{\>\!{\rm arcsec}}\else{arcsec}\fi}
\def\parcs{\ifmmode{\>\!{\rm arcsec}^{-1}}\else{arcsec${-1}$}\fi}
\def\parcss{\ifmmode{\>\!{\rm arcsec}^{-2}}\else{arcsec${-2}$}\fi}

\def\cm{\ifmmode{\>\!{\rm cm}}\else{cm}\fi}
\def\cc{\ifmmode{\>\!{\rm cm}^{3}}\else{cm$^{3}$}\fi}
\def\sqc{\ifmmode{\>\!{\rm cm}^{2}}\else{cm$^{2}$}\fi}
\def\pcc{\ifmmode{\>\!{\rm cm}^{-3}}\else{cm$^{-3}$}\fi}
\def\psc{\ifmmode{\>\!{\rm cm}^{-2}}\else{cm$^{-2}$}\fi}

\def\g{\ifmmode{\>\!{\rm g}}\else{g}\fi}
\def\Msun{\ifmmode{\>\!{\rm M}_{\odot}}\else{M$_{\odot}$}\fi}
\def\hMsun{\ifmmode{\> h^{-1}{\rm M}_{\odot}}\else{$h^{-1}$M$_{\odot}$}\fi}

\def\Zsun{\ifmmode{\>\!{\rm Z}_{\odot}}\else{Z$_{\odot}$}\fi}

\def\rayl{\ifmmode{\>\!{\rm R}}\else{R}\fi}
\def\mR{\ifmmode{\>\!{\rm mR}}\else{mR}\fi}

\renewcommand{\ion}[2]{\hbox{#1\,{\sc #2}}}

\def\lya{\ifmmode{\>\!{\rm Ly}\alpha}\else{Ly$\alpha$}\fi}

\def\Ha{\ifmmode{\>\!{\rm H}\alpha}\else{H$\alpha$}\fi}
\def\Hb{\ifmmode{\>\!{\rm H}\beta}\else{H$\beta$}\fi}

\def\HI{\ifmmode{\> \textnormal{\ion{H}{i}}} \else{\ion{H}{i}}\fi}
\def\HII{\ifmmode{\> \textnormal{\ion{H}{ii}}} \else{\ion{H}{ii}}\fi}
\def\CIV{\ifmmode{\> \textnormal{\ion{C}{iv}}} \else{\ion{C}{iv}}\fi}
\def\SiIV{\ifmmode{\> \textnormal{\ion{S}{iv}}} \else{\ion{Si}{iv}}\fi}

\def\NH{\ifmmode{\> {\rm N}_{\rm H}} \else{N$_{\rm H}$}\fi}
\def\NHI{\ifmmode{\> {\rm N}_{\HI}} \else{N$_{\HI}$}\fi}
\def\MHI{\ifmmode{\> {\rm M}_{ \HI}} \else{M$_{\HI}$}\fi}

\def\mua{\ifmmode{\>\mu_{ \textnormal{\Ha}}}\else{$\mu_{ \textnormal{\Ha}}$}\fi}
\def\alphabha{\ifmmode{\>\alpha_{B}^{(\textnormal{\Ha})}}\else{$\alpha_{B}^{(\textnormal{\Ha})}$}\fi}

\newcommand{\myemail}{tepper@physics.usyd.edu.au}
\newcommand{\ramses}{{\sc Ramses}}
\newcommand{\dice}{{\sc dice}}

\defcitealias{pla14b}{\textcolor{black}{Planck Collaboration} 2014}
\defcitealias{ast13a}{\textcolor{black}{Astropy Collaboration} 2013}

\title[ The puzzle of the leading gas stream ]{ The Magellanic System: the puzzle of the leading gas stream }

\author[T.~Tepper-Garc\'\i{}a et al.]{%
Thor Tepper-Garc\'\i{}a,$^{1,2}$\thanks{E-mail: \myemail} Joss Bland-Hawthorn,$^{1,2}$
Marcel S.~Pawlowski$^{3,4}$\thanks{Hubble Fellow at UCI, then Schwarzschild Fellow at AIP} \newauthor and Tobias K.~Fritz$^{5,6}$\\
$^1$Sydney Institute for Astronomy, School of Physics, University of Sydney, NSW 2006, Australia\\
$^2$ARC Centre of Excellence for All Sky Astrophysics in Three Dimensions (ASTRO-3D)\\
$^3$Department of Physics and Astronomy, University of California, Irvine, CA 92697, USA\\
$^4$Leibniz-Institut f\"ur Astrophysik Potsdam (AIP), An der Sternwarte 16, D-14482 Potsdam, Germany\\
$^5$Instituto de Astrof\'{i}sica de Canarias, Calle Via L\'{a}ctea s/n, E-38206 La Laguna, Tenerife, Spain\\
$^6$Universidad de La Laguna. Avda. Astrof\'{i}sico Fco. S\'{a}nchez, La Laguna, Tenerife, Spain
}

\date{Accepted ---. Received ---; in original form ---}

\pubyear{\date{year}}

\begin{document}
\label{firstpage}
\pagerange{\pageref{firstpage}--\pageref{lastpage}}
\maketitle

\pdfminorversion=5
\begin{abstract}
The Magellanic Clouds (MCs) are the most massive gas-bearing systems falling into the Galaxy at the present epoch. They show clear signs of interaction, manifested in particular by the Magellanic Stream, a spectacular gaseous wake that trails from the MCs extending more than 150\deg\ across the sky. 
Ahead of the MCs is the ``Leading Arm'' usually interpreted as the tidal counterpart of the Magellanic Stream, an assumption we now call into question. We revisit the formation of these gaseous structures in a first-infall scenario, including for the first time a Galactic model with a weakly magnetised, spinning hot corona.
In agreement with previous studies, we recover the location and the extension of the Stream on the sky. In contrast, we find that the formation of the Leading Arm -- that is otherwise present in models without a corona -- is inhibited by the hydrodynamic interaction with the hot component. 
These results hold with or without coronal rotation or a weak, ambient magnetic field. Since the existence of the hot corona is well established, we are led to two possible interpretations: (i) the Leading Arm survives because the coronal density beyond 20 kpc is a factor $\gtrsim 10$ lower than required by conventional spheroidal coronal x-ray models, in line with recent claims of rapid coronal rotation;
or (ii) the `Leading Arm' is cool gas {\it trailing} from a {\em frontrunner}, a satellite moving ahead of the MCs,
consistent with its higher metallicity compared to the trailing stream. Both scenarios raise issues that we discuss.
\end{abstract}


\section{Introduction} \label{sec:intro}

The Magellanic System is the most important source of cold gas accretion within the Galaxy at the present epoch. The total gas mass of the Magellanic Stream \citep[or simply, the Stream][]{wan72a,mat74a}, the Magellanic Bridge \citep[][]{bru05a}, the Magellanic Clouds and the Leading Arm \citep[][]{lu98a,put98a}, is estimated at $2 \times 10^9 ~\Msun \, [r / 55\,\kpc ]^2$ for a median cloud distance $r$ \citep[][]{fox14a}, significantly higher than the mass of all high-velocity clouds (HVCs) around the Galaxy together \citep[$\sim 10^8 \,\Msun$;][]{put12a}. If the gas accretion time onto the Galaxy is fixed by the MC merger time \citep[$\sim 2.5$ Gyr;][]{cau18a}, the gas inflow rate is roughly $1 ~\Msun ~\pyr$, of the same order as the present-day star formation rate of the Galaxy \citep[e.g.][]{rob10a}.

It is however unclear whether and over what timescale the gas will settle onto the Galactic disc. Full simulations of gas-stripped satellites within a realistic Galaxy model that includes a hot corona show that stripped gas may rain onto the Galactic disc on timescales well below 1 Gyr \citep[][]{tep18b}. On the other hand, observations of recombination emission \citep[\Ha;][]{wei96a,rey98a,put03b,mad12a} and ultra-violet (UV) absorption line measurements \citep[][]{fox13a,ric13a,bar17a} indicate that mass budget of the Stream is dominated by ionised gas. A high ionisation fraction may be common to most HVCs \citep[][]{leh11a} in part due to the Galactic UV emission \citep[][]{bla99a,bla01b}, particularly for HVCs close to the disc. For clouds far from the disc, in the absence of alternative ionisation mechanisms \citep[e.g. stellar radiation;][]{ost97a}, the observations are best explained by constrained hydrodynamic models of infalling gas shock-ionised by the interaction with the Galactic corona \citep[`shock cascade' model; q.v.][]{tep15a}. These models indicate that the gas is ionised on a timescale of order a few 100 Myr and will likely evaporate in the corona before reaching the star-forming disc \citep[cf.][]{hei09b}.

The interpretation of the data is complicated by the observation that the Galactic corona is magnetised, with mean field values spanning the range $\sim 0.1 - 10$ \mG\ \citep[][]{sun10a} depending on Galactocentric radius, as is common for spiral galaxies \citep[][]{bec16a}. An ambient magnetic field -- neglected in all models of the Stream to date --
has been long known to provide a mechanism that suppresses the onset of instabilities that ultimately drive the destruction of the infalling gas \citep[][]{mac94a}, thereby increasing its survival chances \citep[for a discussion, see][]{bla09a}.

The Galactic corona, which has been accreted and has evolved over billions of years \citep[][]{van11b}, plays a crucial role in the evolution of gas falling onto the Galaxy. Thus, a full account of its properties is required if we are to arrive at a solid understanding of the Magellanic System. A simple model of the formation of the Magellanic System is based on the paradigm referred to as `first-infall scenario' \citep[][]{bes12a}.\footnote{For a recent review see \citet[][]{don16a}.} However, this model relies on a crude description of the Galaxy; in particular, it neglects the presence of the circumgalactic hot corona.

Earlier models addressed the interaction between gas stripped away from the MCs and the Galactic corona, although using simpler assumptions. Following a purely analytic approach, \citet[][see also \citealt{hel94a,moo94a}]{meu85a} calculated the effect of drag from the Galactic corona on the gas in the Magellanic Bridge, and concluded that this could explain the formation of the Stream. This work was followed by \citet[][]{gar99a}, and extended by \citet[][]{dia11b},  using pure $N$-body models. Both groups modelled the gas drag exerted on the infalling gas by a diffuse, uniform background by adding a drag term to the equation of motion of the otherwise collisionless simulation particles. The first self-consistent $N$-body, hydrodynamical model was presented by \citet[][]{mas05a}, who modelled the interaction between a hot hydrostatic corona around the Galaxy and the LMC, but ignored the presence of the SMC.

Most recently, \citet[][]{ham15a} revisit an idea first expressed by \citet[][]{saw05a}, and argue that the trailing stream and the `Leading Arm' arises from of a convoy of {\em baryon-dominated} tidal dwarf systems, which includes the MCs, that lost gas along their orbit as a result of ram pressure stripping by the intergalactic medium and the Galactic corona. Here, we reconsider this idea as one of the plausible scenarios explaining the leading stream gas, i.e. the gas leads the MCs but does {\it not} emerge from it dynamically.

Even with their adopted simplifications, the previous studies were able to show that the properties of the Magellanic System are sensitive to the presence of an external corona along its orbit. Here we provide a more sophisticated perspective on the evolution of the Magellanic System, in particular of the Stream and the Leading Arm, using an improved model for the Galaxy, which includes -- for the first time -- a magnetised Galactic corona with a range of properties consistent with observations. This is the core of our paper.

We are not attempting to reproduce the properties of the Magellanic System in detail, e.g. the precise location and kinematic properties of the MCs, or the properties of the gas structures such as its total mass, its ionisation state, or the kinematic and structural properties of either the trailing stream or the leading arm. For all the partial success of others in doing so \citep[q.v.][]{don16a}, we maintain that it is futile to pursue such an ambitious task given that the present-day position and kinematics of the MCs, the Stream and the Leading Arm do not unambiguously constrain the required initial conditions, such as the total masses of the MCs and their structure, the inclination of the SMC with respect to its orbit relative to the LMC, their spin, or their full three-dimensional (3D) position and velocity at infall which depend on the unknown Galactic potential. The initial conditions rely on a large number of model-dependent parameters \citep[for a discussion see][]{ham15a}. Since numerical experiments are costly, it is prohibitive to explore in full the available parameter space looking for an optimal solution. Techniques such as genetic algorithms can in fact be used to perform the search more efficiently \citep[][]{gug14a}, but even here the available parameter space needs to be significantly constrained beforehand.

Our approach rests on adapting a recent model \citep[][dubbed `9:1']{par18a} --  based on \citet[][their model 2]{bes12a} -- that broadly reproduces some of the observed properties of the Stream and the Leading Arm. We systematically modify this model to investigate the effect of several factors (dynamical friction, gas drag, confinement by dark matter and magnetic fields) on the evolution of the gas structures associated to the Magellanic System, with particular emphasis on the Leading Arm, in a series of controlled and self-consistent numerical experiments. We argue that the relevant physical processes at play should be captured correctly in our models. Thus, our results should be applicable to other similar systems. 

This paper is organised as follows. In Section \ref{sec:mod}, we describe our adopted base model and variations thereof. In Section \ref{sec:res}, we present the results for each of our models. We discuss the implications of our findings in Section \ref{sec:dis}, and finish with some concluding remarks in Section \ref{sec:con}.

\section{The First-Infall Scenario Revisited} \label{sec:mod}

\begin{table}
\begin{center}
\caption{Structural parameters of the Magellanic Clouds. The parameters values, with exception of $r_{\rm tr}$ and $Z$, are adapted from \citet[][their model 9:1]{par18a}. Column headers are as follows: $M_{\rm t}$, total mass ($10^{9}$~\Msun); $r_{\rm s}$, scalelength (kpc); $r_{\rm tr}$, truncation radius (kpc); $N_{\rm p}$, particle number ($10^{4}$); $Z$, gas metallicity (\Zsun).}
\label{tab:mc}
\begin{tabular}{llccccc}
\hline
\hline
 					& Profile		& $M_{\rm t}$ 	&  $r_{\rm s}$ 	& $r_{\rm tr}$ 	& $N_{\rm p}$  	& $Z$ ~\\
\hline
LMC&&&&&&\\
&&&&&&\\
DM halo$^{\,a}$		& NFW		& 175		& 13			& 75$^{f}$		& 	100		&	--	~\\
Stellar disc$^{\,b}$		& Exp$^{\,e}$	& 2.5			& 2.4	$^{d}$	& 9.6	$^{g}$	&	60		&	--	~\\
Gas disc				& Exp$^{\,c}$	& 2.0			& 9.5			& 38$^{g}$	&	100$^{h}$	&	0.2	~\\
\hline
SMC&&&&&&\\
&&&&&&\\
DM halo$^{\,a}$		& NFW		& 15			& 3.8			& 28$^{f}$		& 	11		&	--	~\\
Stellar disc$^{\,b}$		& Exp$^{\,e}$	& 0.8			& 1.2	$^{d}$	& 4.8$^{g}$	&	6		&	--	~\\
Gas disc				& Exp$^{\,c}$	& 2.0			& 4.7			& 19$^{g}$	&	36$^{h}$	&	0.1	~\\
\hline
\end{tabular}
\end{center}
\begin{list}{}{}
\item {\em Notes}. NFW, \citet[][]{nav97a} profile; Exp, Exponential profile\\
$^{a\,}$Mass enclosed within $r_{\rm tr}$ is $\sim 1.3\times10^{11}$~\Msun\ (LMC) and $\sim 1.3\times10^{10}$~\Msun\ (SMC).\\
$^{b\,}$The stellar metallicity is ignored as it is of no relevance for our study.\\
$^{c\,}$Exponential profile in $R$. Scaleheight set by vertical hydrostatic equilibrium, initially at $T = 10^4$ K.\\
$^{d\,}$Scaleheight set to 0.48 kpc (LMC) and 0.25 kpc (SMC).\\
$^{e\,}$Exponential profile both in $R$ and $z$.\\
$^{f\,}$Corresponding tidal radius at $d = 220$ kpc from a  $1.5\times10^{12}$~\Msun\ host.\\
$^{g\,}$Extension set to 4$\times$ the scalelength.\\
$^{h\,}$Number of particles used to sample the density profile before mapping onto the simulation grid.
\end{list}
\end{table}

\subsection{The base model} \label{sec:mod1}

Our fiducial model follows the 9:1 model by \citet[][]{par18a}. This model is constructed essentially in two steps: i) The LMC and SMC are evolved for $\sim 6$ Gyr as a binary pair in isolation starting from some prescribed initial conditions; ii) The evolved binary system is placed into a Galaxy-like potential at one Galactic virial radius ($\sim 220$ kpc), and evolved further for roughly 1 Gyr (see Appendix \ref{sec:ver} for further details). The gravitational effect of the Galaxy on the MCs is approximated by a {\em static} dark matter (DM) host halo parametrised by a spherical \citet[][NFW]{nav97a} profile with a virial mass of $\sim 1.5\times10^{12}$~\Msun, and a scale radius of $\sim 20$ kpc (corresponding to a concentration of 12). In contrast, the MCs are modelled as fully live, multi-component galaxies, each consisting of a stellar disc and a gas disc embedded in a DM host subhalo. The structural parameters of the MCs are summarised in Table \ref{tab:mc}. 

In this model, the MCs have -- by design -- large initial gas fractions, implying that higher gas masses are deposited onto the structures that later form the Stream and the Leading Arm, compared to models with lower initial gas fractions. Yet the column densities along the Stream and the Leading Arm in this model are lower than observed \citep[see][for a discussion]{par18a}. We note that our choice of simulation code (grid-based rather than SPH; see Section \ref{sec:res}) forces upon us the need to define an arbitrary truncation radius for each model component in order for them fit within the simulation volume.\footnote{The truncation radius needs to be chosen with care, given that orbital history of the SMC around the LMC is very sensitive to their masses (dynamical friction), and the truncation radius affects the {\em actual} mass of the corresponding component. We have chosen values that most closely reproduce \citet[][]{par18a}'s 9:1  model. } We refer to the base model as `Static DM halo' model.

\subsection{Extended models} \label{sec:mod2}

We consider, in addition, the following stepwise improvements of the Galaxy model with respect to the starting point described above.\footnote{These modifications only affect the evolution of the Magellanic System at infall; in other words, the evolution of the binary pair in isolation is identical in all the models considered here.} These `extended' models are summarised in Table \ref{tab:mw}.

First, we remove the restriction of a {\em static} DM host halo and approximate the gravitational effect of the Galaxy onto the MCs using a fully {\em live} DM host halo with identical structural properties. This modification allows to study the effect of dynamical friction on the orbital history of the MCs and its impact on the formation of the Stream and the Leading Arm. This model is dubbed `Live DM halo' model.

As an extension of the latter model, we include a static (i.e. non-rotating), spheroidal Galactic corona. The inclusion of this component in our model makes it possible to explore the impact of gas drag along the orbit of the MCs on the infalling gas within a realistic Galaxy model in a self-consistent manner.

\begin{figure}
\centering
\includegraphics[width=0.45\textwidth]{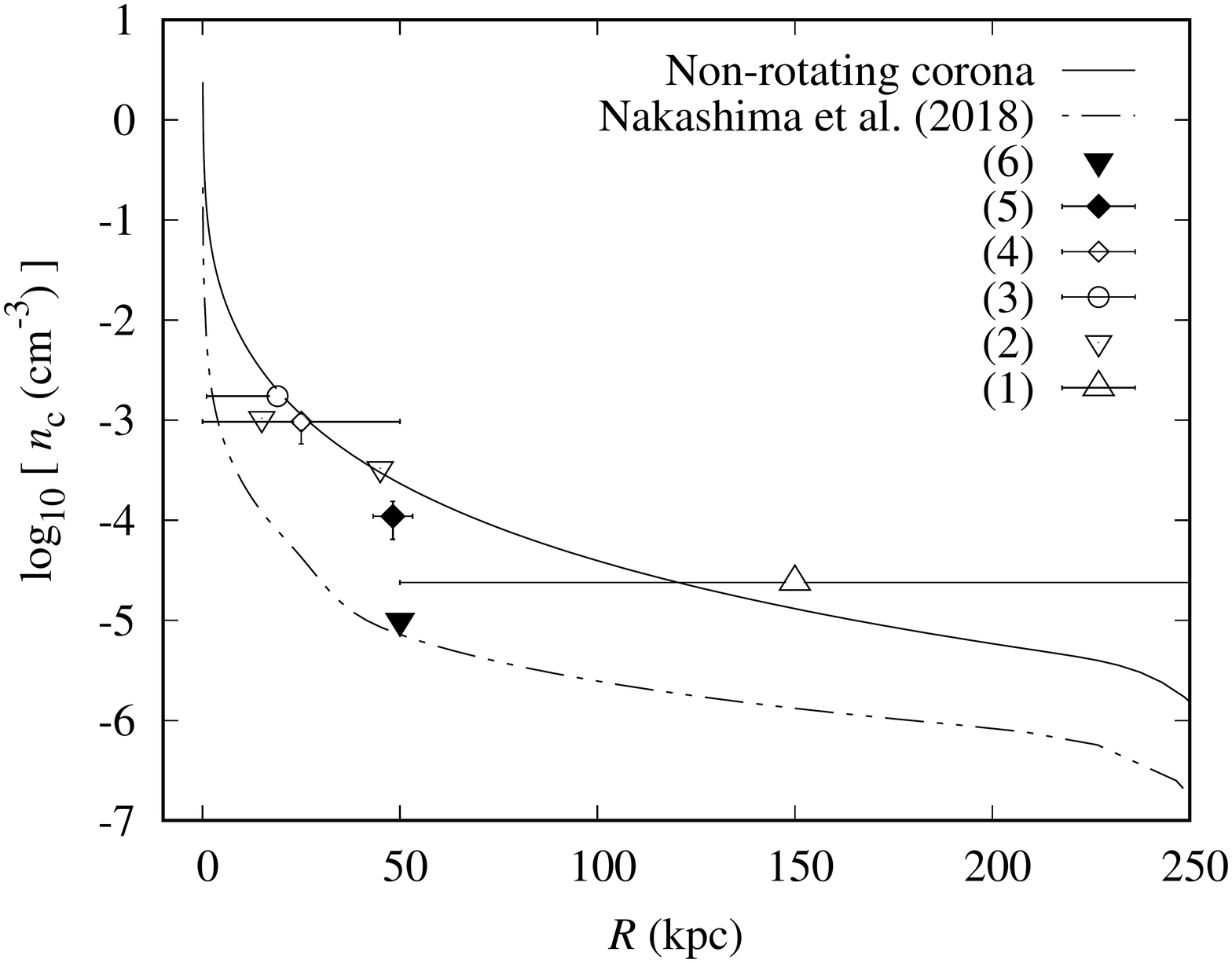}
\vspace{-5pt}
\caption[ Coronal density profile ]{ Density profile of the Galactic corona (at $t_{\rm sim} = 0$ Gyr; solid curve). The initial profile for the `Slow corona' and the `Fast corona' models is identical to the `Non-rotating corona' model and are therefore omitted here. Data points and corresponding reference key list is as follows. 1: \citet[][]{bli00a}; 2: \citet[][]{sta02a}; 3: \citet[][]{bre07b}; 4: \citet[][]{and10a}; 5: \citet[][]{sal15a}; (6) \citet[][]{mur00a}. Note that downward (upward) pointing triangles indicate upper (lower) limits. The double dotted-dashed curve corresponds to the model by \citet[][see Section \ref{sec:dis}]{nak18a}. }
\label{fig:dens}
\end{figure}

\begin{figure}
\centering
\includegraphics[width=0.45\textwidth]{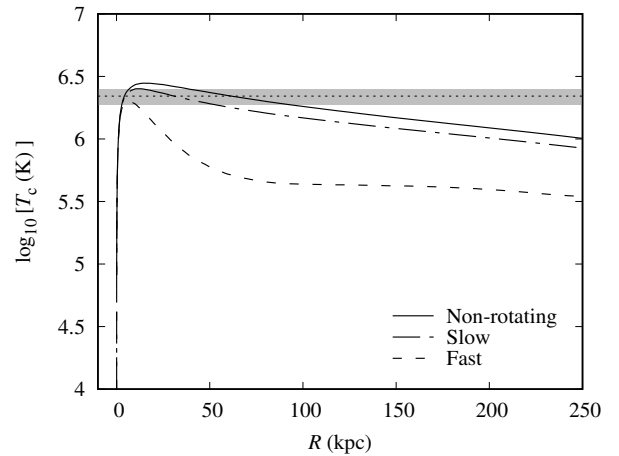}
\vspace{-5pt}
\caption[ Coronal density profile ]{ Temperature profile of the Galactic corona (at $t_{\rm sim} = 0$ Gyr) in our Galaxy models. The dotted horizontal line and the shaded area indicate the mean temperature of the Galactic corona and its associated uncertainty, $\sim (2.2\pm0.3)\times10^6$ K, inferred from observations \citep[][]{hen13e}. }
\label{fig:temp}
\end{figure}

\begin{figure}
\centering
\includegraphics[width=0.45\textwidth]{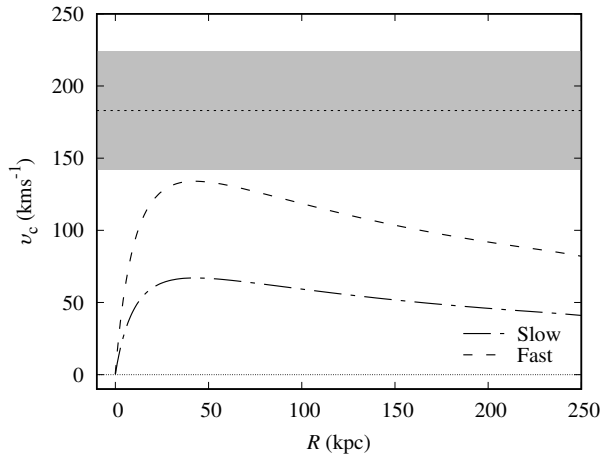}
\vspace{-5pt}
\caption[ Coronal density profile ]{ Rotation velocity profile of the Galactic corona (at $t_{\rm sim} = 0$ Gyr) in our Galaxy models. The net rotation speed in the model featuring a static corona is trivially zero everywhere and is therefore omitted here. The dotted horizontal line and the shaded area indicate the mean velocity of the Galactic corona and its associated uncertainty inferred from observations \citep[$181\pm41$ \kms;][]{hod16a}. The rotation curves indicate the speed in the plane of the Galaxy ($z = 0$); the rotation speed decreases with distance from the plane (not shown). }
\label{fig:vrot}
\end{figure}

Our model corona follows initially the same profile and extension as the DM host halo, adopting identical structural parameters. It has a total initial mass of $\sim 3\times10^{10}$~\Msun\ within 250 kpc, and properties (density profile, temperature profile) in broad agreement with the properties of the Galactic corona as inferred from a number of observations \citep[Figures \ref{fig:dens} - \ref{fig:vrot}; q.v.][]{bla16a}. It is worth noting that our coronal model is consistent with the most recent model of the Galactic hot halo by \citet[][]{bre18a}. We refer to this model as `Non-rotating corona'. 

Recent kinematic models based on x-ray spectra suggest that the Galactic corona is spinning fast \citep[$181\pm41$ \kms\ out to $R \approx 50$ kpc;][]{hod16a}. Such a rapidly rotating corona can be accommodated within barotropic \citep[][]{pez17a} and baroclinic \citep[][]{sor18a} equilibrium models within the context of cosmologically motivated angular momentum distributions. Therefore, we improve upon our `Non-rotating corona' model by allowing the corona to have net rotation over a range of radii, $v_{\rm c}(r)$. Thus we are able to explore for the first time the effect of the torque exerted by the corona on the gas streams associated with the Magellanic System. 

\begin{table}
\begin{center}
\caption{Overview of Galaxy models.}
\label{tab:mw}
\begin{tabular}{lcccc}
\hline
\hline
 	Name			& DM halo		& Corona	& Peak $v_{\rm c}$$^a$	& $B_{\rm c}$$^b$	\\
\hline
Static DM	halo			& Static		& --		& --					& --			\\
Live DM halo			& Live		& --		& --					& --			\\
Non-rotating corona		& Live		& Yes	& 0					& 0			\\
Slow corona			& Live		& Yes	& 70					& 0			\\
Fast corona			& Live		& Yes	& 135				& 0			\\
Magnetised corona$^c$	& Live		& Yes	& 0					& 0.1			\\
\hline
\end{tabular}
\end{center}
\begin{list}{}{}
\item {\em Notes}. Live DM haloes are sampled with $5\times10^5$ particles. Galactic coronae follow initially the same profile as the DM halo (see text).\\
$^a$In \kms. The full rotation curve is displayed in Figure \ref{fig:vrot}.\\
$^b$Initial, uniform magnetic field (in \mG) directed along the $z$ axis.\\
$^c$Non-rotating corona
\end{list}
\end{table}

We set the rotation speed of the corona according to the specific angular momentum of the Galactic DM halo, which follows roughly its cumulative mass profile \citep[][]{bul01b}, adopting a fixed value for the spin parameter $\lambda$. We consider two variants: a `Slow corona' model and a `Fast corona' model, with peak rotation speeds in the plane of the Galaxy\footnote{In our models, the rotation speed decreases with distance from the plane \citet[cf.][]{sor18a}.} ($z = 0$) of $v_{\rm c} \approx 70$ \kms\ and $v_{\rm c} \approx 135$ \kms, corresponding to $\lambda = 0.08$ and $\lambda = 0.16$, respectively  These values are at the high end of the typical range for $\lambda$ found at $z = 0$ in $N$-body, hydrodynamic cosmological simulations of structure formation \citep[][]{zju17a}. The peak rotation speed in the `Fast corona' model is marginally consistent with the low end of the Galactic coronal rotation speed inferred from observations (Figure \ref{fig:vrot}). A higher rotation speed would lead to a coronal temperature that is too low to be consistent with the temperature inferred from x-ray data. This is a consequence of the condition of thermal equilibrium imposed on the model corona, which naturally leads to lower mean temperatures for higher net rotation speeds (Figure \ref{fig:temp}). 

The radial velocity dispersion of the gas, implied by the spherically symmetric \citet[][]{jea15a}'s equation with isotropic velocity distribution, is given by
\begin{equation}
	\sigma^2_r(r) = \frac{ 1 }{ \rho_{\rm c}(r) }	\int_r^\infty \rho_{\rm c}(r') \frac{ \partial \Phi(r') }{ \partial r' }d r' \, ,
\end{equation}
where $\rho_{\rm c}(r) =  \mu ~m_p ~n_{\rm c}(r)$ describes the mass density profile of the corona, $n_{\rm c}$ is the corresponding particle density, and $\Phi$ is the total potential of the Galaxy.
The specific internal energy of the gas, corrected for its net rotation speed $v_{\rm c}(r)$, is given by
\begin{equation}
	e \equiv \frac{ 1 }{ \gamma -1} \frac{ k T_{\rm c}}{ \mu m_p }=  \frac{ 1 }{ (\gamma -1)  } ( \sigma_r^2 - v_{\rm c}^2 ) \, .
\end{equation}
Thus, for a fixed $\rho_c$ and $\Phi$, there is a maximum rotation speed such that $T_{\rm c}$ is consistent with observations. A higher rotation speed requires either: i) abandoning the assumption of thermal equilibrium, which makes it significantly harder to model the corona; or ii) adopting a more massive DM host halo for the Galaxy, ruled out by a number of observations \citep[e.g.][]{pif14a,kaf14a,die17b}.

The last extended model we consider corresponds to a weakly magnetised version of our `Non-rotating corona' model, referred to as the `Magnetised corona' model.  We adopt a magnetic field initially uniform along the $z$ axis set initially to 0.1 \mG\ everywhere. We are aware that this is not entirely consistent with the structure of the magnetic field around the Galaxy \citep[e.g.][]{jan12a}. In particular, the field is stronger (weaker) compared to the observed field far away (close) to the Galactic Centre. However, our goal is not to model the evolution of the field \citep[see e.g.][]{rie16a} but to explore the effect of an initially weak ambient magnetic field on the infalling gas with the simplest possible configuration.\footnote{We note that the magnetic pressure, $P_B = B^2/8 \pi$, is negligible compared to the thermal pressure in the corona, $P_{\rm c}$, i.e. $\beta  = P_{\rm c} / P_B \gg 1$, such that the presence of a magnetic field does not affect the thermal properties of the model corona. } The use of a more realistic field structure is left for future work.

It is worth emphasising that while the gas in and around the MCs is known to be magnetised \citep[q.v.][]{kac17b}, we have chosen to ignore this circumstance in order to cleanly isolate the effect of a magnetised medium (the Galactic corona) on non-magnetised infalling gas clouds (the gas associated to the MCs).

For the current work, it was unnecessary to adopt more sophisticated Galaxy models (e.g. models that include a stellar bulge, a stellar disc, or a gas disc) as we have done in previous work \citep[][]{tep18a,tep18b}. While these baryonic components may generally dominate the potential close to the Galactic Centre, their gravitational effect is marginal at the mean distance of the MCs today \citep[$r_{\rm MC} \approx 55$ kpc;][]{wal12a,gra14a}. 
Nonetheless, the gas disc may have an important effect: it may negatively affect the lifetime of gas structures that interact with it \citep[e.g.][]{gal16a}. In view of our results (yet to be discussed), we opt to ignore this component for now, and leave more complex Galaxy models for future work.

%

%

%

\subsection{Numerical setup} \label{sec:set}

The initial conditions (i.e. the particle positions and their velocities; and additionally the internal energy for the gas components) for each of our models described above are generated with the \dice\ code \citep[][]{per14c}.\footnote{The sequence of geometric and kinematic transformations used to map the evolved, isolated MC binary pair onto the Galaxy is described in Appendix \ref{sec:ver}. See also footnote \ref{foo:code}.}

The evolution of the MCs in isolation and its subsequent infall into the Galaxy are calculated from the corresponding initial conditions using \ramses, an $N$-body, magneto-hydrodynamic (MHD) grid code with adaptive mesh refinement \citep[AMR; version 3.0 of the code last described by][]{tey02a}. For the run adopting a magnetised corona, we exploit the MHD capabilities of \ramses, which implements the constrained transport, divergence-preserving (to numerical precision) MHD solver of \citet[][see also \citealt{fro06a}]{tey06a}. The contribution to the overall gravitational field of every component (gaseous or collisionless) is taken into account at all times by solving the Poisson equation\footnote{The source term is given by the sum of the individual component densities at each point. } on the AMR grid using the multi-grid scheme of \citet[][]{gui11a}.  The simulations of the isolated MCs and the composite system (MCs and Galaxy) are run in a cubic volume of 500 kpc and 1 Mpc length per side, respectively.

The base grid level is set at 7, corresponding to a physical spatial resolution of roughly 4 kpc (8 kpc) in our 500 kpc (1 Mpc) box. We adopt a maximum refinement level of 13 when calculating the evolution of the MCs in isolation, and 14 for the composite system. The physical spatial resolution is $\gtrsim 60$ pc in either case. The minimum cell size we adopt is thus well below the typical linear sizes of clouds along the Stream inferred from radio observations \citep[][]{hsu11a}.

The grid is refined (coarsened) at runtime whenever the gas mass in a cell exceeds (falls short of) $\sim 2.5 \times 10^6 ~\Msun$ or the number of collisionless particles is larger (smaller) than 40. Also, cell refinement is applied in order for the \citet[][]{jea15a} length to be resolved at all times with at least four cells \citep[][]{tru97a}.

In addition to the above refinement criteria, we have used a set of fixed, nested, spherical grids centred on the model Galaxy of increasing size (with radii ranging from 0.4 kpc to 450 kpc) and decreasing resolution (by a factor 2 from grid to grid) moving from the centre outwards.\footnote{This approach is similar to the so-called {\em Enhanced Halo Resolution} (EHR) used by \citet[][]{hum18a}. } The innermost level has a resolution equal to our maximum adopted refinement level (14). This approach was necessary in order to roughly maintain the same resolution across the simulation volume in models with a live DM halo compared to the model with a static DM halo.  For the latter, the AMR grid is refined using a density-based approach using the {\em analytic} DM density profile. In simulations with a live DM halo, the refinement strategy associated with collisionless components is governed by the number of particles within a given voxel, which depends on the actual $N$-body realisation of the DM density profile and the adopted mass resolution.\footnote{The setup files used to create initial conditions as well as those containing the details such as the hydro solver, the refinement technique, etc. used to run our simulation are freely available upon request to the corresponding author (TTG). \label{foo:code}}\\

We include radiative gas cooling by hydrogen, helium, and heavier elements following \citet[][]{the98b}. The gaseous components are all taken to be monoatomic ($\gamma = 5/3$), ideal gases characterised by an hydrogen fraction $\XH = 0.76$ and an initially uniform metallicity $Z$ that varies from component to component (Table \ref{tab:mc}). The metallicity of the Galactic corona is set to $Z = 0.3$~\Zsun\ \citep[][]{mil15a}. The average gas metallicity of the MCs today (as given by the abundance of $\alpha$ elements) is $Z \approx 0.3$~\Zsun\ (SMC) and $Z \approx 0.4$~\Zsun\  \citep[LMC;][]{rus92a}. Chemical evolution models predict that the gas phase abundances in these galaxies 1 - 2 Gyr ago were roughly 0.2 -- 0.3 dex lower than today \citep[][]{har04a,har09a}. Therefore, we set the initial gas metallicity of the SMC (LMC) to $Z = 0.1~\Zsun$ (0.2~\Zsun).\footnote{We ignore the value of the gas metallicity adopted by \citet[][]{par18a}. } Our approach neglects the chemical enrichment of the gas via stellar feedback. A higher metallicity would enhance the cooling of the bound gas allowing the gas to sink further into the potential well of the dwarfs, making stripping more difficult. Energetic stellar feedback -- which we ignore; see below -- would have the opposite effect. For the time being we simply assume that these effects roughly balance out. We note that this assumption does not affect the outcome of our experiments as we shall discuss later.
 
In contrast to \citet[][]{par18a}, we do not include star formation (SF) in our models. However, this process does not appear to have a significant impact on the overall properties of the Magellanic System in their model. Indeed, as we show in Section \ref{sec:res} and in Appendix \ref{sec:ver}, we are able to broadly replicate their results in spite of our ignoring this process. Since we do not include SF, we neglect the associated feedback, as do \citet[][]{par18a} and \citet[][]{bes12a}. While stellar feedback and their associated outflows in the LMC subject to the ram pressure of the Galactic corona may contribute to the formation of the Stream, its relative importance in terms of mass is minor \citep[][]{sal15a,bus18a}.

We neglect the ionising effect of UV radiation field of the Galaxy or the MCs on the gas. This is justified since we are not interested here in the ionisation state of the gas, which has negligible impact on the magnitude of gas drag (see Section \ref{sec:gas}). Also, the heating of the ionising radiation on the gas is expected to be small \citep[e.g.][]{nic11a}.\\

In a nutshell, our experiments consist of full 3D, $N$-body, magneto-hydrodynamic, grid-based (AMR), non-adiabatic, numerical calculations of the infall of the MCs into the Galaxy that take into account the self-gravity of each component.

\begin{figure*}
\centering
\includegraphics[width=0.33\textwidth]{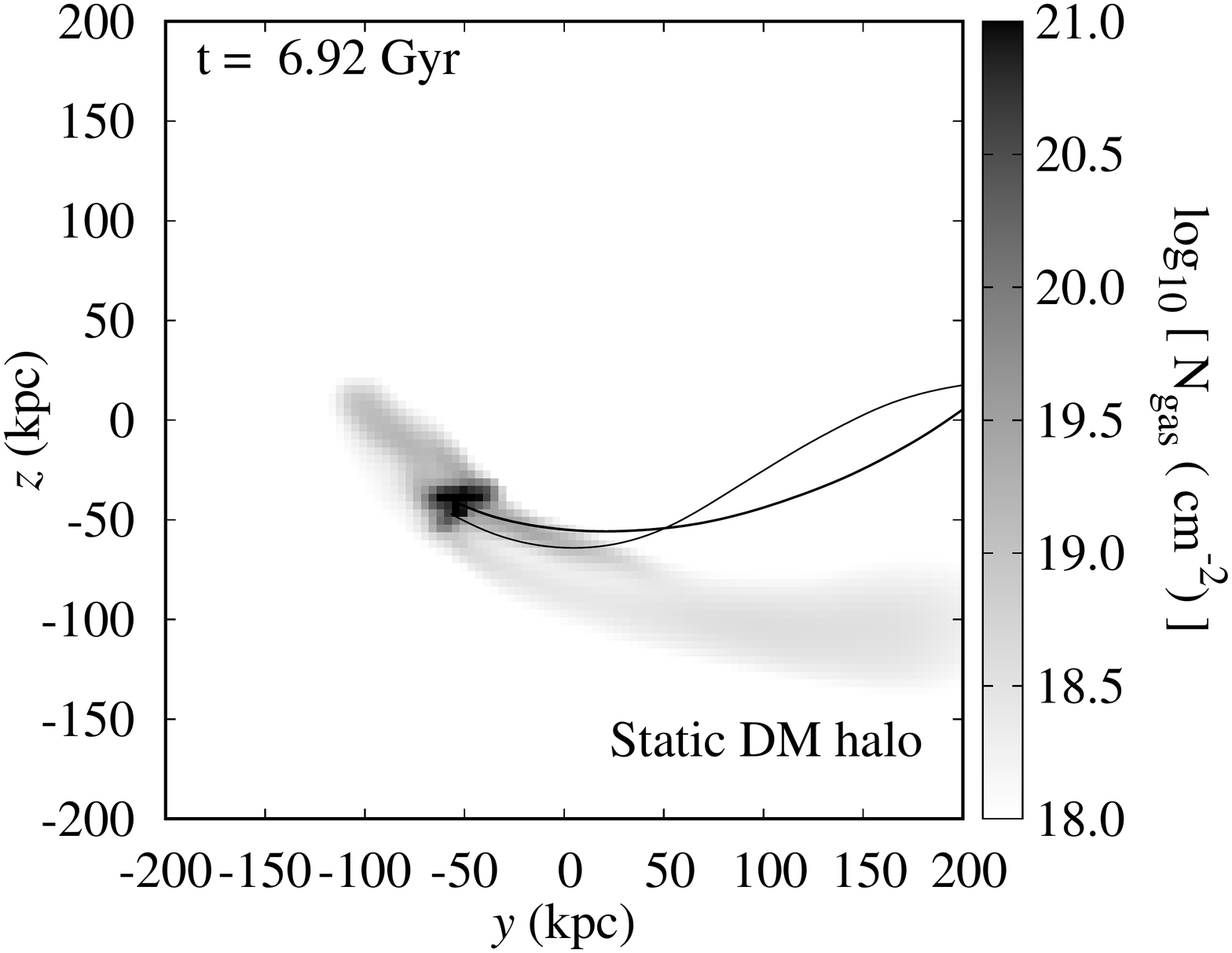}
\includegraphics[width=0.33\textwidth]{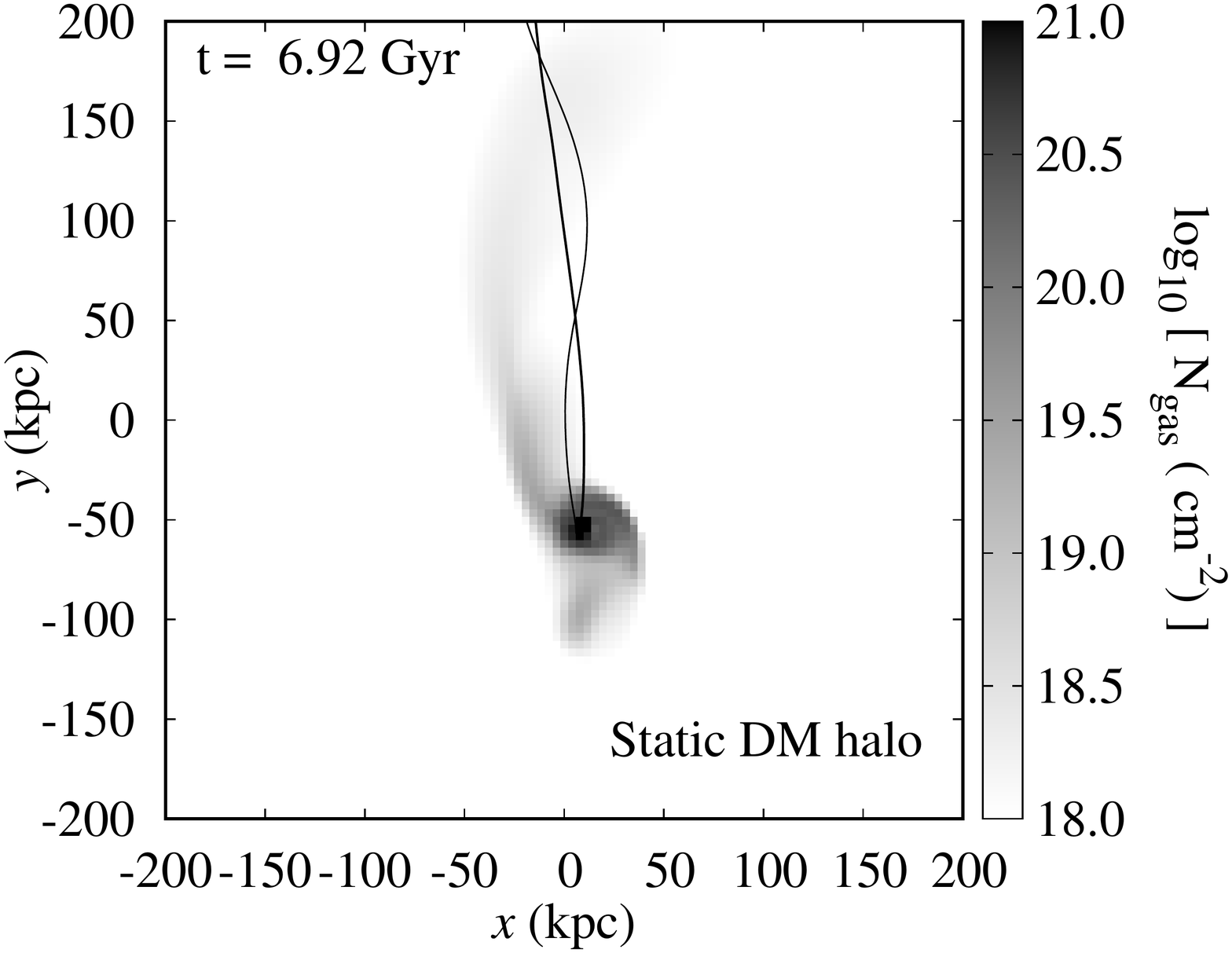}
\includegraphics[width=0.33\textwidth]{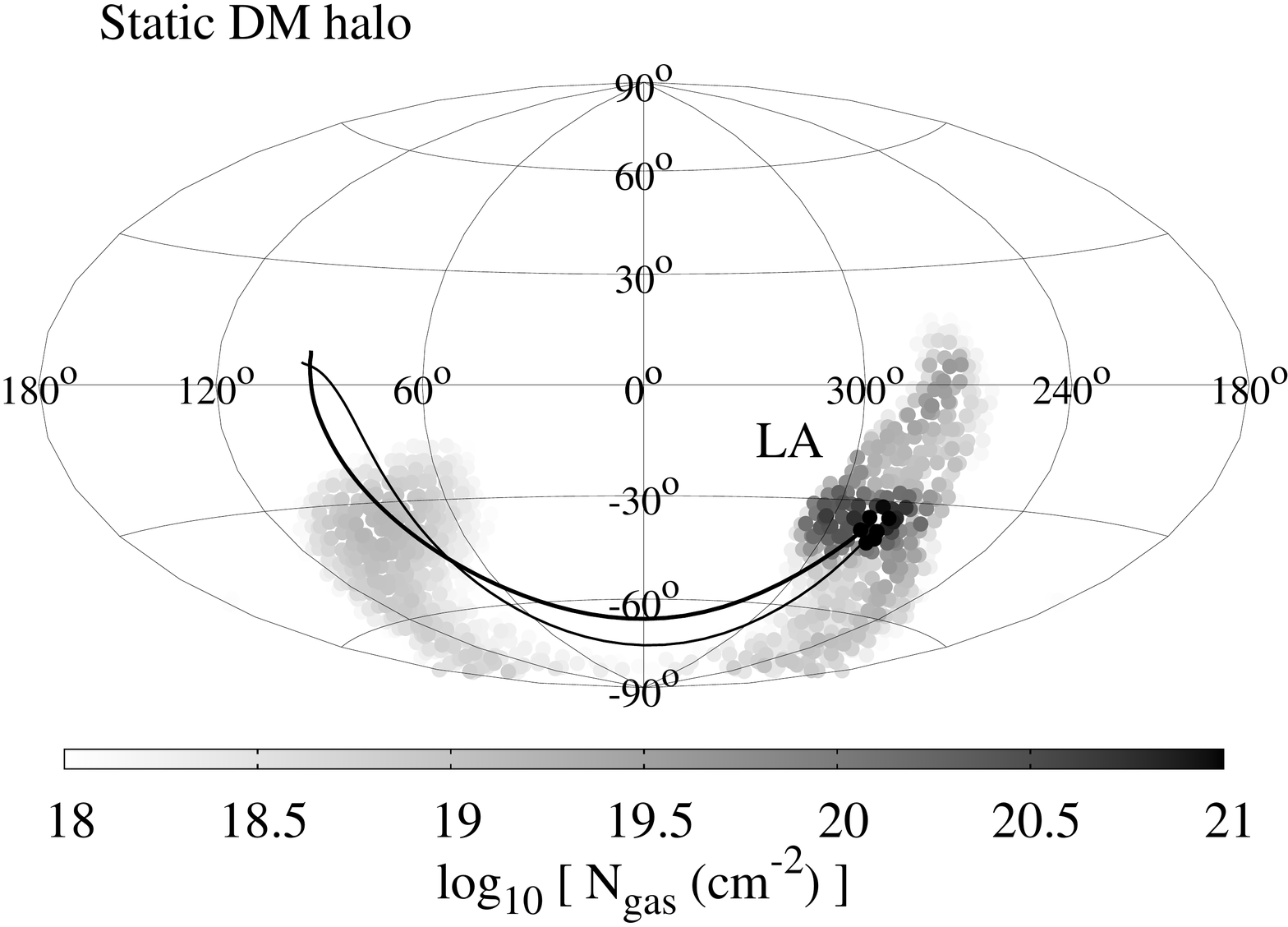}\\
\includegraphics[width=0.33\textwidth]{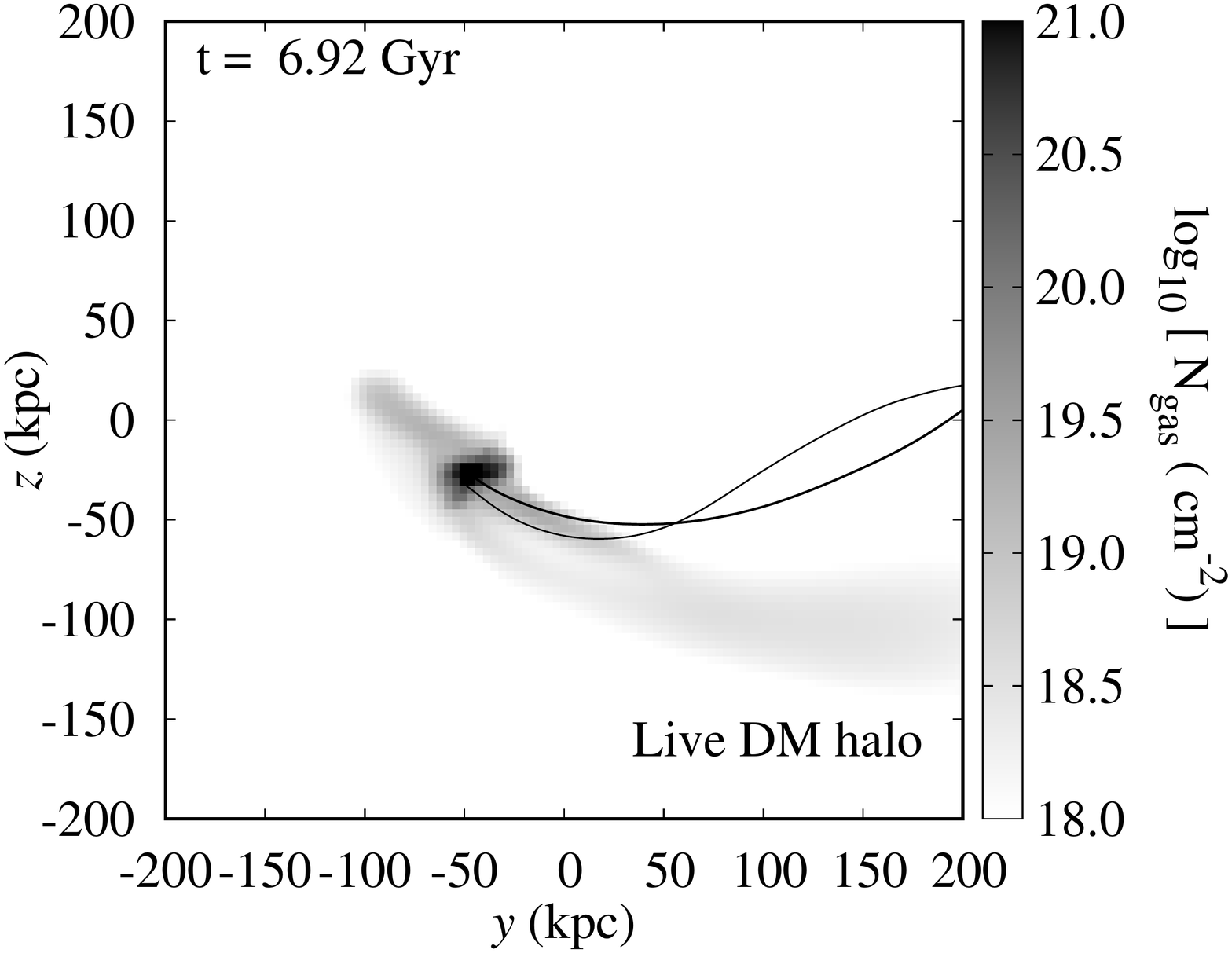}
\includegraphics[width=0.33\textwidth]{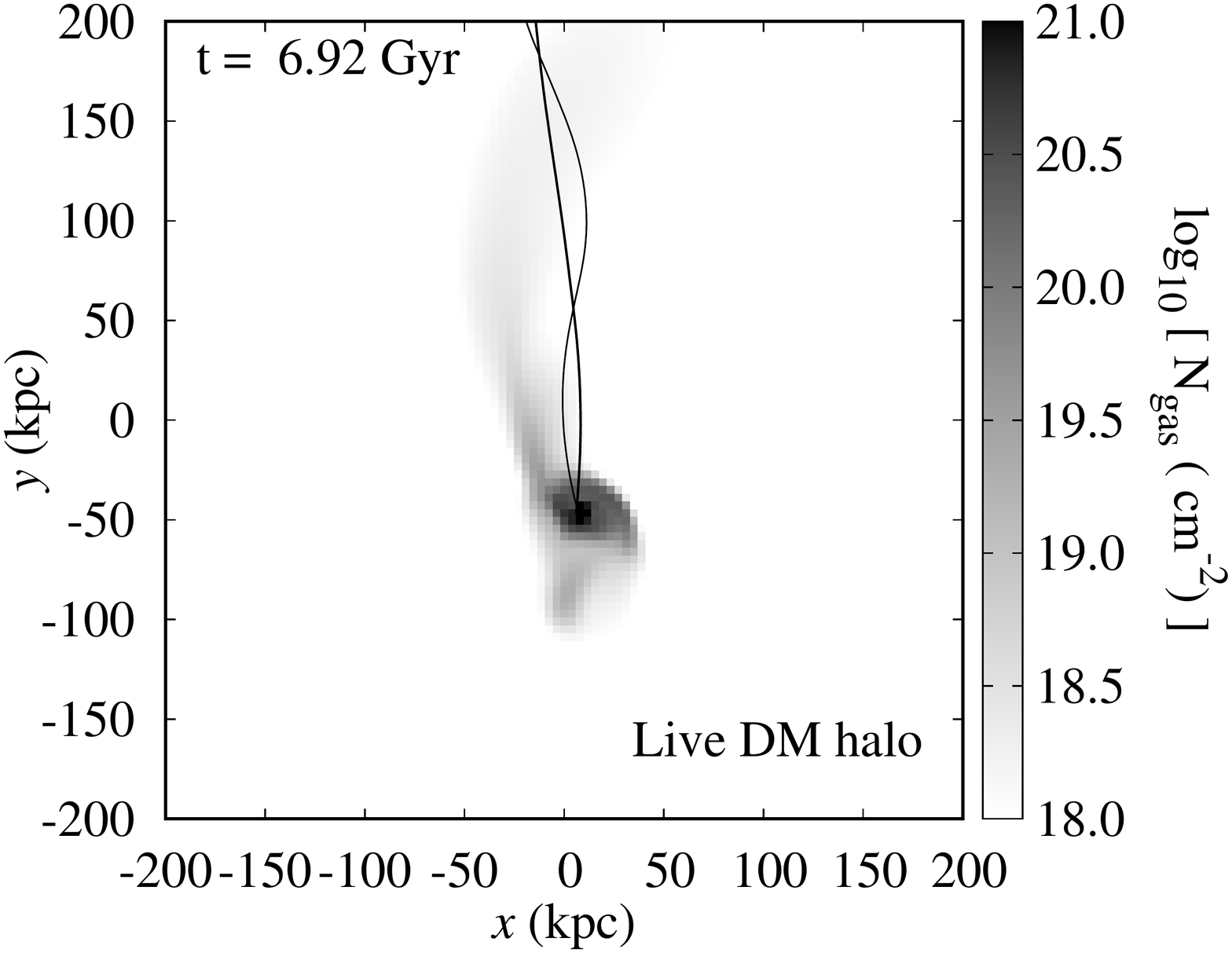}
\includegraphics[width=0.33\textwidth]{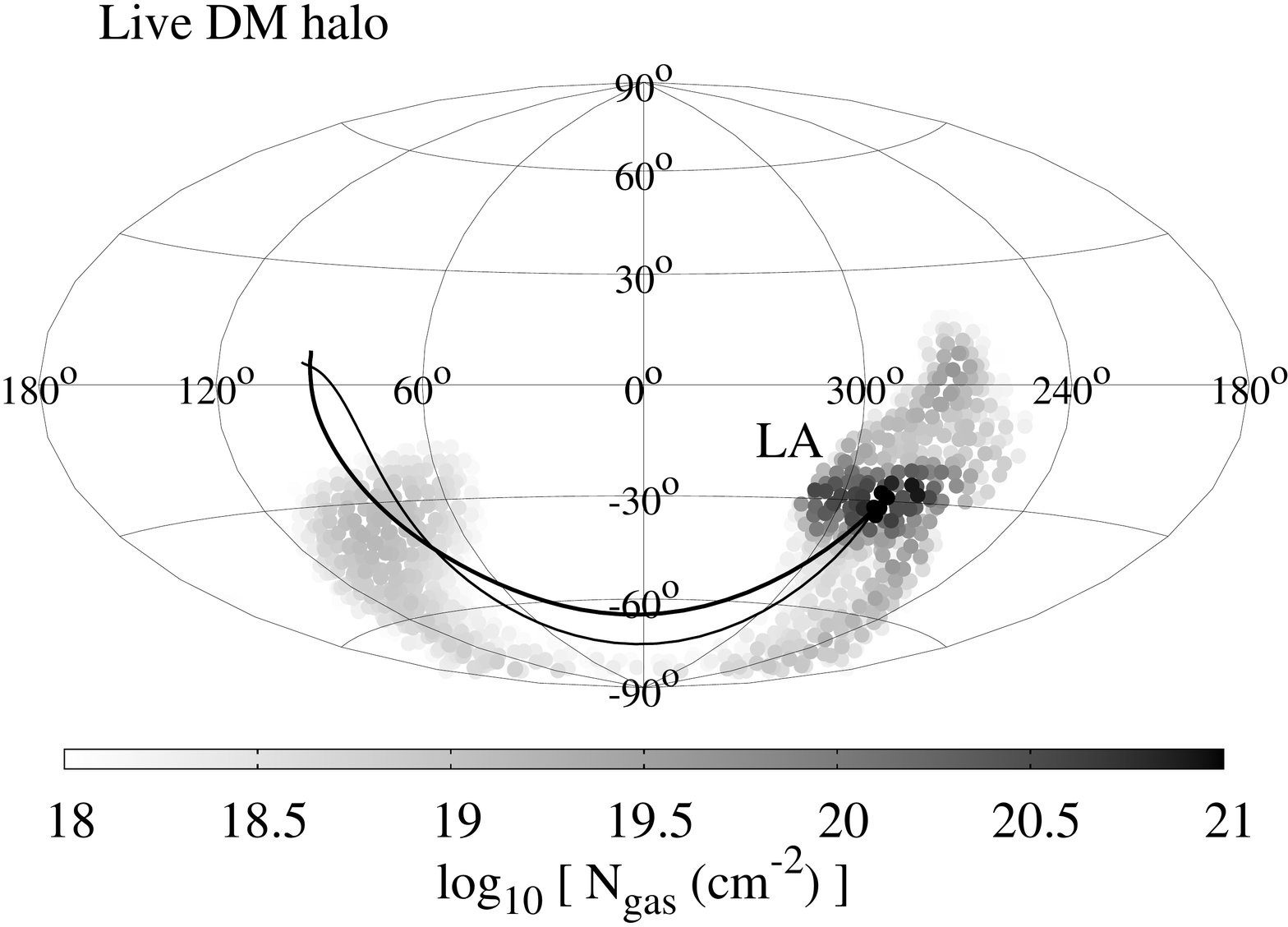}\\
\caption[  ]{ Distribution of gas in the the Magellanic System at the present epoch, i.e. $\sim 1$ Gyr after infall into the Galaxy, in the {\em absence} of a hot corona. Each row corresponds to a different model. Left~/~centre: Total gas column density in physical space on a face-on (left) and edge-on (centre) projection with respect to the Galactic plane. The Galactic Centre (GC) in model `Static DM halo is at (0,0,0) kpc, by construction, while it is roughly at (0,-10,0) kpc in the `Live DM halo' as a result of the Galaxy's recoil. Right: Total column density projected on the sky as seen from the Sun in Galactic coordinates $(l, b)$. The GC is at $(0,0)\deg$ in both models; the MCs', at $\approx (300, -30)\deg$; their space orbits are indicated by the thin / thick solid curves in all panels. Apparently, dynamical friction does not significantly affect the orbit of the MCs and consequently the formation of the Magellanic System.  }
\label{fig:ms}
\end{figure*}

\section{The evolution of the Magellanic System} \label{sec:res}

Quite generally, we find in all the models considered here that the MCs have present-day positions and velocities broadly consistent with their observed values, in agreement with \citet[][]{par18a}. Similarly, we find that the Stream roughly matches its observed extension and position across the sky. However, we are led to some striking conclusions with respect to the formation of the Leading Arm.

Our main findings\footnote{In order to keep the image file sizes manageable, all the physical space figures have been created using up to 8 (out of 14) AMR refinement levels; this implies that all the quantities in the simulation cube (gas density, magnetic field, etc.) are in fact slightly higher than shown the corresponding figures.} are summarised in Figures \ref{fig:ms}, \ref{fig:ms2}, and \ref{fig:ms3}. There, we show the distribution of the gas associated with the Magellanic System, i.e. initially bound to the MCs, at the present epoch for each our models. Each row corresponds to a different different Galaxy model, and consists of three panels, which display, respectively, the gas distribution both in physical space projected along two orthogonal directions, and in the observed space. It is worth emphasising that the column densities shown presented here correspond to {\em total} gas densities, and thus put a strict upper limit on the expected \HI\ column density of the gas (which we do not consider here). The simulation results in physical space are mapped onto observed space by assuming the Sun is located at $(x,y,z)=(-8,0,0)$ kpc at all times, and the North Galactic Pole points along the $z$ direction. We do not adopt the exact solar coordinates \citep[cf.][]{bla16a} since we are not interested here in reproducing the precise location of the Magellanic System and its components, as stated above.

In all models, we find that the total gas mass of the trailing gas stream {\em alone} is $\lesssim 10^9$ \Msun, at the low end of the range of masses of the Magellanic Stream estimated using ultra-violet absorption spectroscopy of bright background sources \citep[][]{fox14a}. Our result is agreement with the models by \citet[][]{bes12a} and \citet[][]{par18a}.

We now discuss in detail the results for each of our models and their differences where relevant.

\begin{figure*}
\centering
\includegraphics[width=0.24\textwidth]{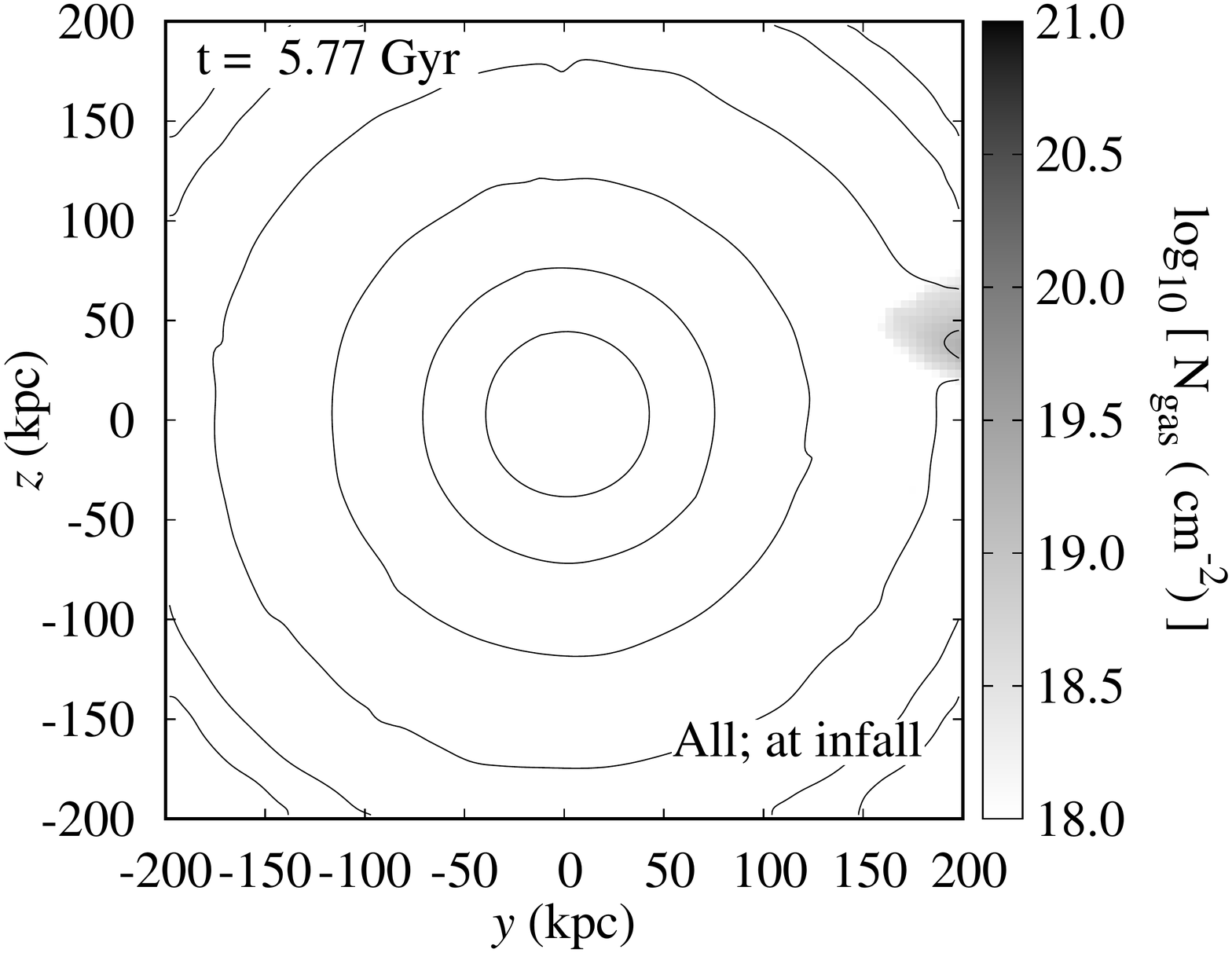}
\includegraphics[width=0.24\textwidth]{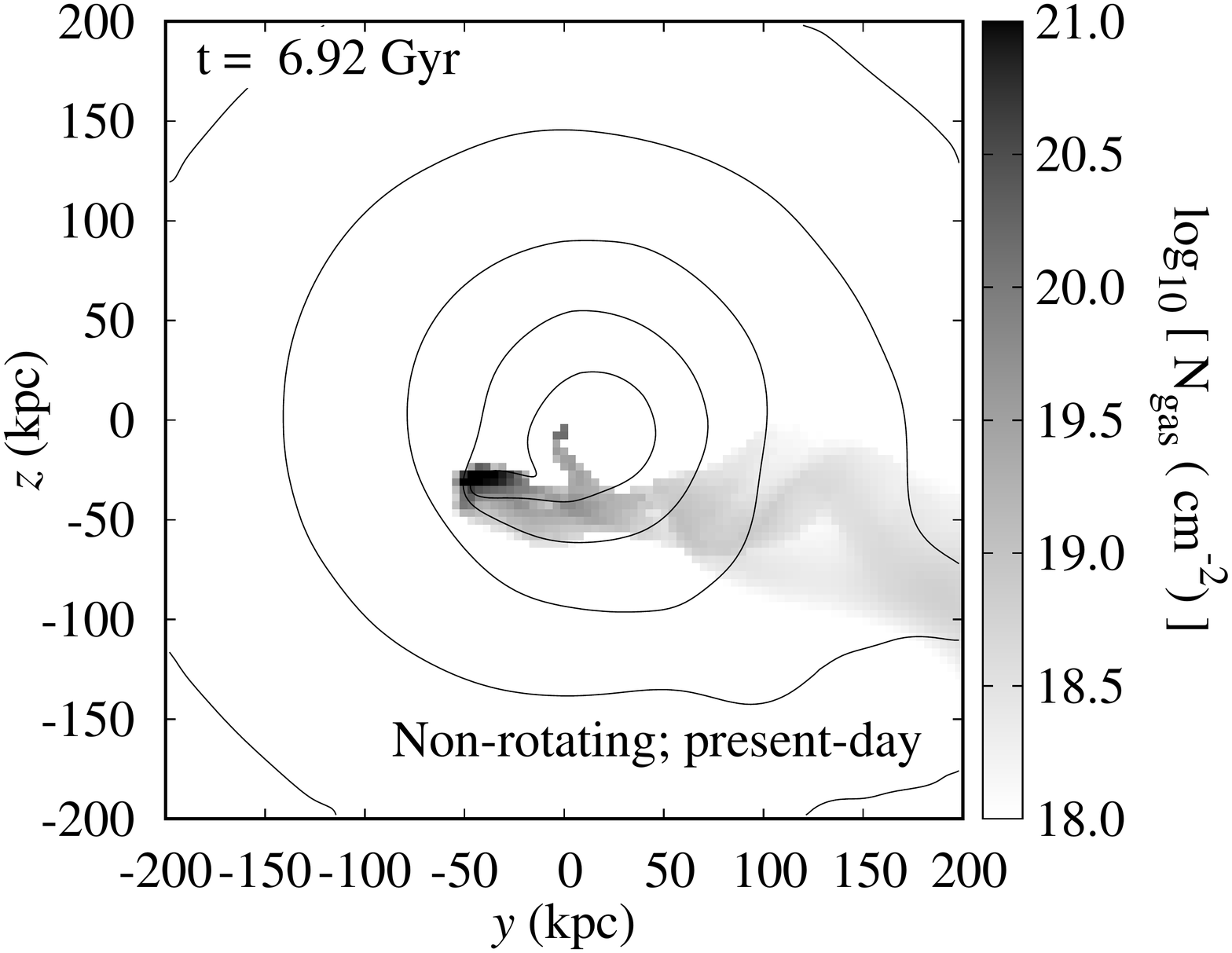}
\includegraphics[width=0.24\textwidth]{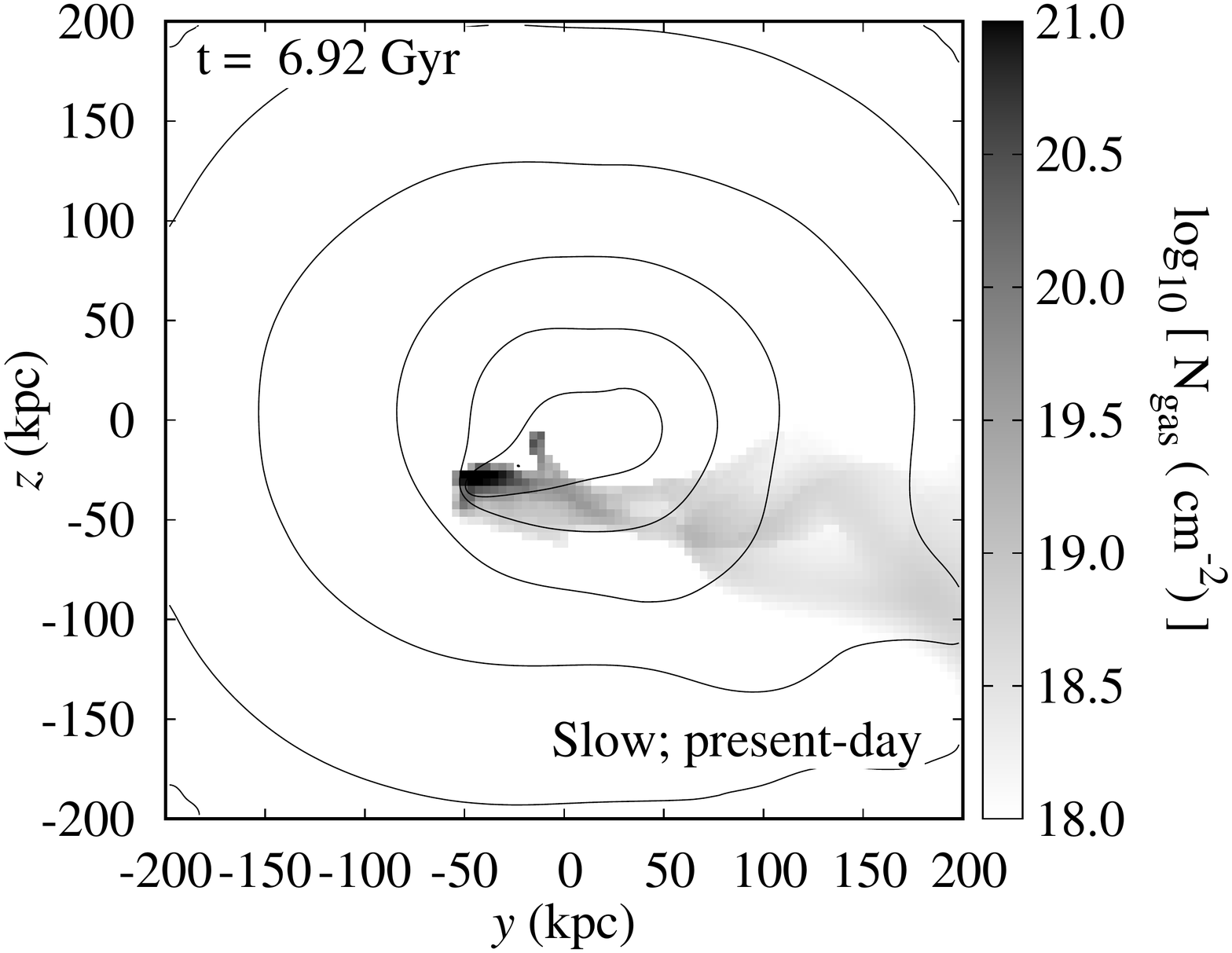}
\includegraphics[width=0.24\textwidth]{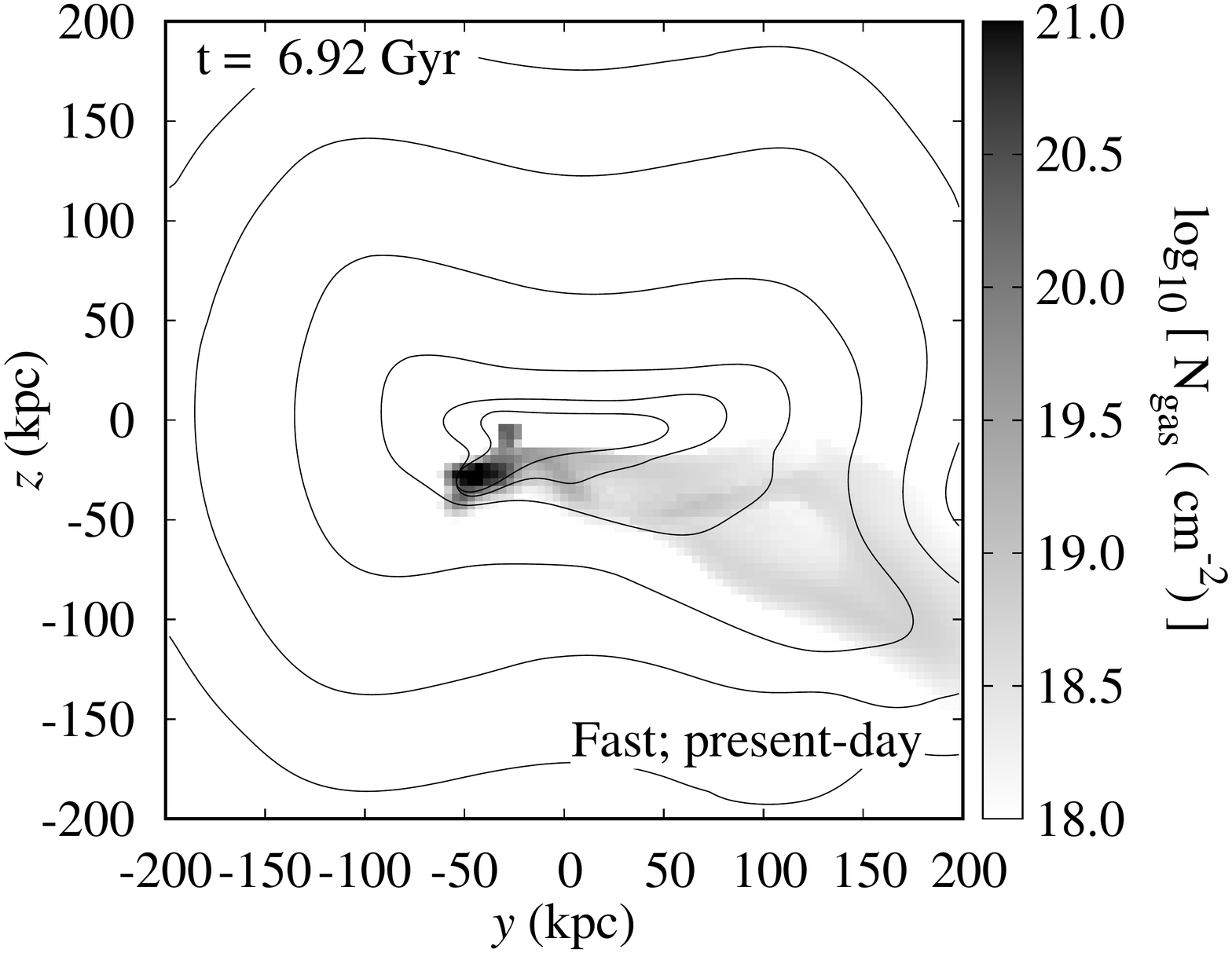}
\caption[  ]{ Density structure of the model Galactic corona. The contours outline the projected coronal gas density on an edge-on view with respect to the Galactic plane, with values in the range $\log_{10} \left[ {\rm N}_{\rm gas} (\psc) \right] = (18,20)$ in steps of 0.4 dex, overlaid onto the gas associated to the Magellanic System. The face-on view (i.e. $x$ projection) is fairly circular in all cases and is therefore omitted. The left panel displays the density structure at infall, which is identical in all models. The other panels display the density structure in each of the models at the present epoch. The deviation from an otherwise approximately symmetric profile at $z < 0$ is caused by the perturbation induced by the presence there of the Magellanic System. The flattening of an initially spherical, spinning corona is apparent, even more so the higher the net rotation speed. See also Figure \ref{fig:dens3}. Note that the final states of the non-magnetised, non-rotating corona and the magnetised, non-rotating corona are similar, and thus the latter is omitted here.  }
\label{fig:dens2}
\end{figure*}

\begin{figure}
\centering
\includegraphics[width=0.45\textwidth]{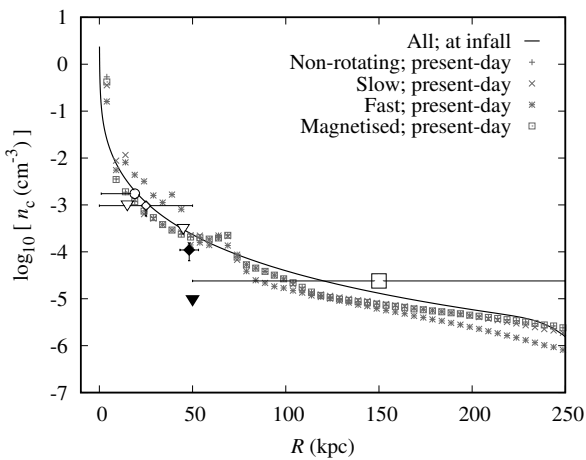}
\vspace{-5pt}
\caption[ Coronal density profile ]{ Mean density profile of the Galactic corona in each of our models at the present epoch (grey symbols). The solid curve corresponds to the initial density profile. Note that the solid curve and data points are identical to Figure \ref{fig:dens}. See also Figure \ref{fig:dens2}. }
\label{fig:dens3}
\end{figure}
%

\subsection{The effect of dynamical friction} \label{sec:dyn}

As noted above, our model `Static DM halo' is, by design, identical to the model 9:1 by \citet[][]{par18a}. As we show here and in Appendix \ref{sec:ver}, our results agree remarkably well with theirs. It is worth stressing that this agreement is not at all trivial, given the significant difference in the codes and methods used by each group both to generate the initial conditions and to calculate the evolution of the system.

The effect of dynamical friction on the orbit of the MCs can be studied by comparing the `Static DM halo' and `Live DM halo' models (Figure \ref{fig:ms}). Overall, we find no significant difference in the large-scale distribution of the Magellanic System between these two models. However, there are some noticeable differences in the orbital history of the MCs in the presence of a live DM host halo. In this case, the pericentric distance of the MCs are smaller, and their corresponding space velocities lower, compared to their infall into a static DM host halo (Figure \ref{fig:ver2}).

In addition to generating dynamical friction, a live DM host halo responds as a whole to the gravitational interaction with a massive system such as the LMC-SMC binary pair, i.e. it is free to `recoil'. The recoil of the Galactic DM halo may have important consequences on the orbital history of infalling systems \citep[][]{gom15a}. In the presence of a live Galaxy, as the MCs fall in, they and the Galaxy move around their common centre of mass. A fraction of the orbital energy of the MCs is absorbed by the motion of the Galaxy as a whole, and therefore the MCs move slower and sink further. We find that the recoil of the Galactic DM host halo in our models is non-negligible, on the order of 10 kpc when the MCs roughly reach their present-day position. This has a noticeable effect on the positions and kinematics of the MCs (Appendix \ref{sec:ver}), but it does not affect substantially the overall position and extension of the Stream.

There is a subtlety inherent in our choice of simulation code. Like any grid-based code, \ramses\ does not allow the existence of a true vacuum within the simulation volume. At initialisation, each cell is `filled' with a fluid defined by its density, pressure, and kinetic energy, however small. In our `Static DM halo' and `Live DM halo' models, we have chosen the initial density of the cells {\em not} associated to the MCs gas to a very low value, and chosen the temperature to match the mean temperature of the gas associated to the MCs at infall. The fact that we are able to broadly reproduce \citet[][]{par18a}'s results indicates that these choices have not affected our results in any significant way.

\subsection{The effect of gas drag} \label{sec:gas}

\begin{figure*}
\centering
\includegraphics[width=0.33\textwidth]{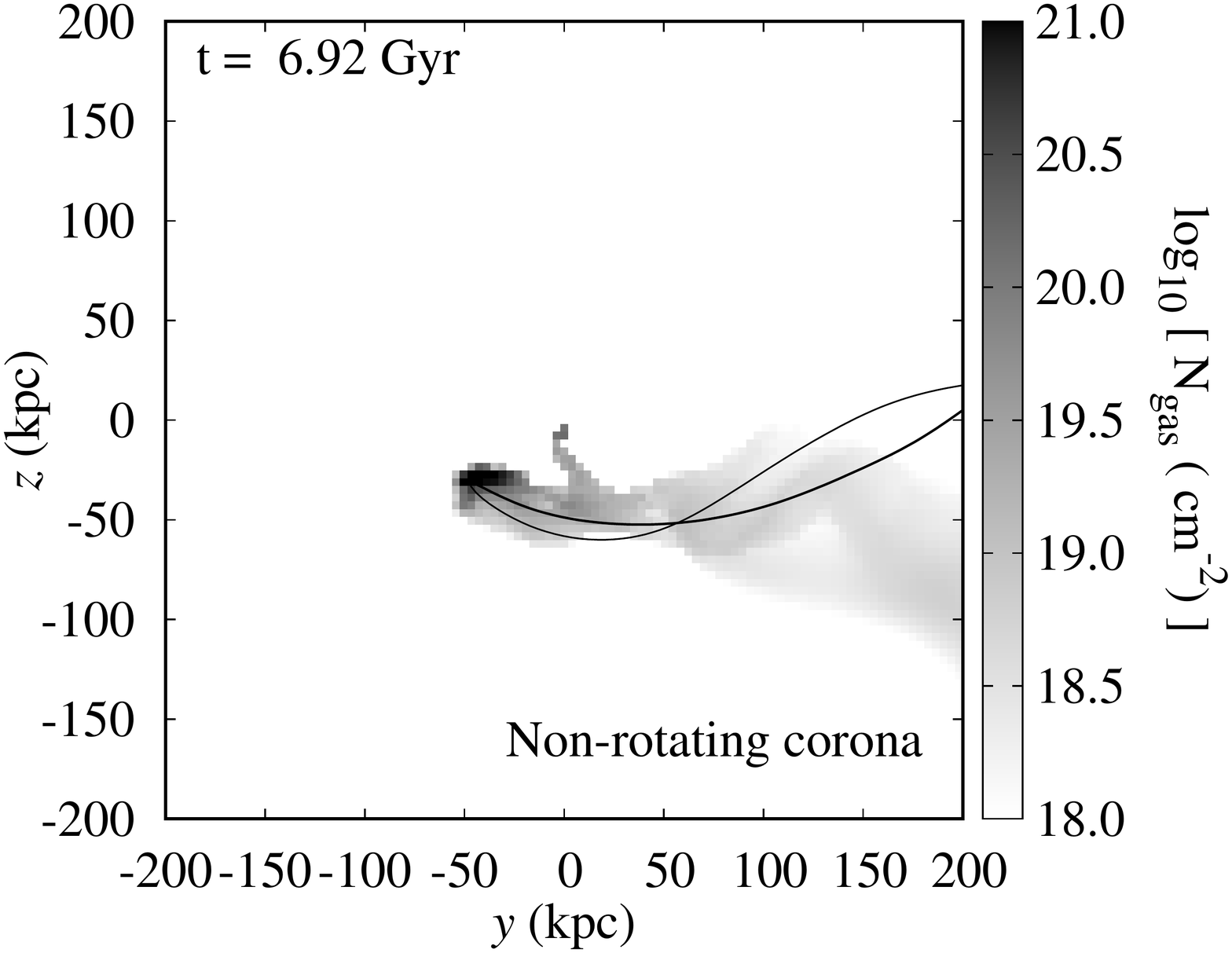}
\includegraphics[width=0.33\textwidth]{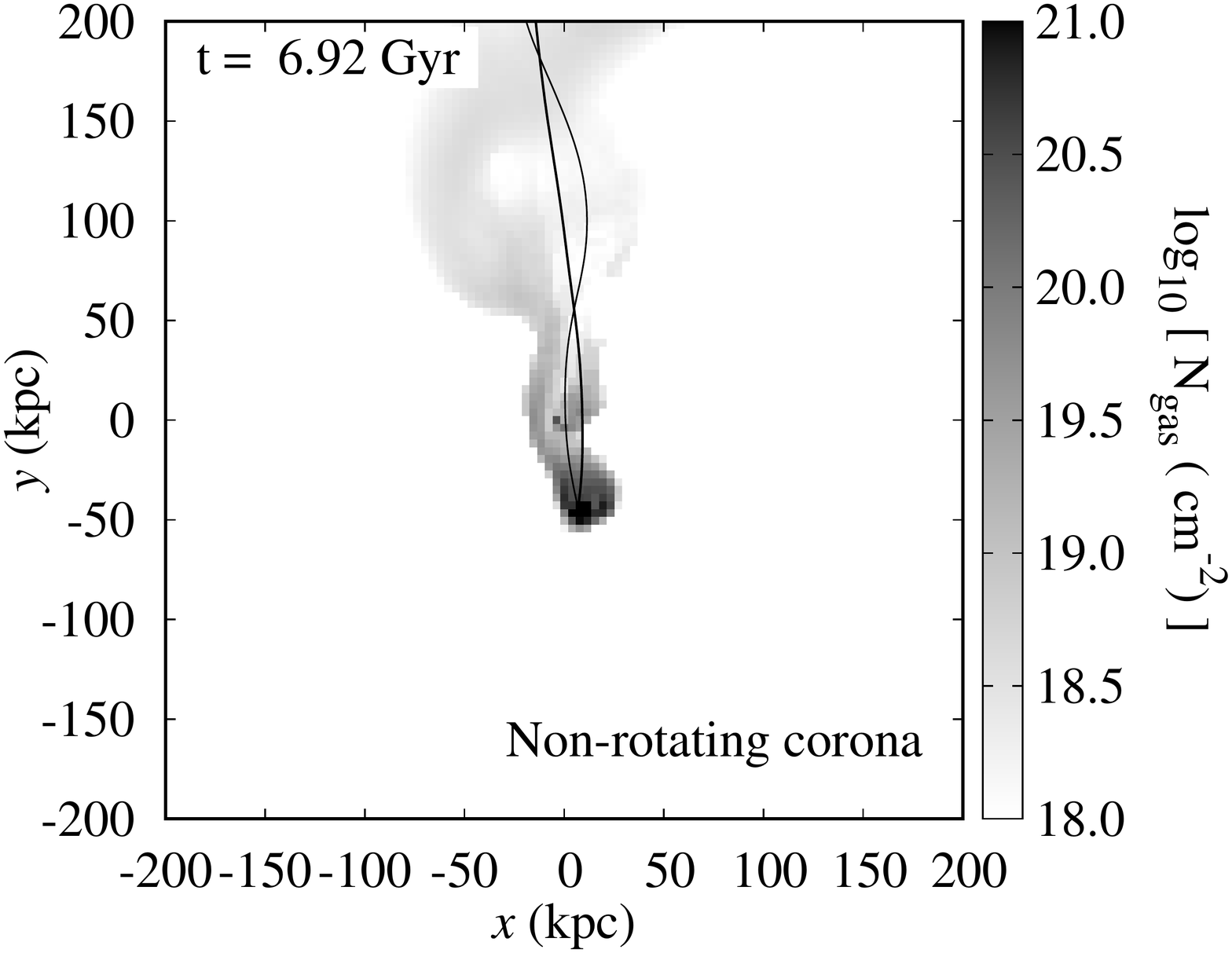}
\includegraphics[width=0.33\textwidth]{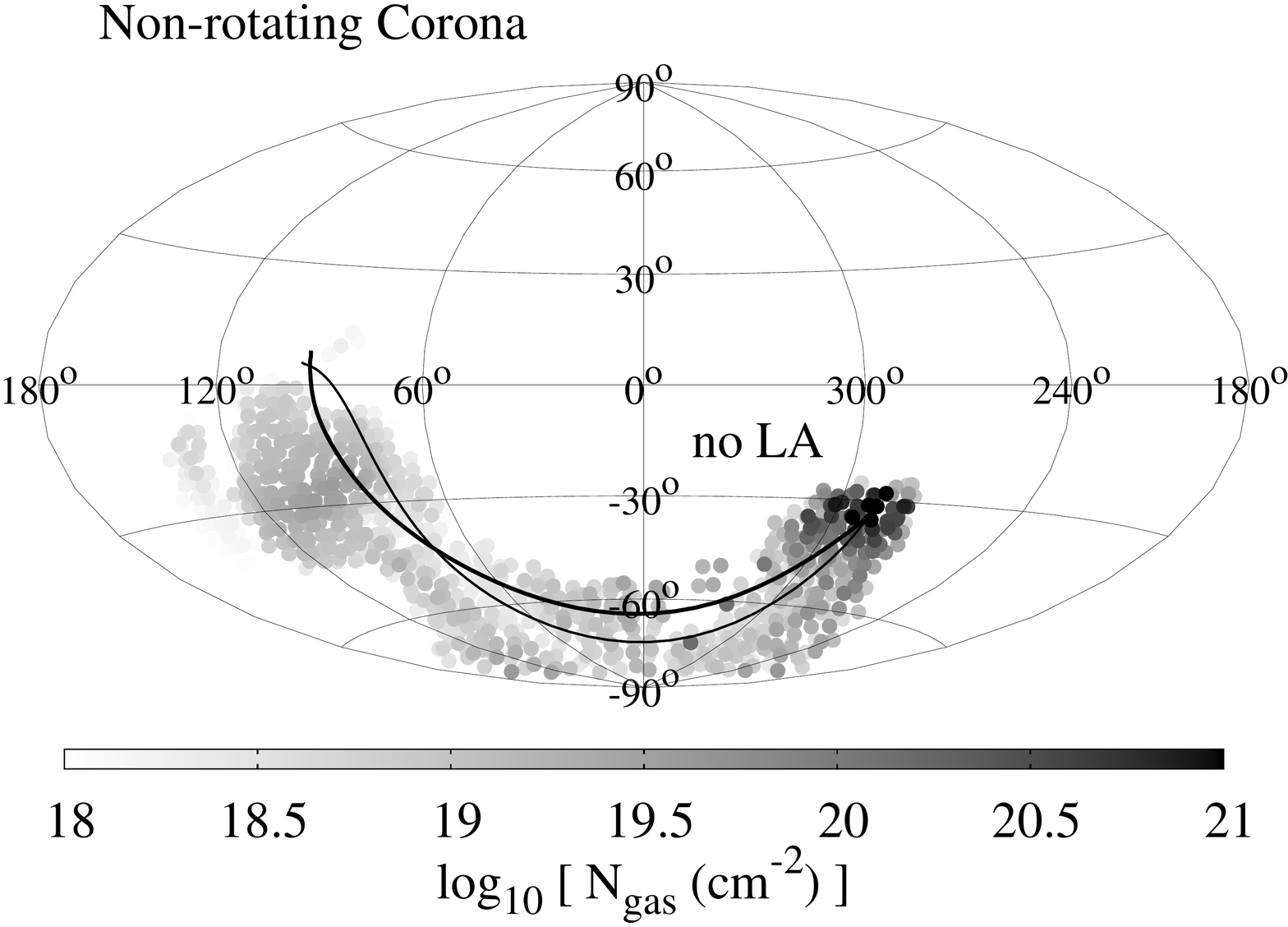}\\
\includegraphics[width=0.33\textwidth]{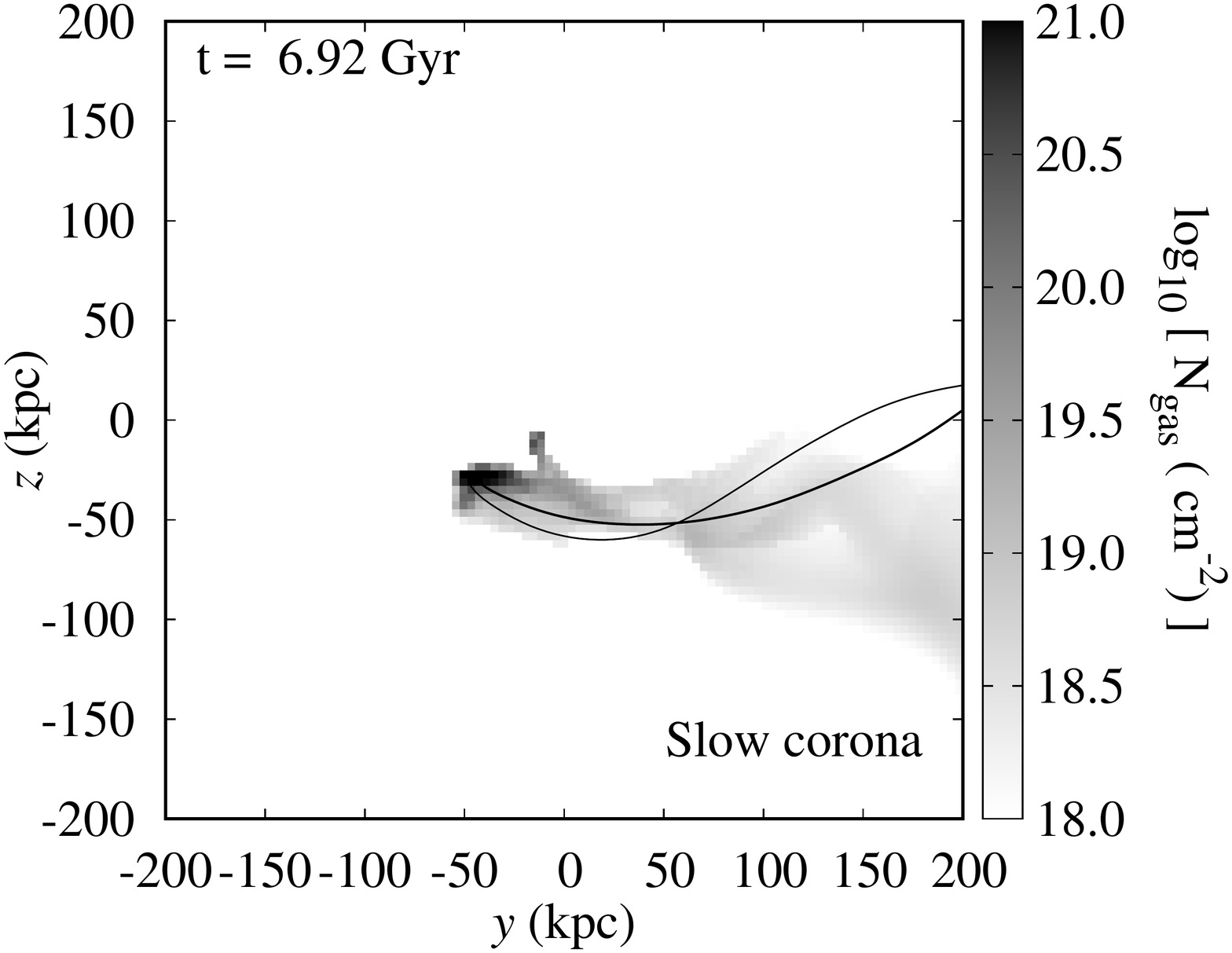}
\includegraphics[width=0.33\textwidth]{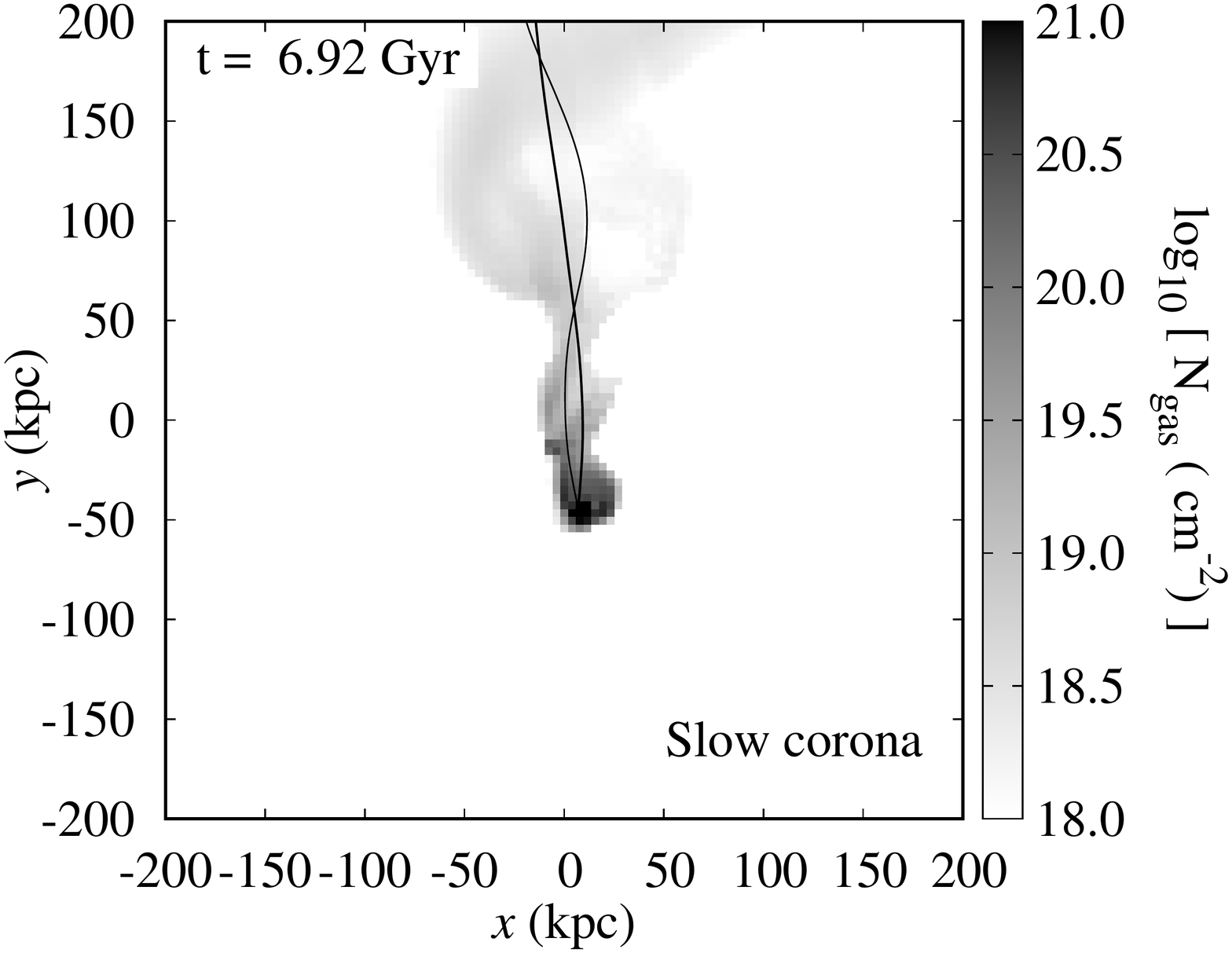}
\includegraphics[width=0.33\textwidth]{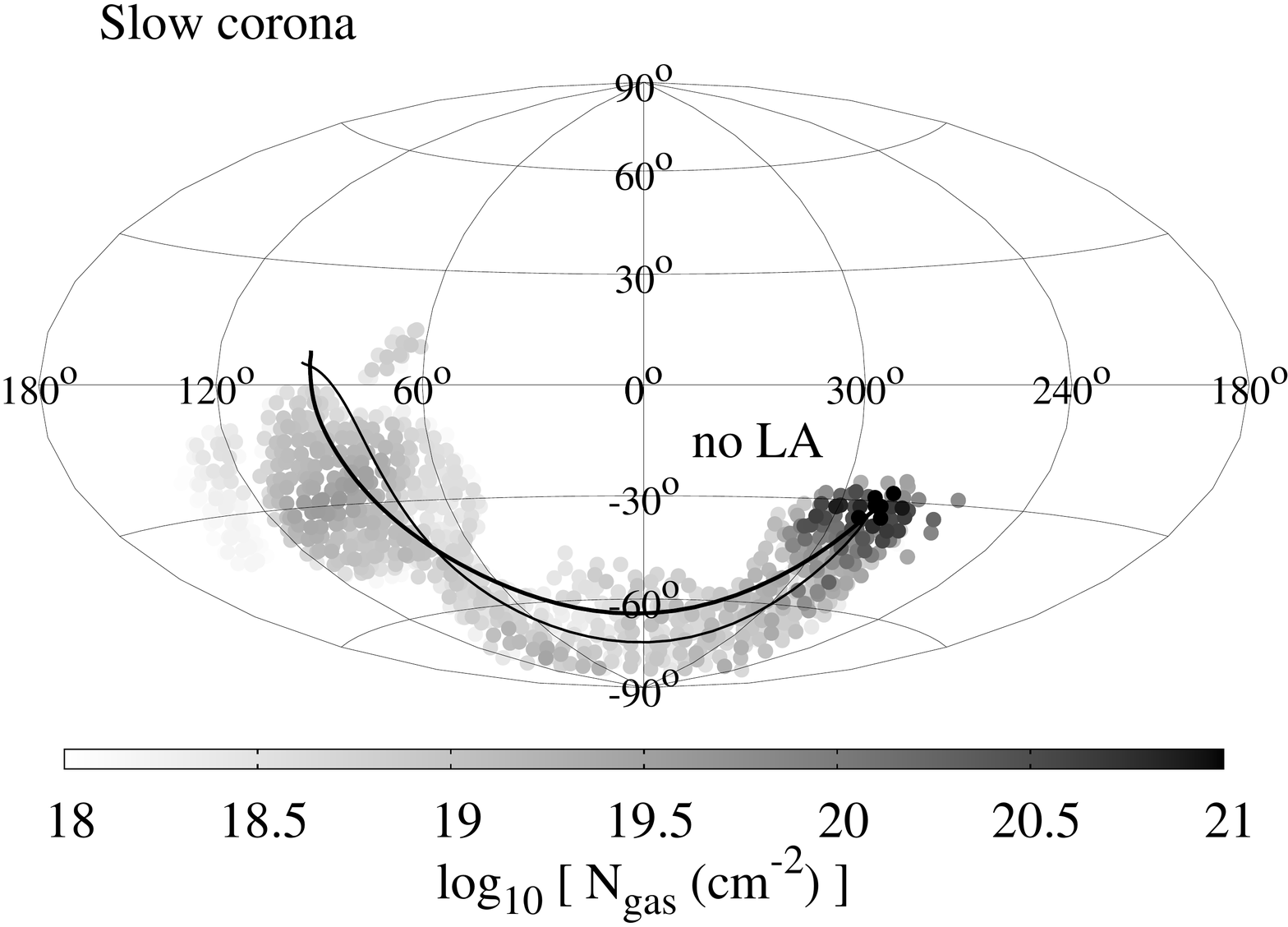}\\
\includegraphics[width=0.33\textwidth]{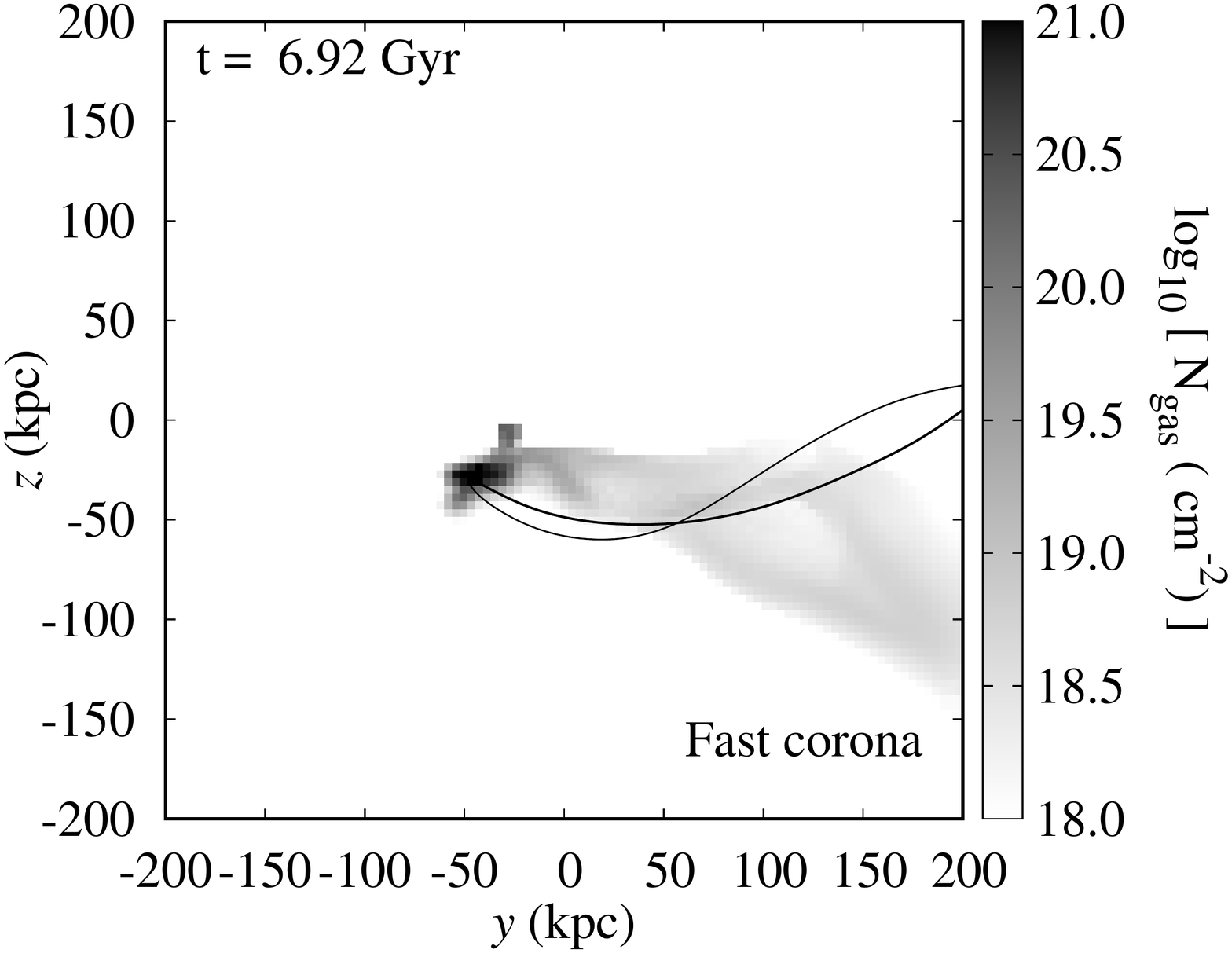}
\includegraphics[width=0.33\textwidth]{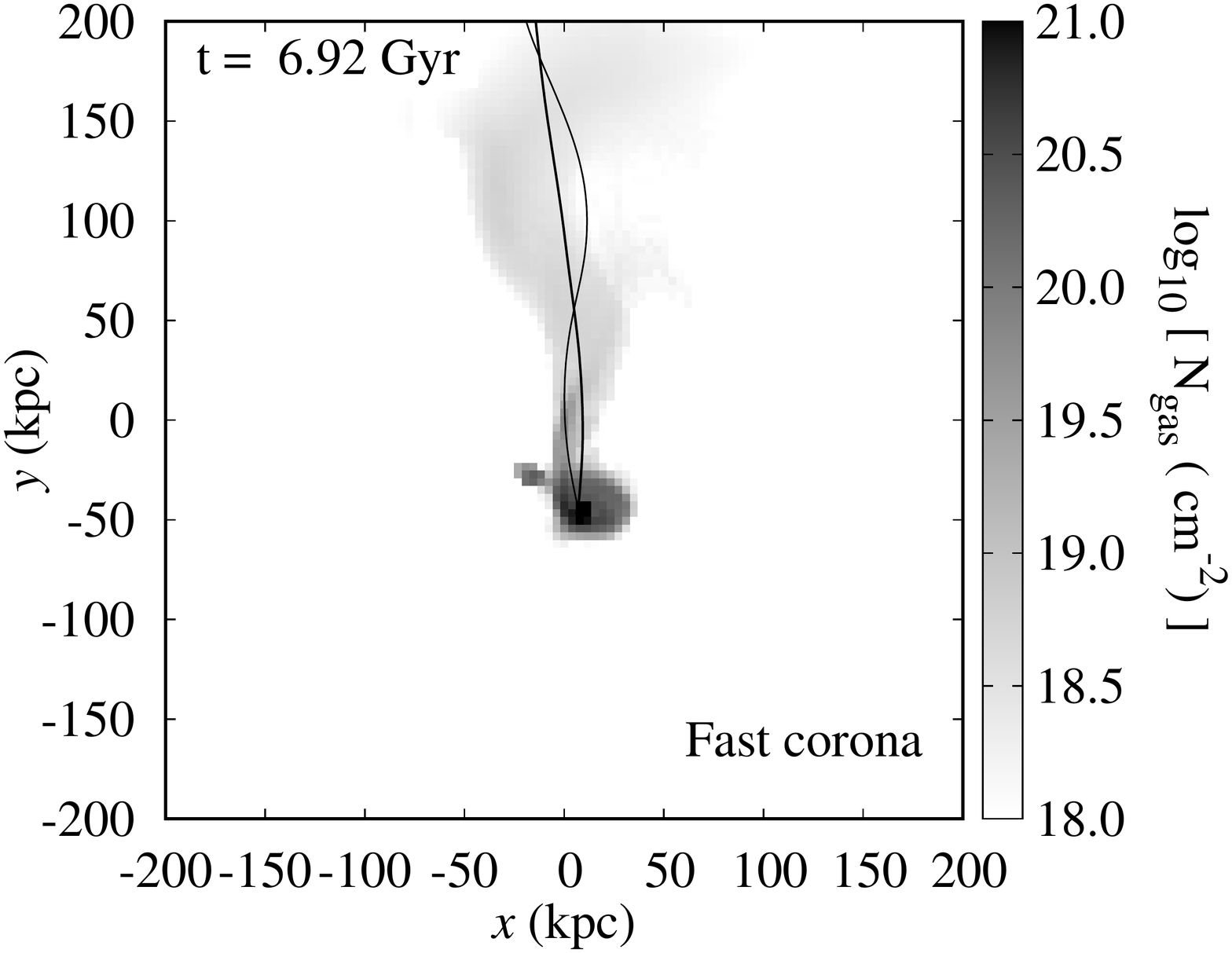}
\includegraphics[width=0.33\textwidth]{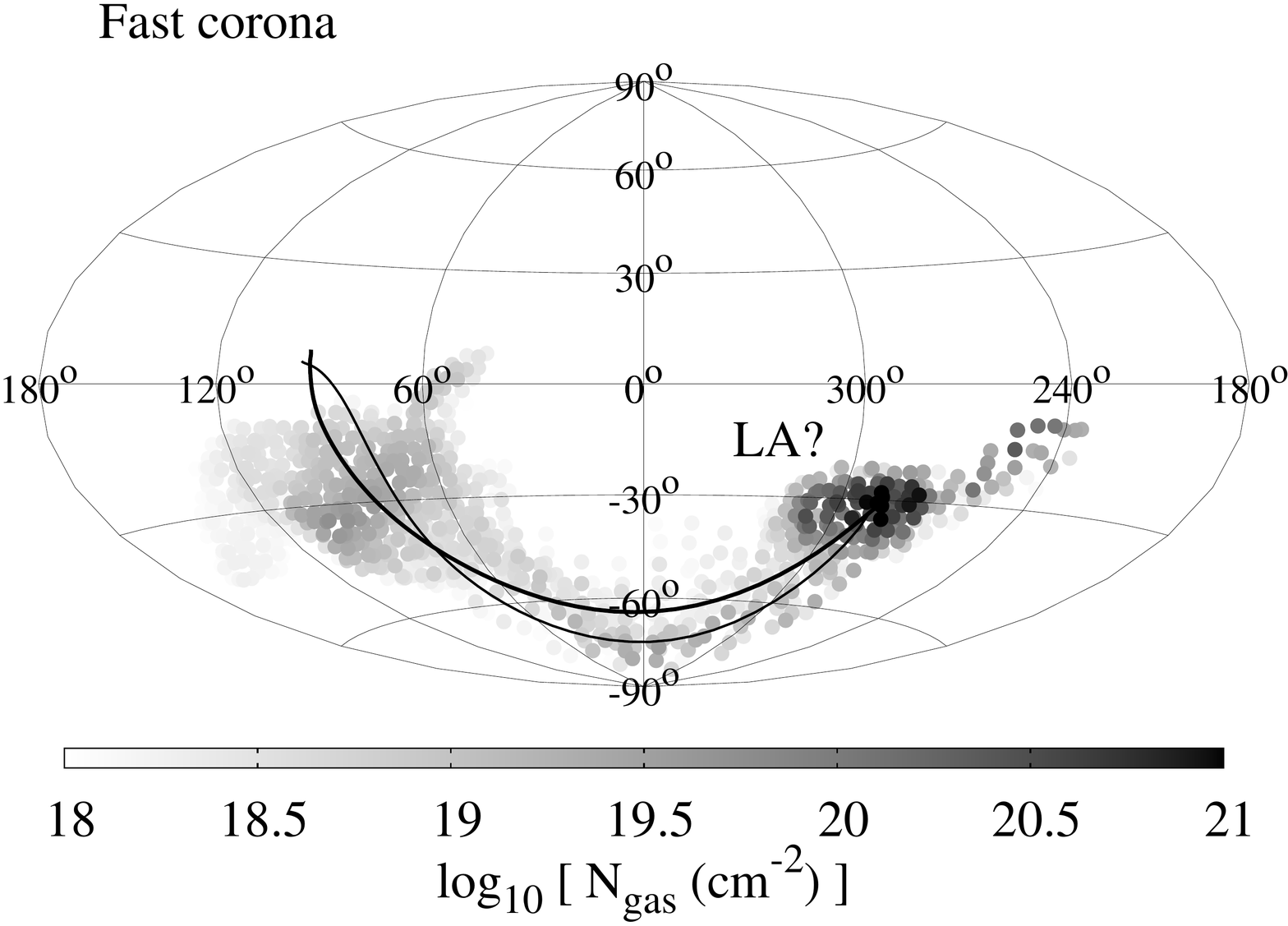}
\caption[  ]{ Distribution of gas in the the Magellanic System at the present epoch, i.e. $\sim 1$ Gyr after infall into the Galaxy, in the {\em presence} of a hot corona. Each row corresponds to a different model. Left~/~centre: Total gas column density in physical space on a face-on (left) and edge-on (centre) projection with respect to the Galactic plane. The Galactic Centre (GC) is roughly at (0,-10,0) kpc. Right: Total column density projected on the sky as seen from the Sun in Galactic coordinates $(l, b)$. The GC is at $(0,0)\deg$; the MCs', at $\approx (300, -30)\deg$; their space orbits are indicated by the thin / thick solid curves in all panels. The presence of the hot corona around the Galaxy clearly inhibits the formation of a leading gaseous stream, regardless of the corona's kinematic properties. Note that the gas displayed in each panel corresponds to gas with a tracer value $\geq 5$, and therefore excludes the overwhelmingly majority of the coronal gas.  }
\label{fig:ms2}
\end{figure*}

\begin{figure*}
\centering
\includegraphics[width=0.33\textwidth]{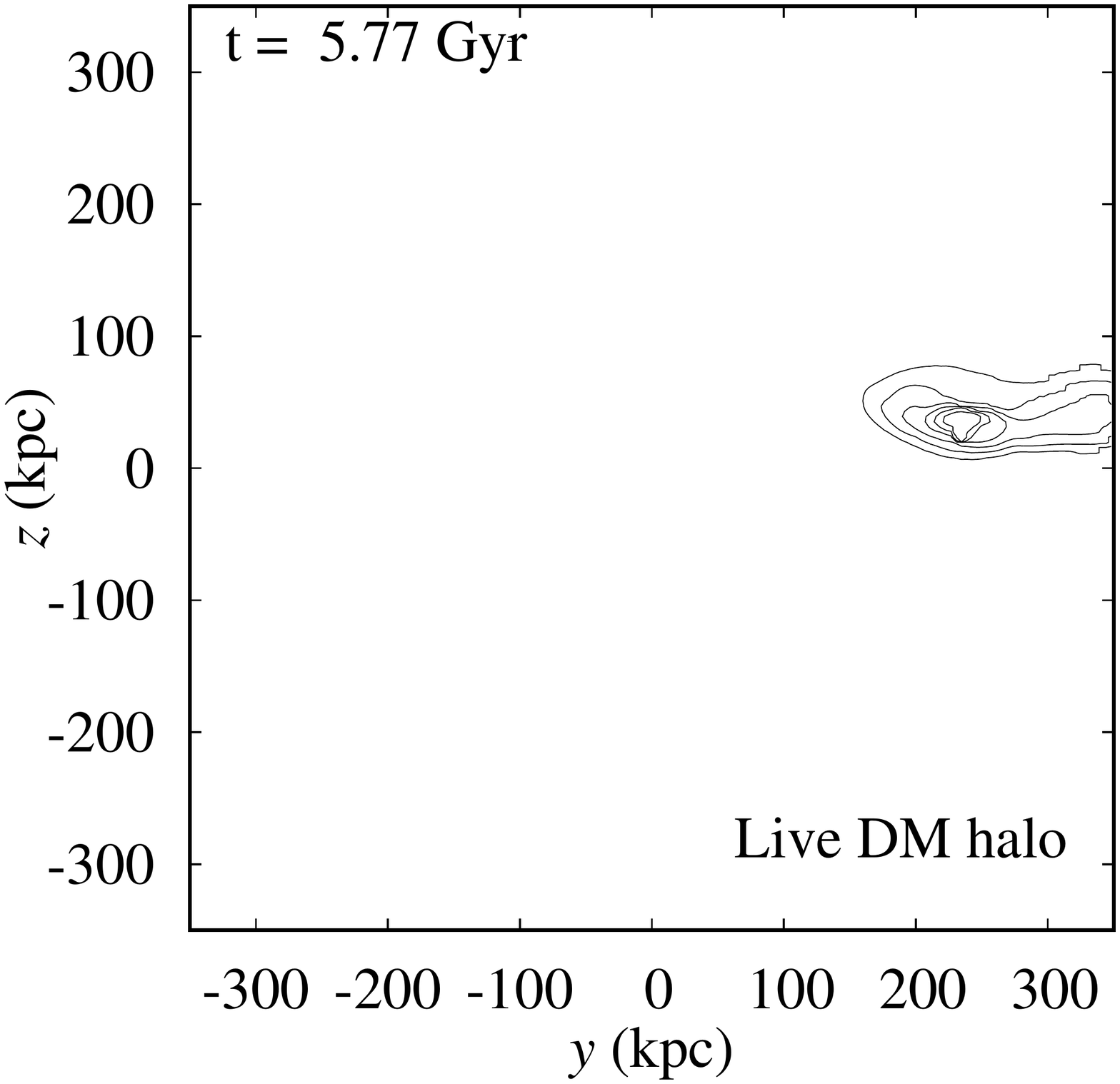}
\includegraphics[width=0.33\textwidth]{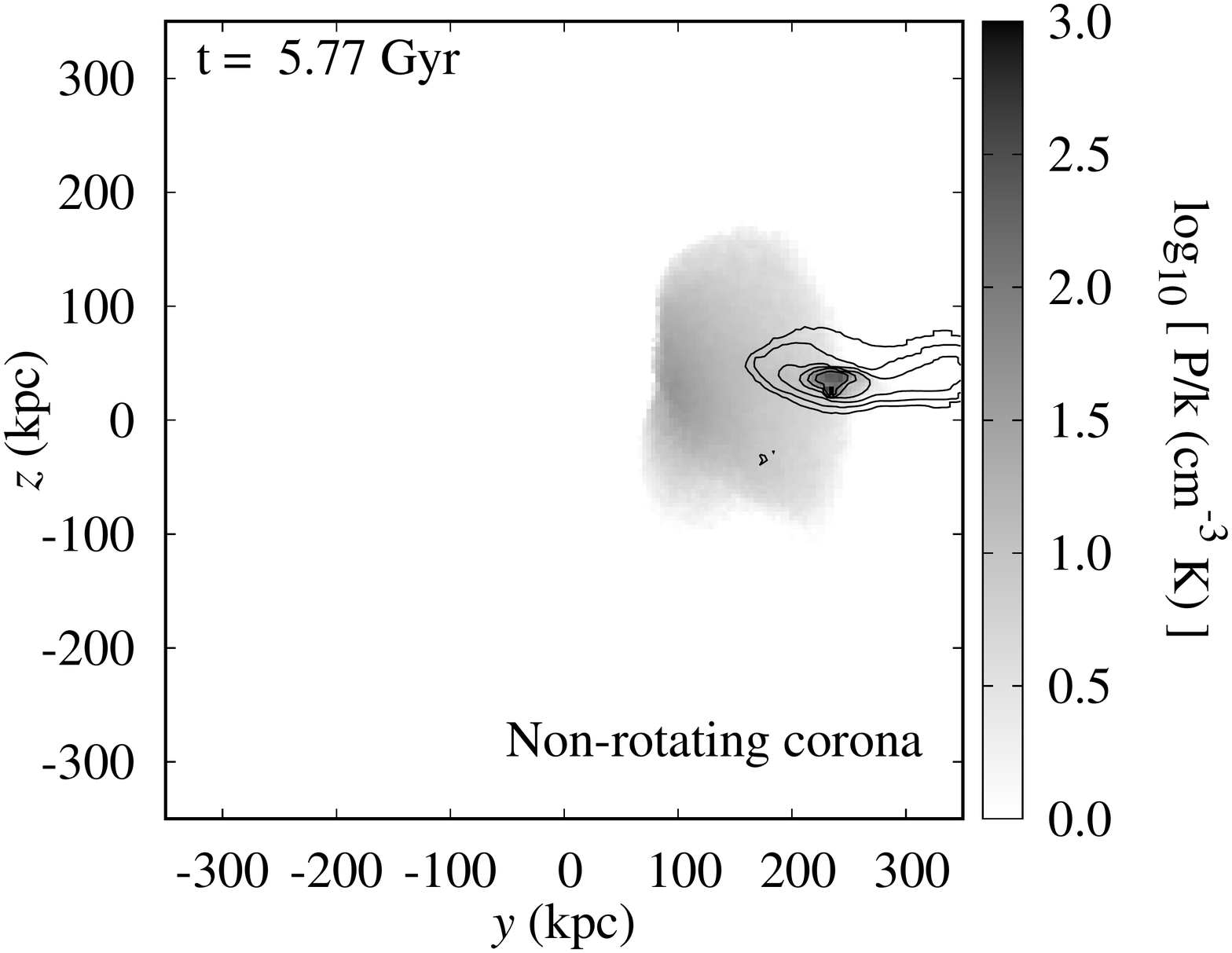}
\includegraphics[width=0.33\textwidth]{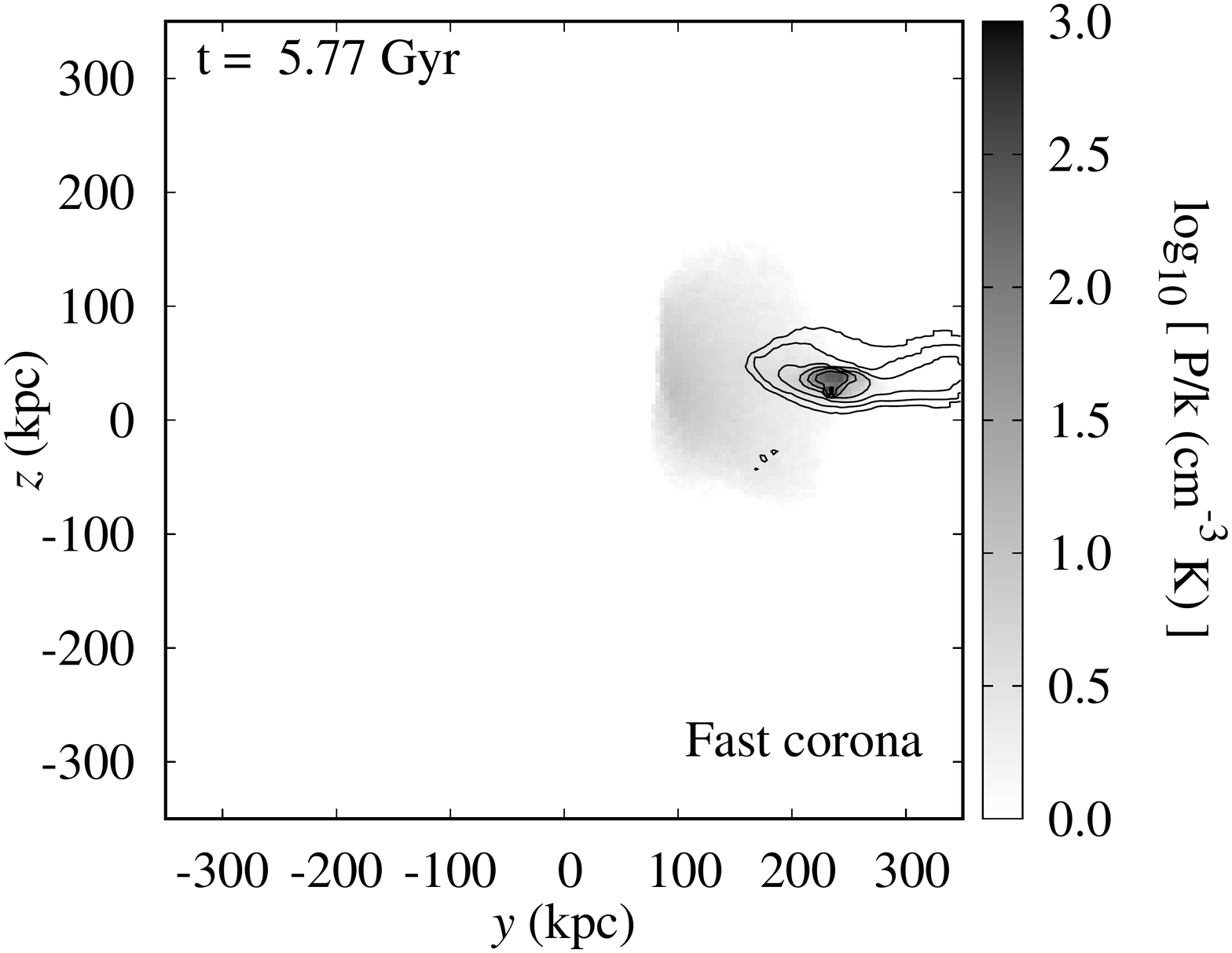}
\includegraphics[width=0.33\textwidth]{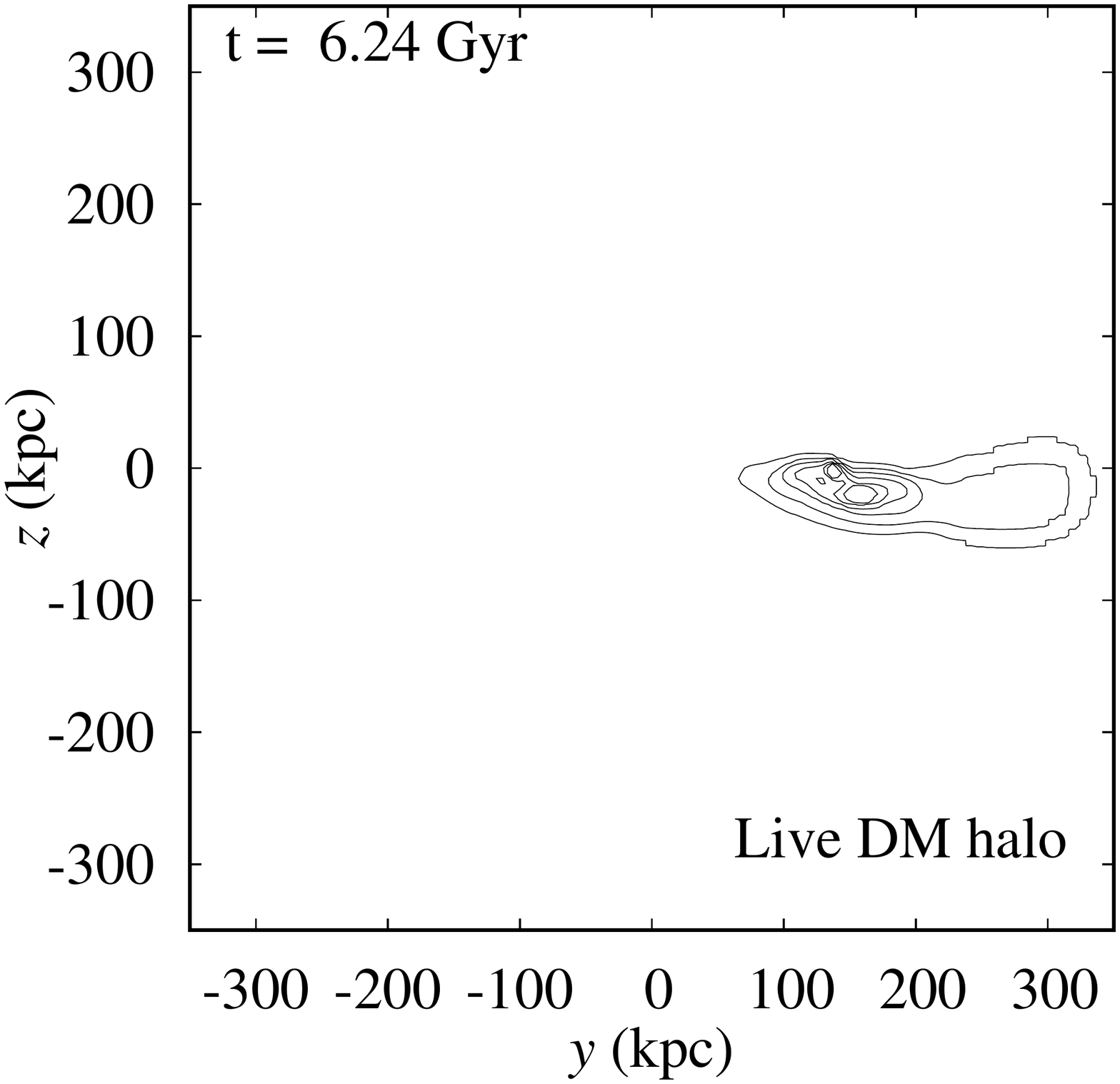}
\includegraphics[width=0.33\textwidth]{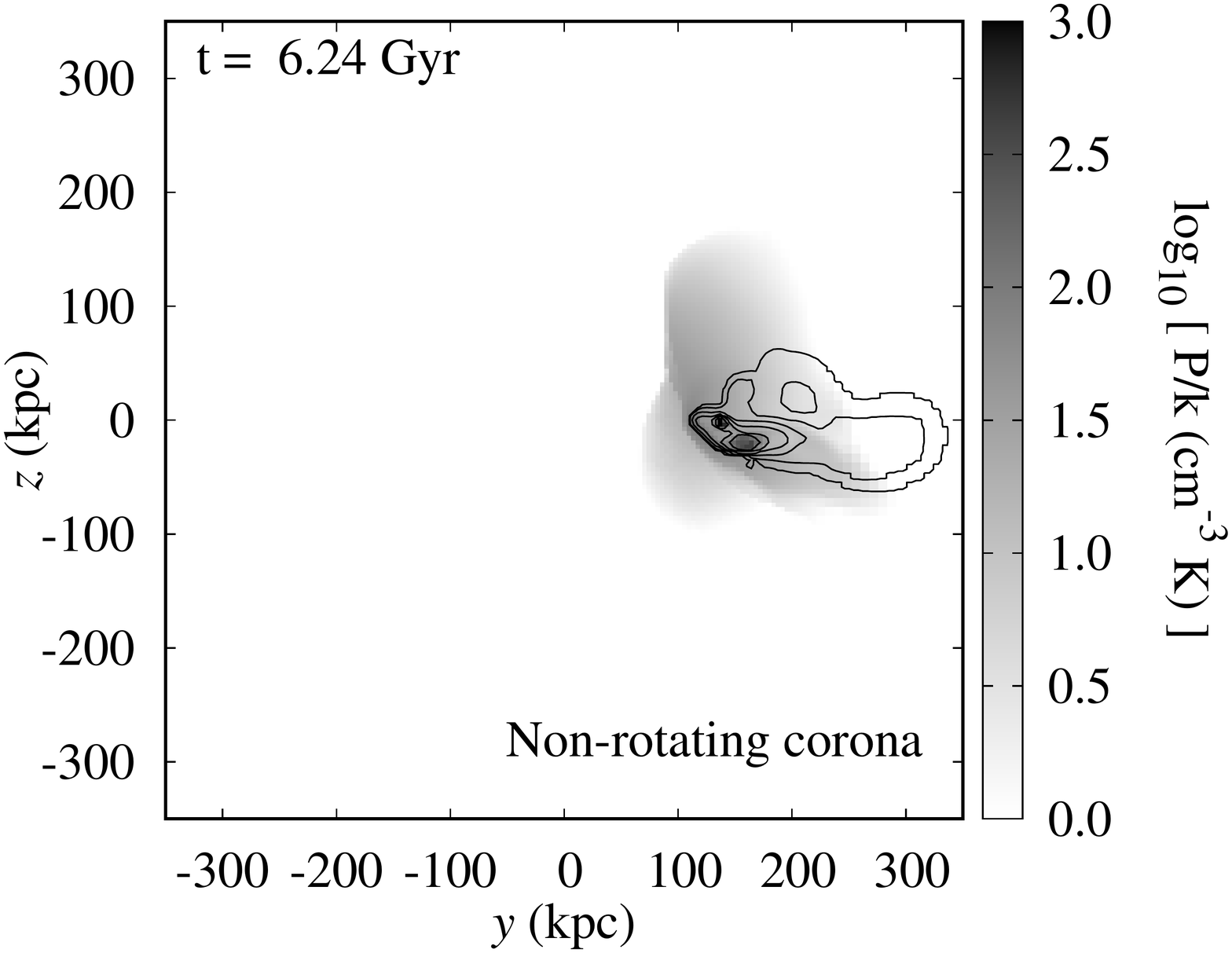}
\includegraphics[width=0.33\textwidth]{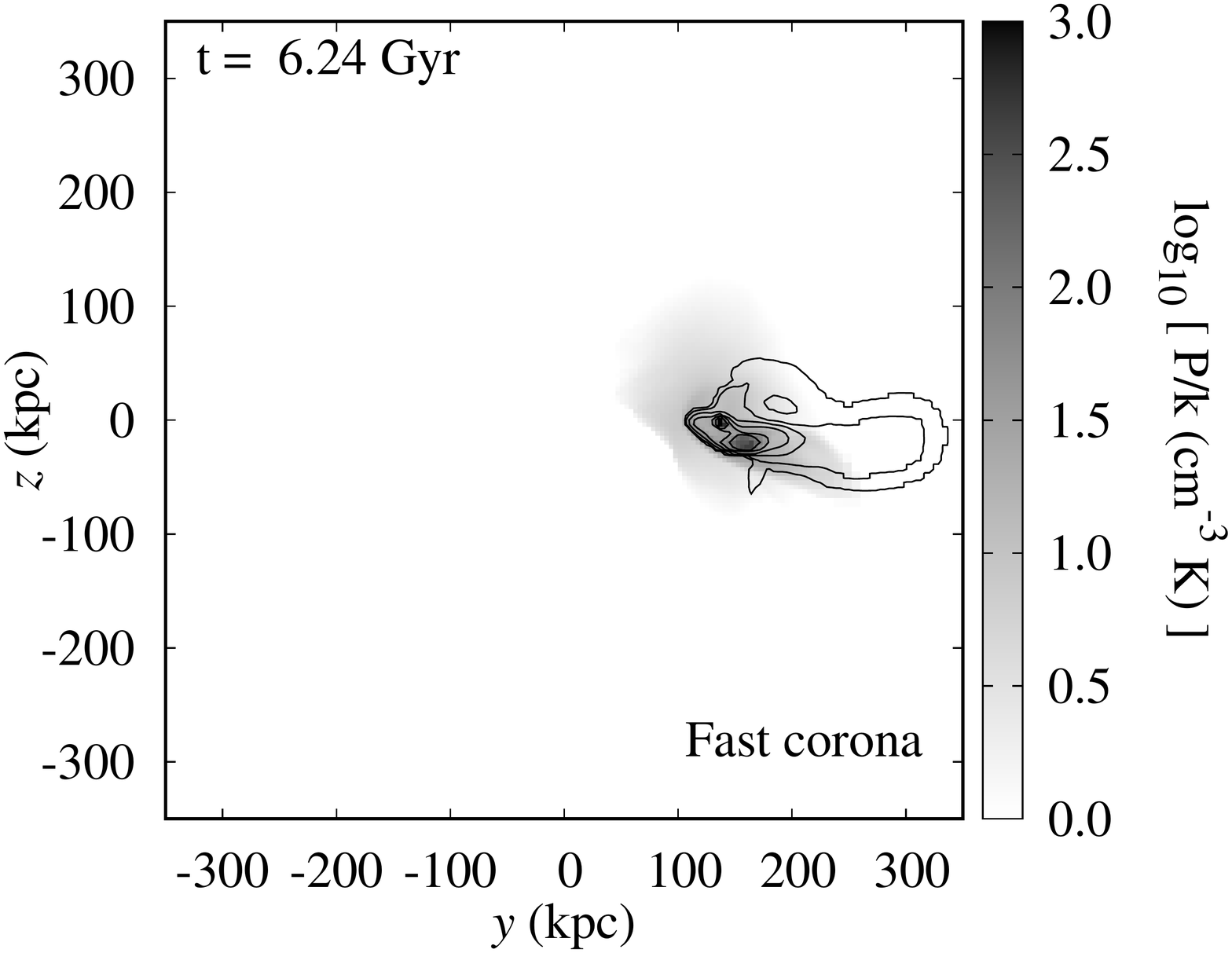}
\includegraphics[width=0.33\textwidth]{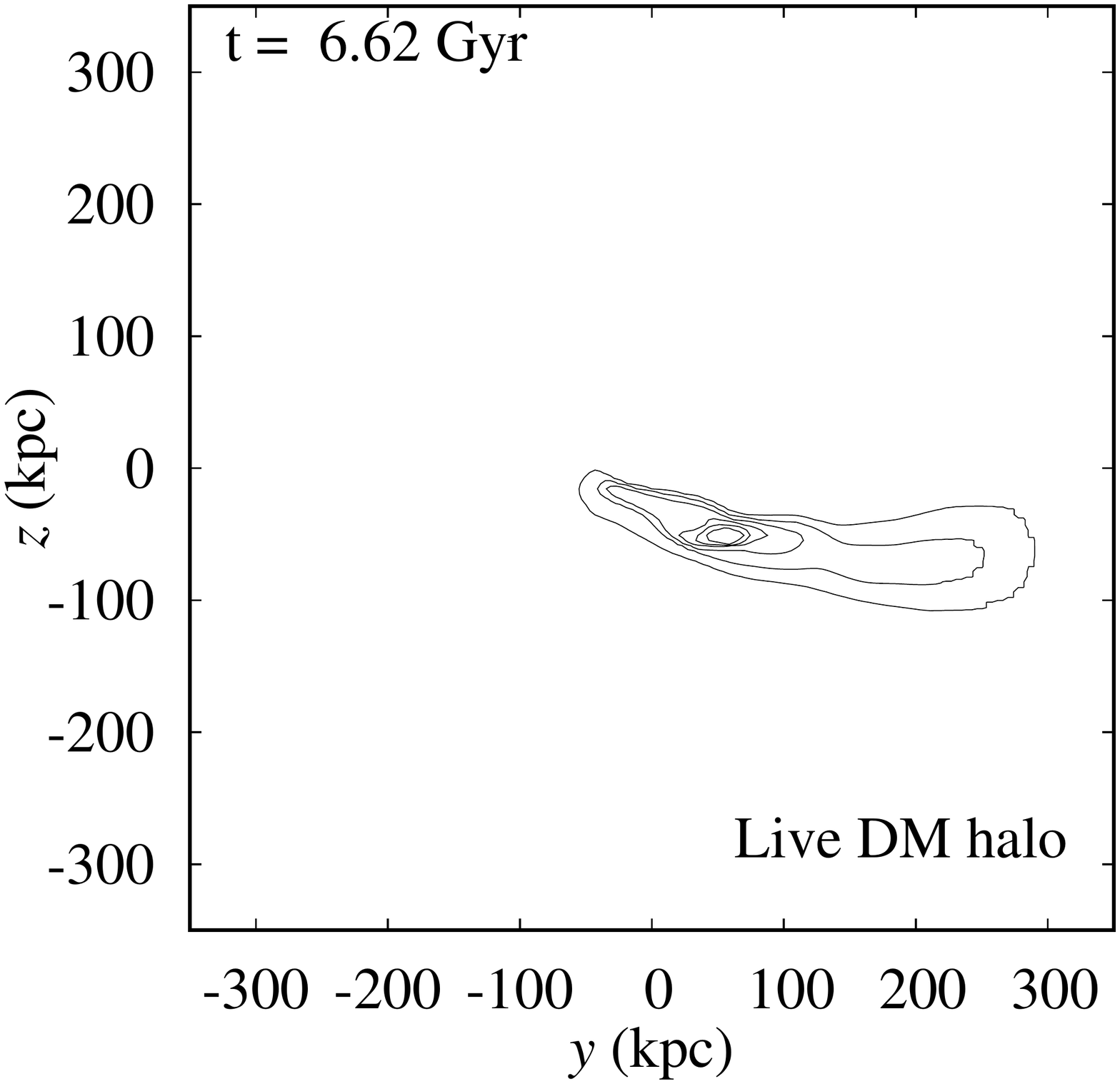}
\includegraphics[width=0.33\textwidth]{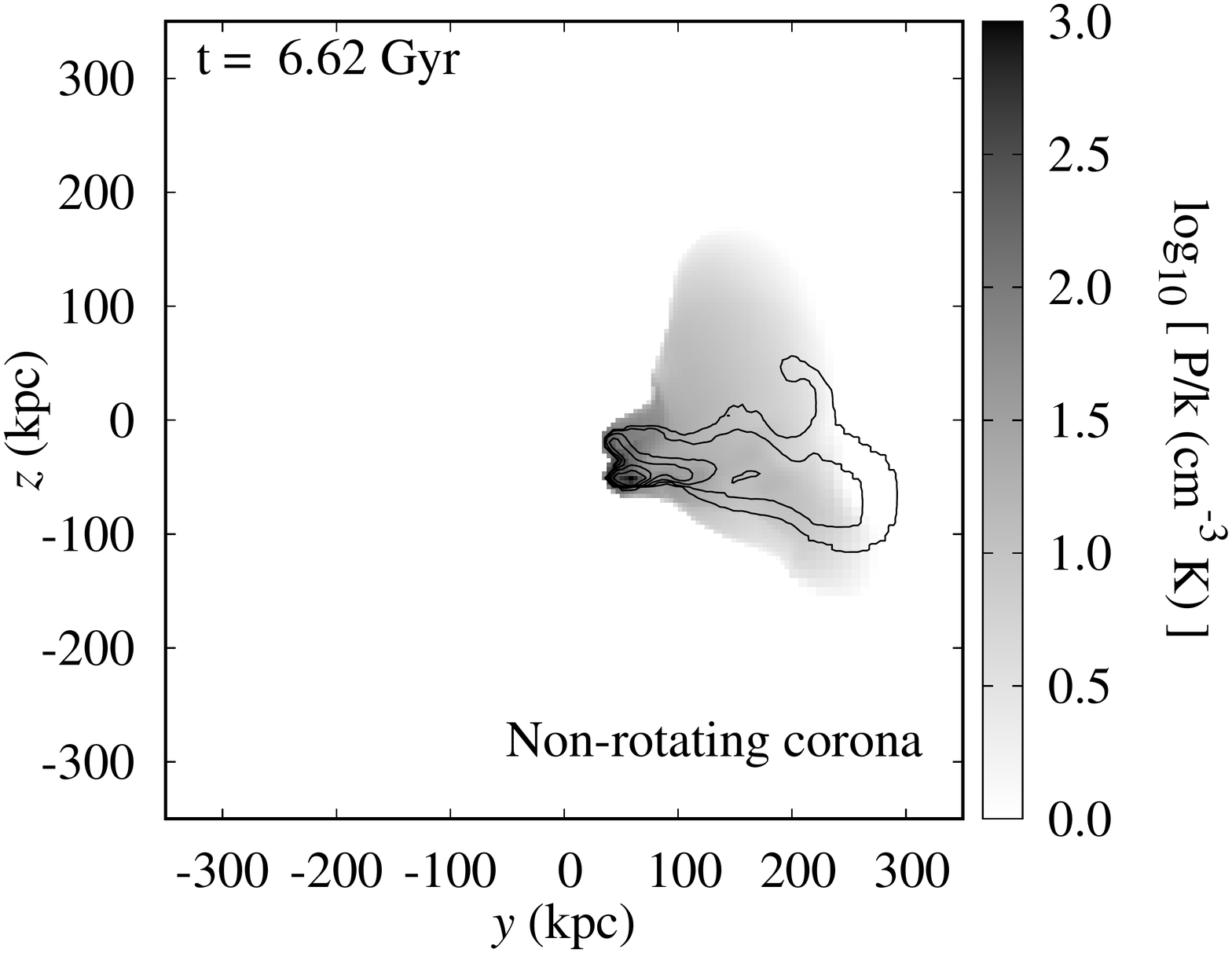}
\includegraphics[width=0.33\textwidth]{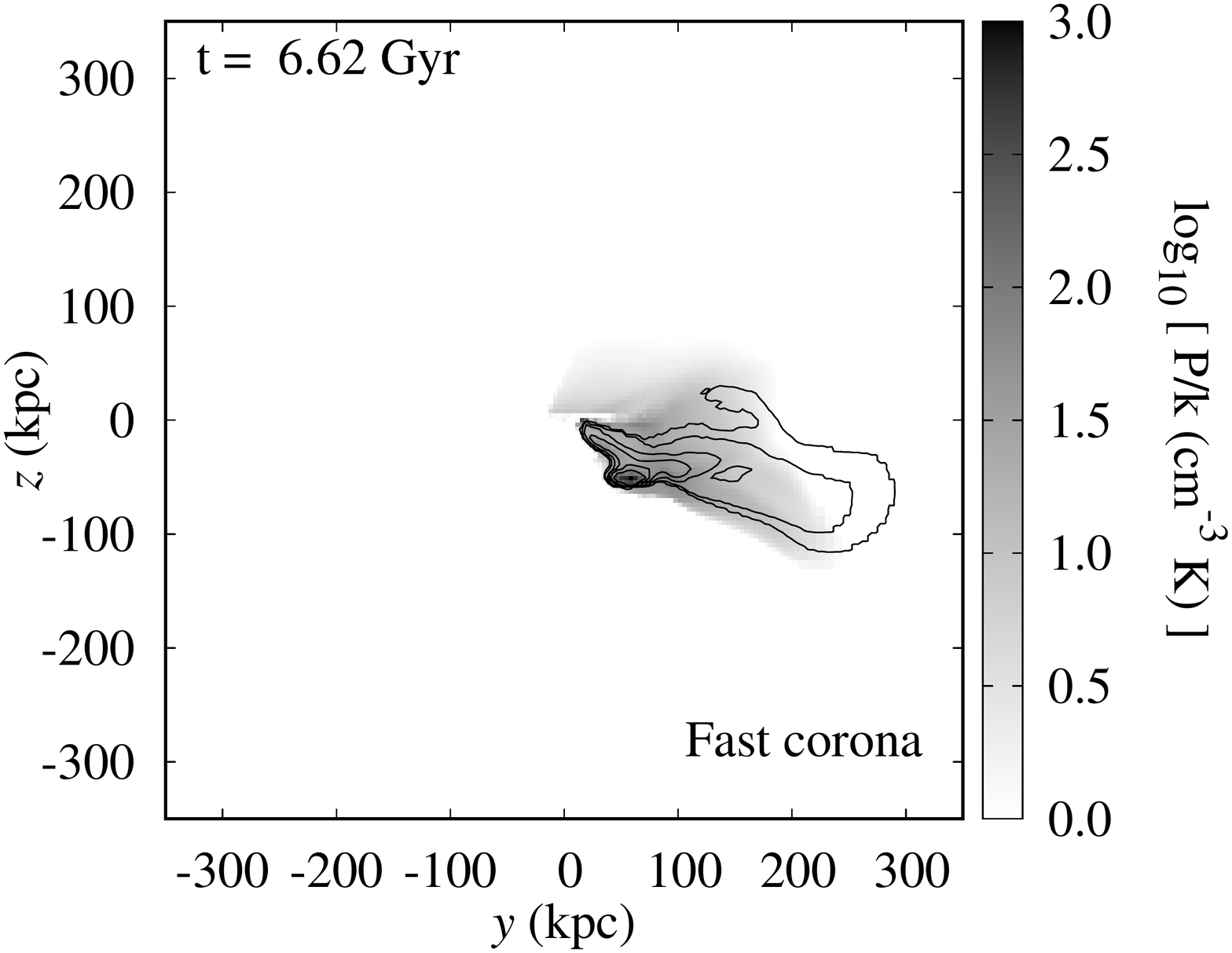}
\includegraphics[width=0.33\textwidth]{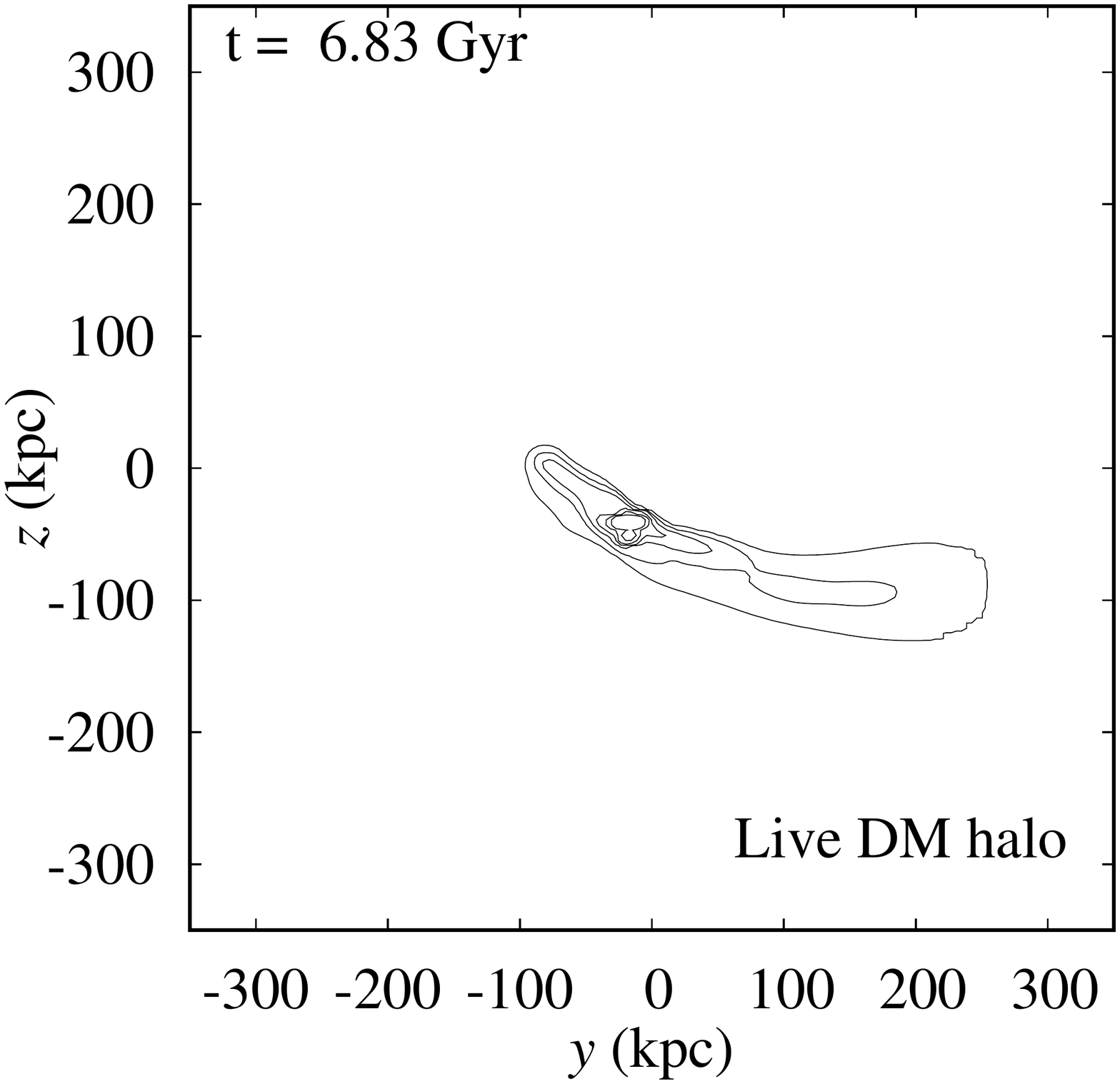}
\includegraphics[width=0.33\textwidth]{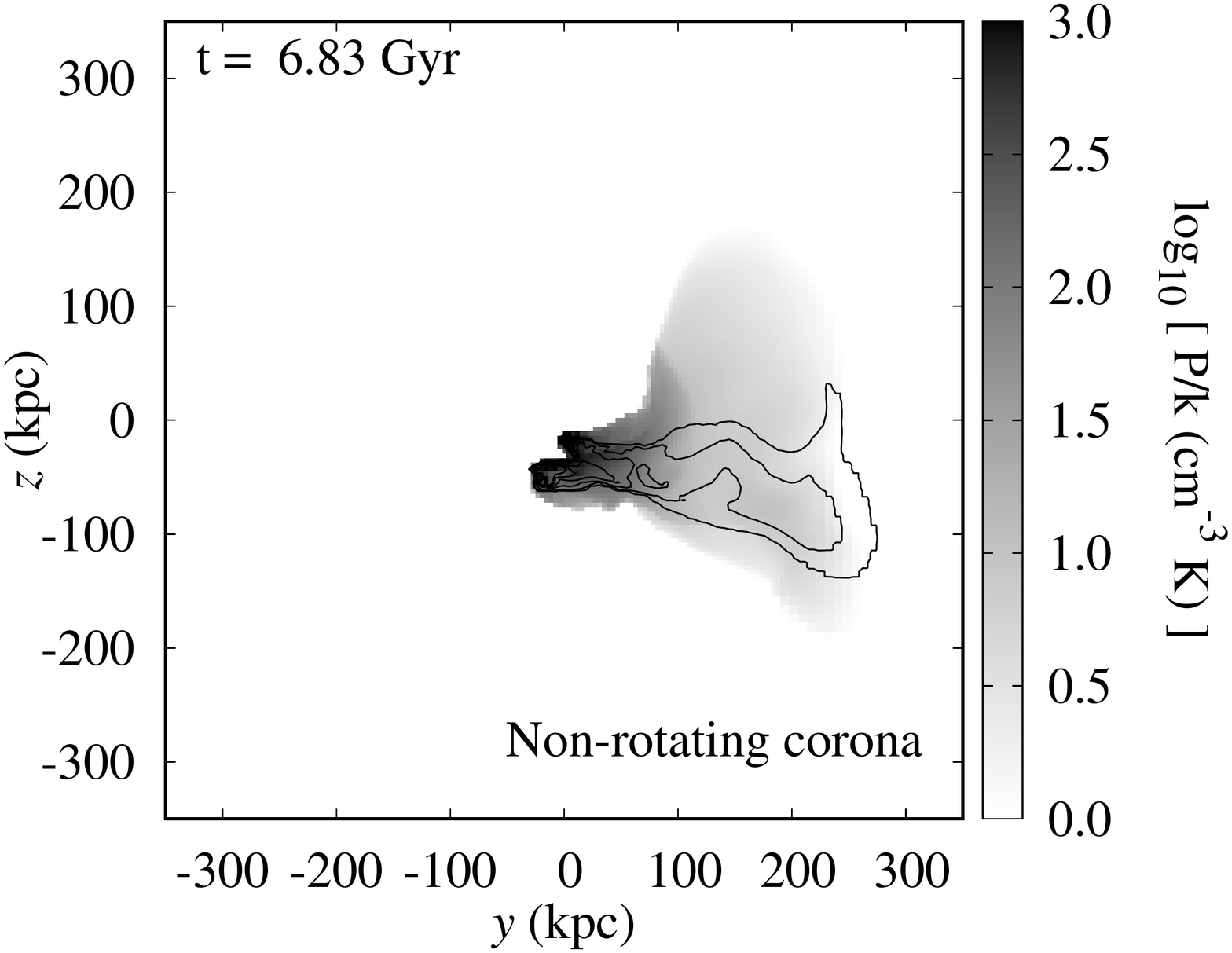}
\includegraphics[width=0.33\textwidth]{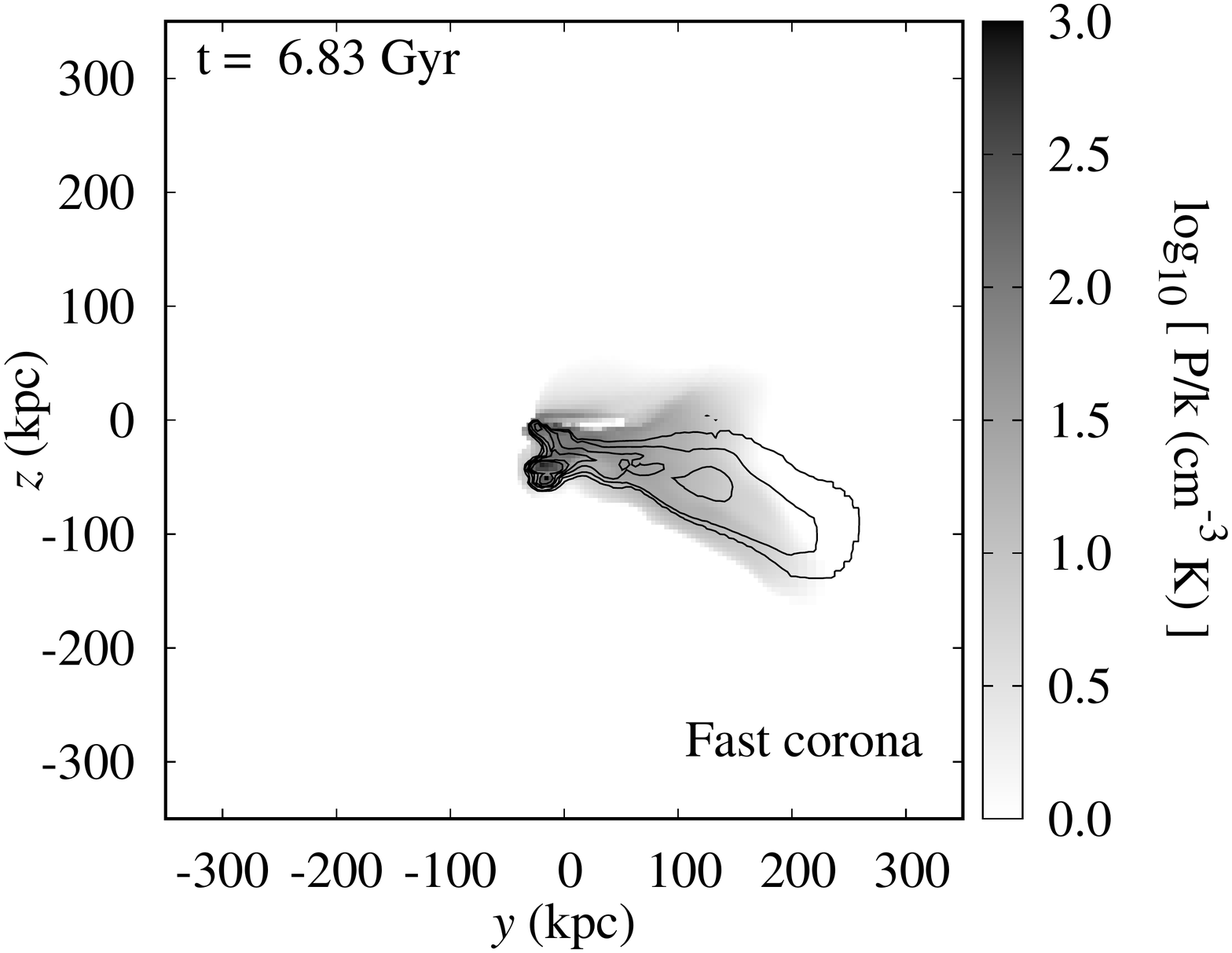}
\caption[ ]{ Effect of ram pressure on the evolution of the Magellanic System.  Each row compares the structure of the infalling gas on an edge-on projection with respect to the Galactic plane in a model without a Galactic corona (left column), a model with a non-rotating corona (centre), and a model with a fast spinning corona (right) across a sequence of snapshots spanning a timeline from the infall of the MCs (top) up to $\sim100$ Myr before the present day (bottom). The gas pressure $P / k$ is indicated by the grey scale. The contours indicate the total gas column density of the gas associated to the Magellanic System with values in the range $\log_{10} \left[ {\rm N}_{\rm gas} (\psc) \right] = (18,20.5)$ in steps of 0.5 dex. The build-up of ram pressure ahead of the MCs and the resulting gradient along the gas streams as they falls in are apparent. Note the reduced ram pressure in the presence of a fast spinning corona compared to a non-rotating one. The formation of a clear leading stream in the presence of a corona is inhibited early on. The orbit of the MCs have been omitted here for clarity. The ram pressure is calculated adopting a gas tracer value of 0.01 to include not only the gas associated to the Magellanic System but also the coronal gas around the MCs. }
\label{fig:pram}
\end{figure*}

Our model series allows us for the first time to study the impact of the Galactic corona on the infalling gas in a systematic and self-consistent manner. A key requirement of our models is that the properties of the model corona -- notably its mean density distribution -- be consistent with the observed properties Galactic corona during the infall of the MCs.

Given the strong differences between each of the coronal models in terms of their internal kinematics, their structure may evolve in time in a noticeable way. This is illustrated in Figure \ref{fig:dens2}. When $v_{\rm c} > 0$ \kms, a corona that is initially spherically symmetric by design, eventually flattens along its spin axis. The effect is more severe for higher rotation speeds. Such a flattening is expected in equilibrium models \citep[][]{pez17a,sor18a}.\footnote{We caution against a misinterpretation of the word `flattening'. At times, the term is used to indicate a shallower, albeit spherically symmetric, density profile of the hot corona compared to some reference profile \citep[e.g.][]{bre18a}. Here, the term is used to indicate that the hot corona is `squashed' along its spin axis, with the result that its 3D structure is no longer spherically symmetric. \label{foo:flat} }

In spite of the evolution and the consequent difference in structure between the corona models, their {\em mean} density profile is comparable at all times, and does not change significantly over the course of our simulations. In Figure \ref{fig:dens3}, we compare the initial (i.e. at infall) and final (i.e at the present epoch) density profiles of the corona in each of our relevant models well beyond the present-day distance of the MCs. We find that the model corona evolves away from its initial state but remains well within the available observational constraints. In particular, the coronal density drops slightly at $r > 50$ kpc such that the gas drag is slightly reduced along the MCs' orbit.

The inclusion of a Galactic corona in our simulations introduces a technical difficulty when using AMR codes. In contrast to the models that lack a corona, it is no longer straightforward to isolate (and visualise) the gas initially associated with the MCs. In order to overcome this difficulty, we attach a passive scalar (colour tracer) to the gas initially associated with the MCs prior to placing the evolved binary system at the Galactic virial radius. Specifically, we set the value of the gas tracer associated with the MCs to 20, and to 0 everywhere else. Note that these choices are arbitrary and do not affect in any way our results.

The presence of a hot corona embedded within a live DM host halo does not affect in any significant way the orbital history of the MCs compared to the model that features a live DM host halo but no corona (see Figure \ref{fig:ver3}). But it does have a noticeable impact on the evolution of the infalling gas.

A visual comparison between models `Live DM halo' (second row in Figure \ref{fig:ms}) and `Non-rotating corona' (first row in Figure \ref{fig:ms2}) readily reveals our more striking result: {\em In the presence of a Galactic corona, the Leading Arm is absent from the final configuration that otherwise resembles the observed Magellanic System}. In contrast, the Stream appears slightly more extended (by roughly 30\deg) compared to models that lack a Galactic corona.

The absence of a leading gaseous stream is common to all the models that include a Galactic corona, regardless of the corona's rotation speed, as can be seen by comparing models `Non-rotating corona', `Slow corona', and `Fast corona'. Note that the `Fast corona' model features a gas structure that, projected on the sky, could be reminiscent of the Leading Arm. Such a feature results from the gas clouds that are decelerated by the drag force, thus falling onto lower orbits and eventually leading the MCs \citep[e.g.][]{moo94a}. However, the distribution of gas in physical space in our `Fast corona' model reveals that it is not a truly leading gas stream (cf. bottom row of Figures \ref{fig:ms2} and \ref{fig:rare}).

\begin{figure*}
\centering
\includegraphics[width=0.33\textwidth]{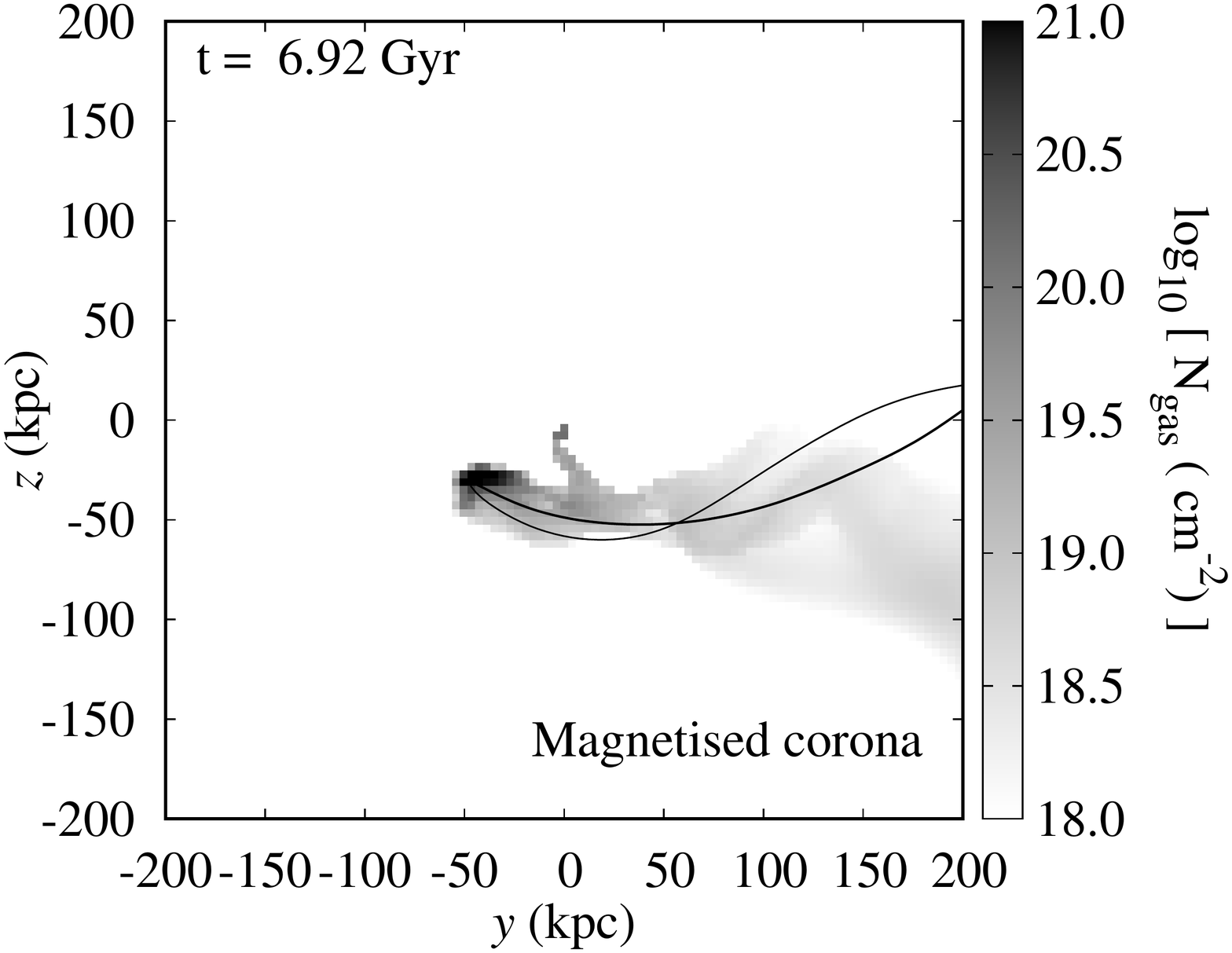}
\includegraphics[width=0.33\textwidth]{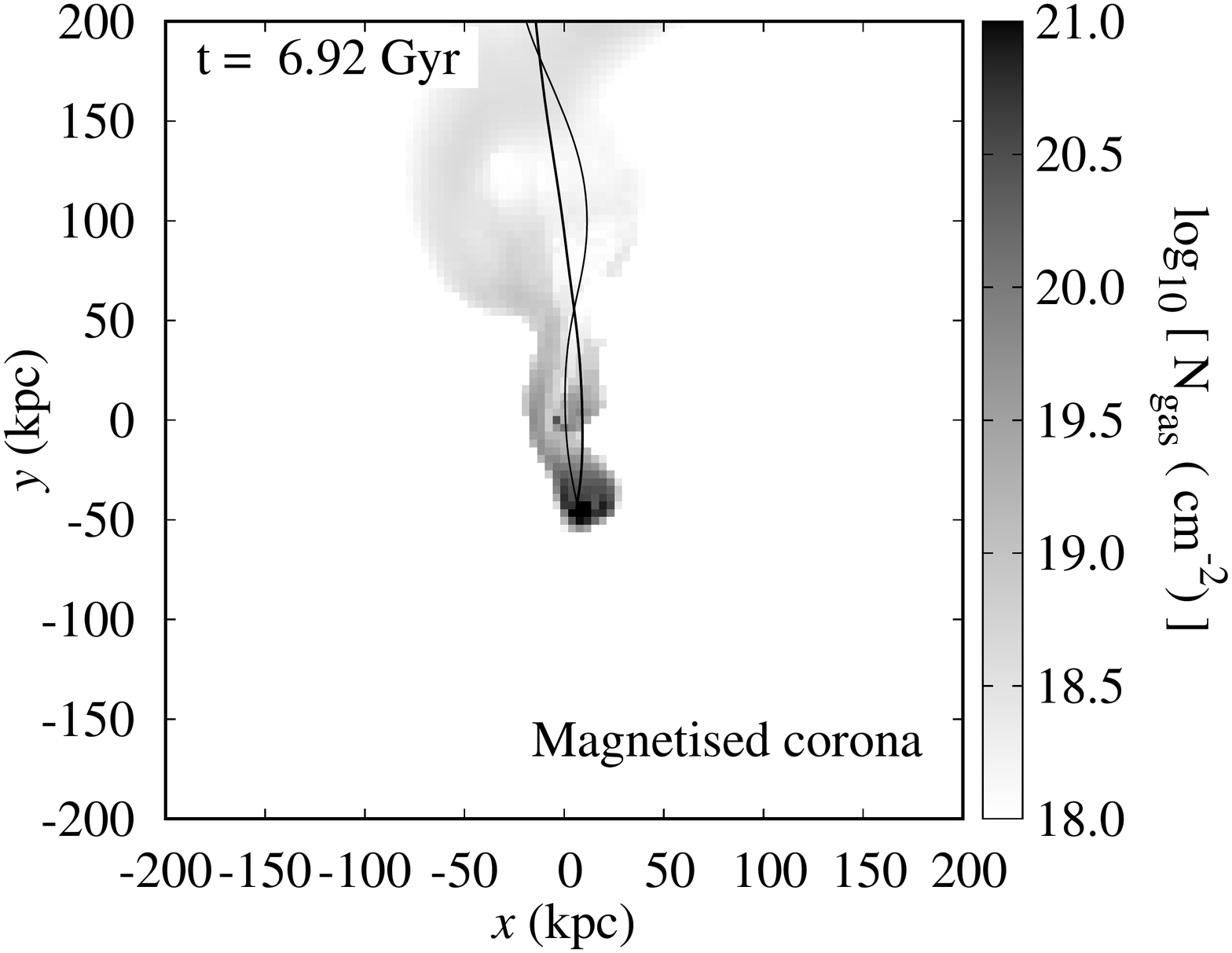}
\includegraphics[width=0.33\textwidth]{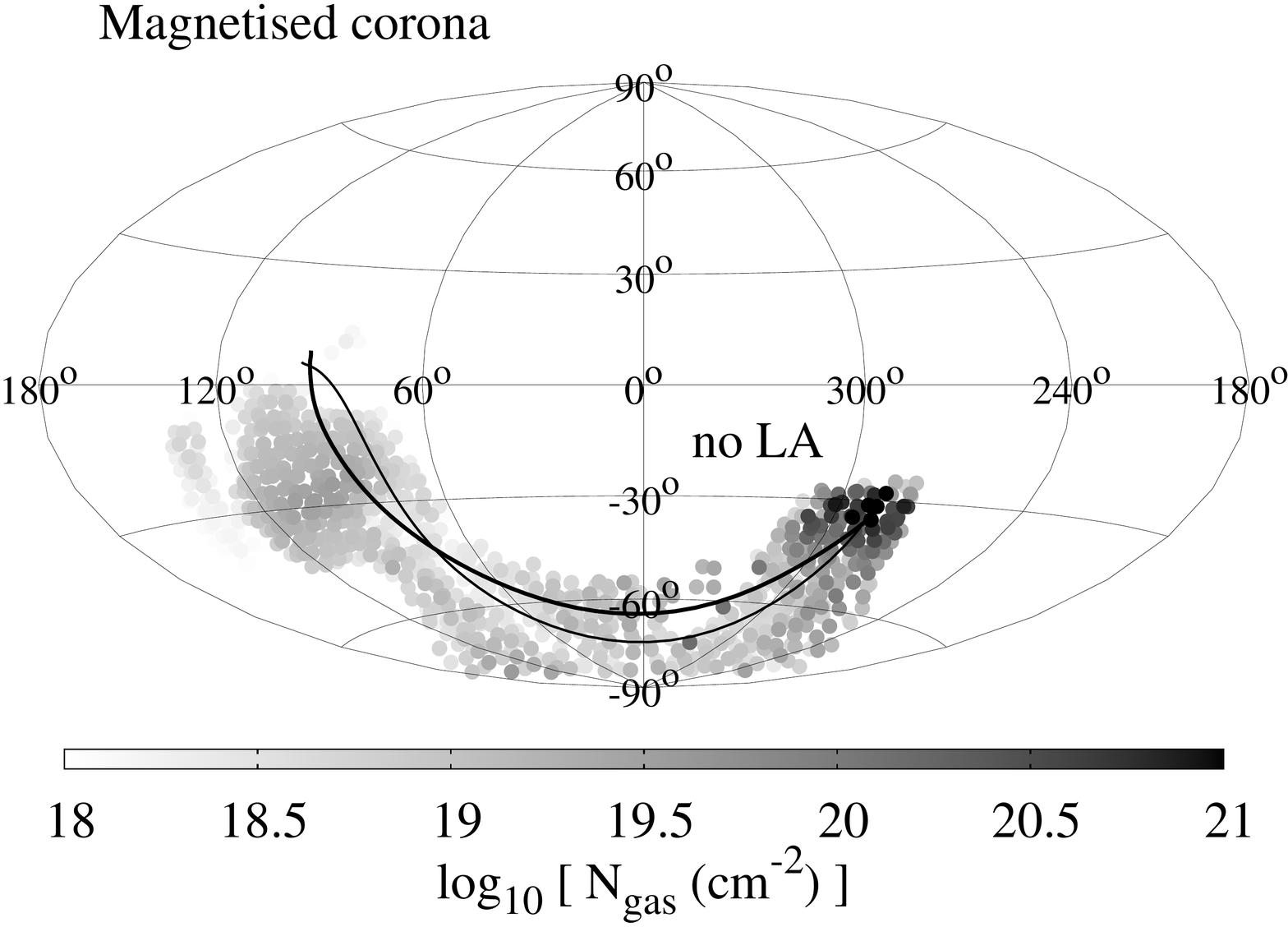}
\vspace{-10pt}
\includegraphics[width=0.33\textwidth]{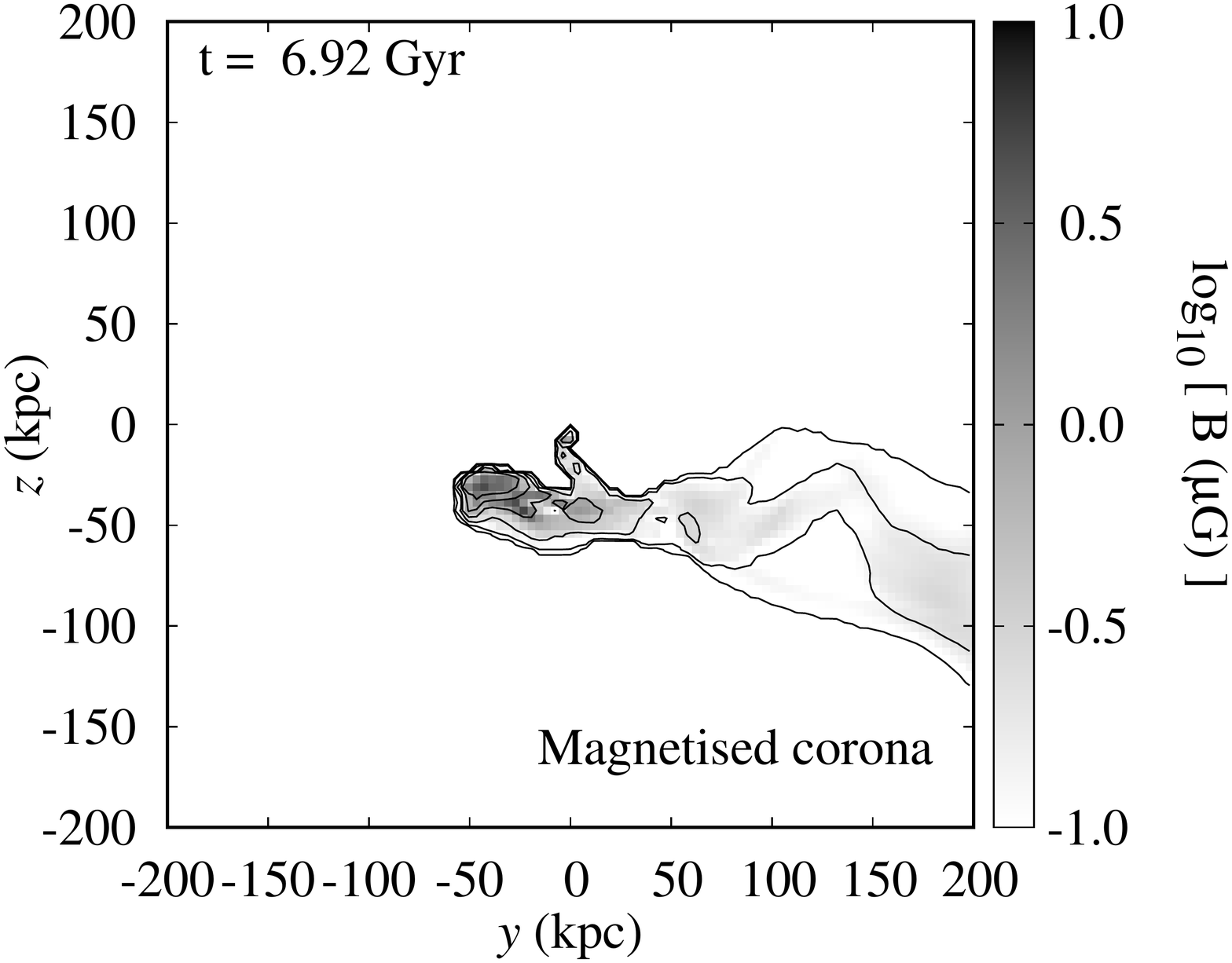}
\includegraphics[width=0.33\textwidth]{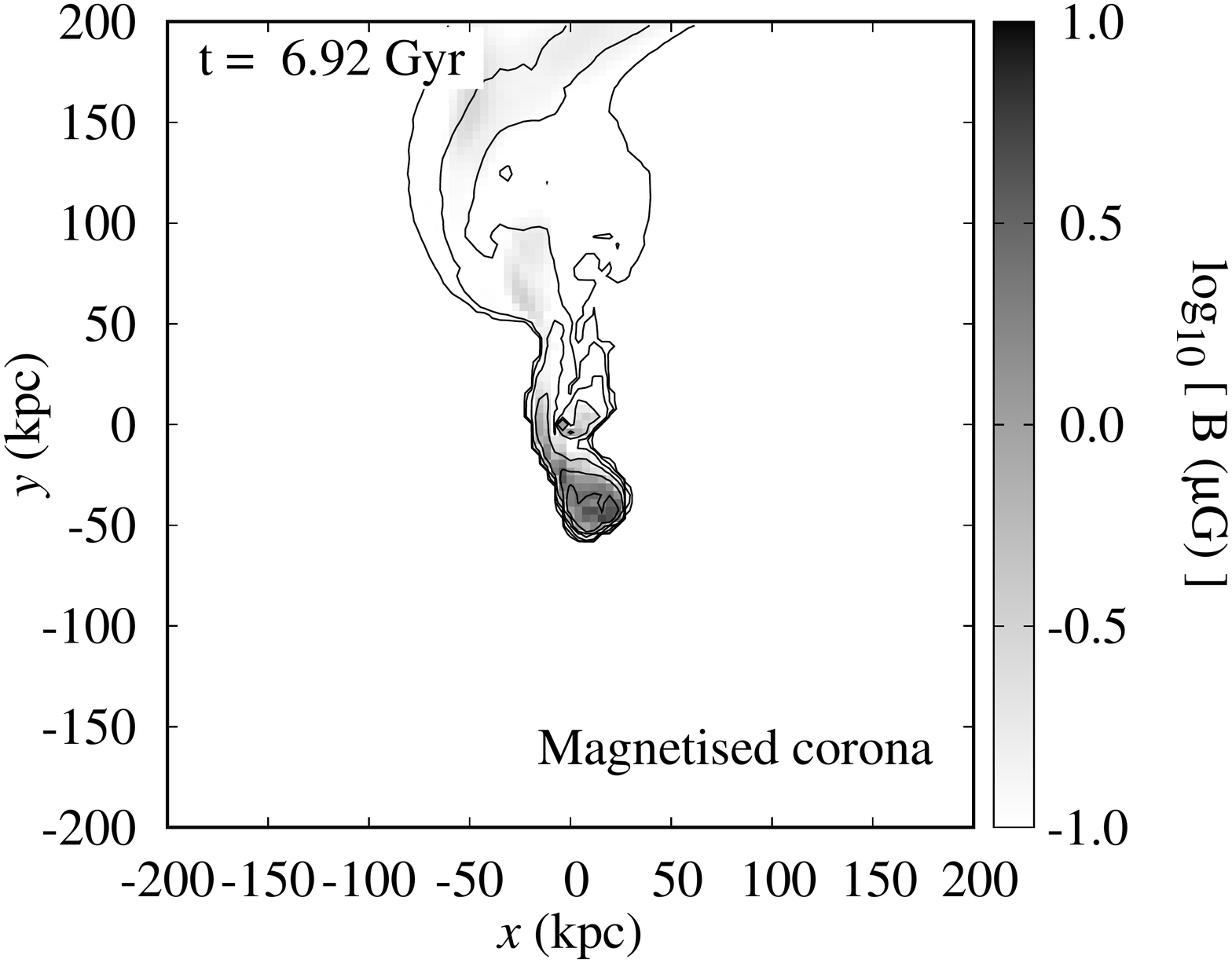}
\includegraphics[width=0.33\textwidth]{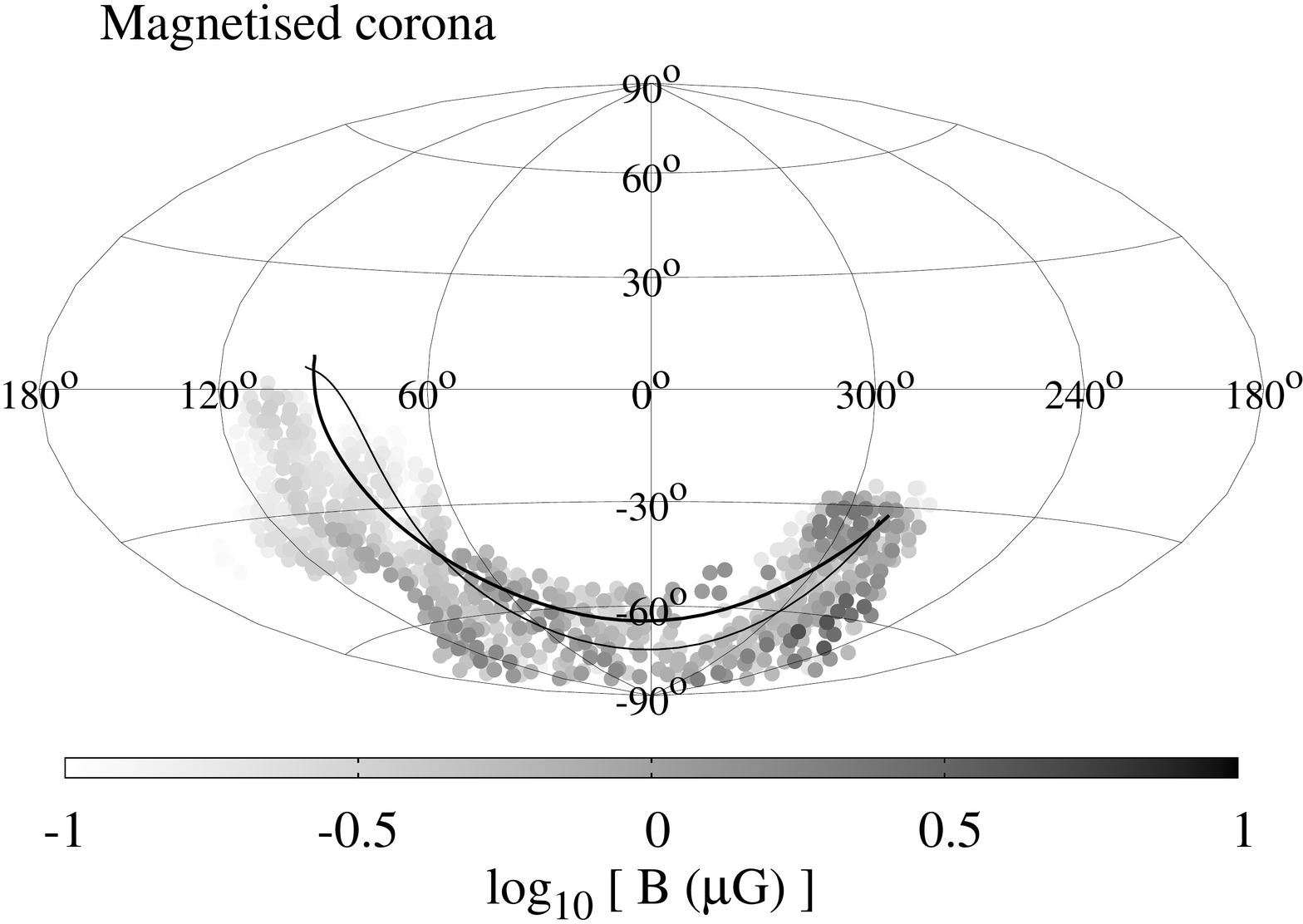}
\caption[  ]{ Distribution of the gas in the Magellanic System at the present epoch, $\sim1$ Gyr after infall into the Galaxy in the presence of a {\em magnetised}, non-rotating Galactic corona (cf. top panel in Figure \ref{fig:ms2}). Top: Total gas column density. Bottom: Magnetic field strength. The left and central columns correspond, respectively to an edge-on and face-on projection with respect to the Galactic plane in physical space. The contours in the bottom panels outline the total gas column density levels as shown in the top panels. The Galactic Centre (GC) is roughly at (0,-10,0) kpc. The right column shows the corresponding projection on the sky as seen from the Sun in Galactic coordinates $(l, b)$. The GC is at $(0,0)\deg$; the MCs', at $\approx (300, -30)\deg$. Their space orbits are indicated by the thin / thick solid curves in all panels, but they have been omitted from the bottom left and right panels for clarity. Apparently, the presence of an ambient magnetic field does not hinder the destruction of the leading gaseous stream. Note that the gas displayed in each panel correspond to gas with a tracer value $\geq 5$, and therefore excludes the overwhelmingly majority of the coronal gas.  }
\label{fig:ms3}
\end{figure*}

The absence of a leading gas stream in all models that include a Galactic corona can be understood in terms of drag forces. The magnitude of the drag force, or equivalently the ram pressure $P_{\rm ram}$, of the corona on the infalling gas associated to the Magellanic System depends on the mass density of the corona $\rho_{\rm c}$ and the infall velocity of the gas $\vec{v}_{\rm g}$ relative to the velocity of the corona $\vec{v}_{\rm c}$ as
\begin{equation} \label{eq:pram}
	P_{\rm ram} = \frac{1}{2} C_{\rm D} ~\rho_{\rm c} \left( \vec{v}_{\rm g} - \vec{v}_{\rm c} \right)^2 \,.
\end{equation}
The drag coefficient $C_{\rm D}$ is a parametrisation of the effective momentum transfer between the infalling gas and the corona \citep[][]{ben97a}. 

Both $\rho_{\rm c}$ and $v_{\rm g} \equiv \vert \vec{v}_{\rm g} \vert$ increase as the MCs approach the Galaxy, and so does the destructive effect of the Galactic corona on the infalling gas. Since the MCs move on a slightly prograde orbit with respect to corona's rotation,\footnote{The orbit is in fact near-polar.} the effective ram pressure exerted by the corona on the infalling gas is expected to be somewhat smaller in models with a faster spinning corona compared to models with a non-rotating corona.\footnote{Both these statements are easily proven for any $\vec{v}_{\rm g}$ and $\vec{v}_{\rm c}$ such that $v_{\rm c} < v_{\rm g}$, a condition that is satisfied in our models, with aid of the reverse triangle inequality, $$ \left\vert \vec{v}_{\rm g} - \vec{v}_{\rm c} \right\vert \geq \left\vert v_{\rm g} - v_{\rm c} \right\vert \, .$$ In the first case, consider a fixed $v_{\rm c}$. Trivially, the ram pressure increases because $v_{\rm g}$ increases. In the second case, consider a fixed $v_{\rm g}$. Here the ram pressure {\em decreases} because $v_{\rm c}$ {\em increases}. } These processes are all illustrated in Figure \ref{fig:pram}. There, we compare the evolution of the infalling gas in a model without a Galactic corona (left column), a model with a non-rotating corona (centre), and a model with a fast spinning corona (right). Each row corresponds to a different snapshot from the infall of the MCs up to $\sim100$ Myr before the present day. The pressure of the gas is indicated by the grey scale. Overlaid are a series of contours corresponding to total gas column density of the gas associated to the Magellanic System with values in the range $\log_{10} \left[ {\rm N}_{\rm gas} (\psc) \right] = (18,20.5)$ in steps of 0.5 dex.

\begin{figure*}
\centering
\includegraphics[width=0.24\textwidth]{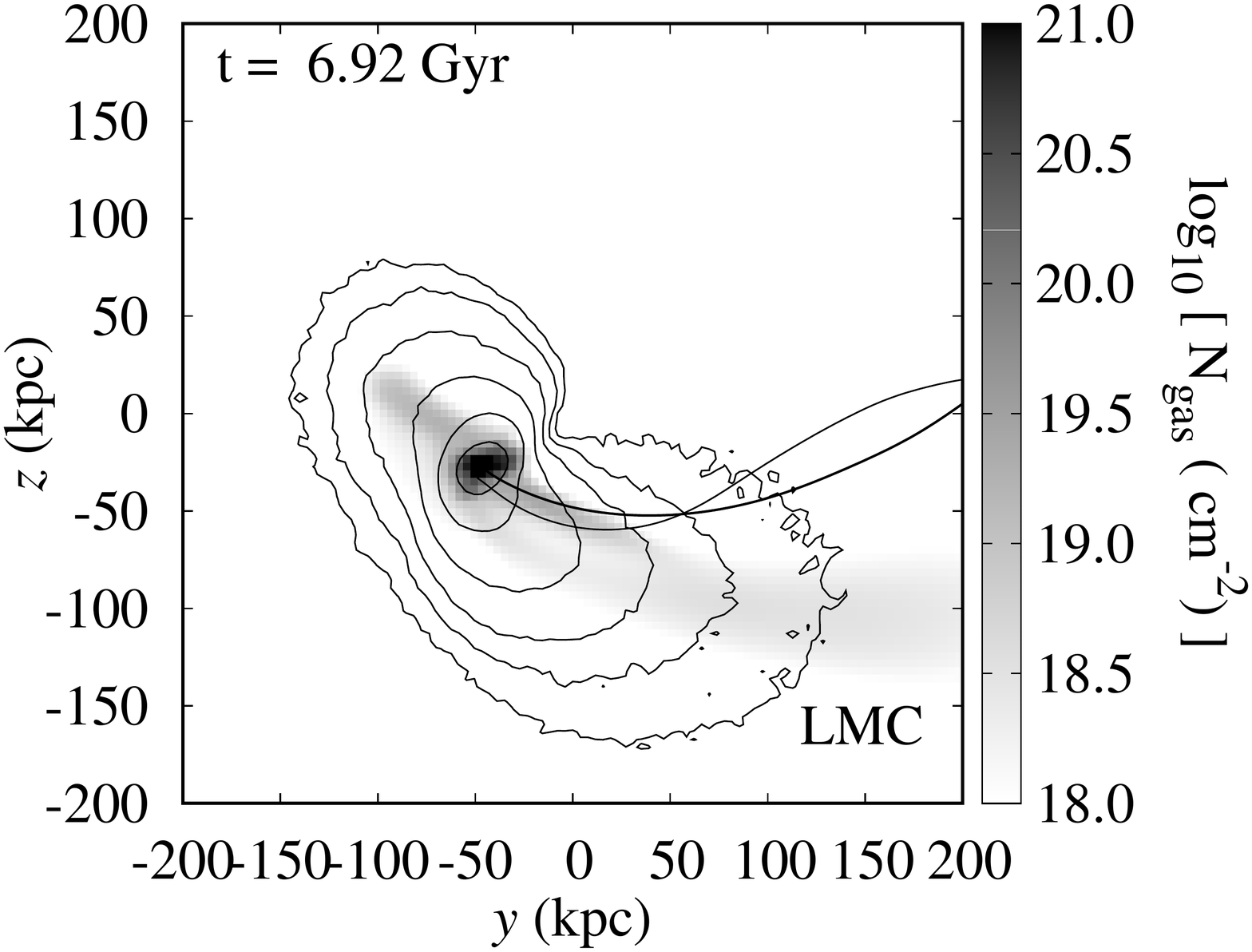}
\includegraphics[width=0.24\textwidth]{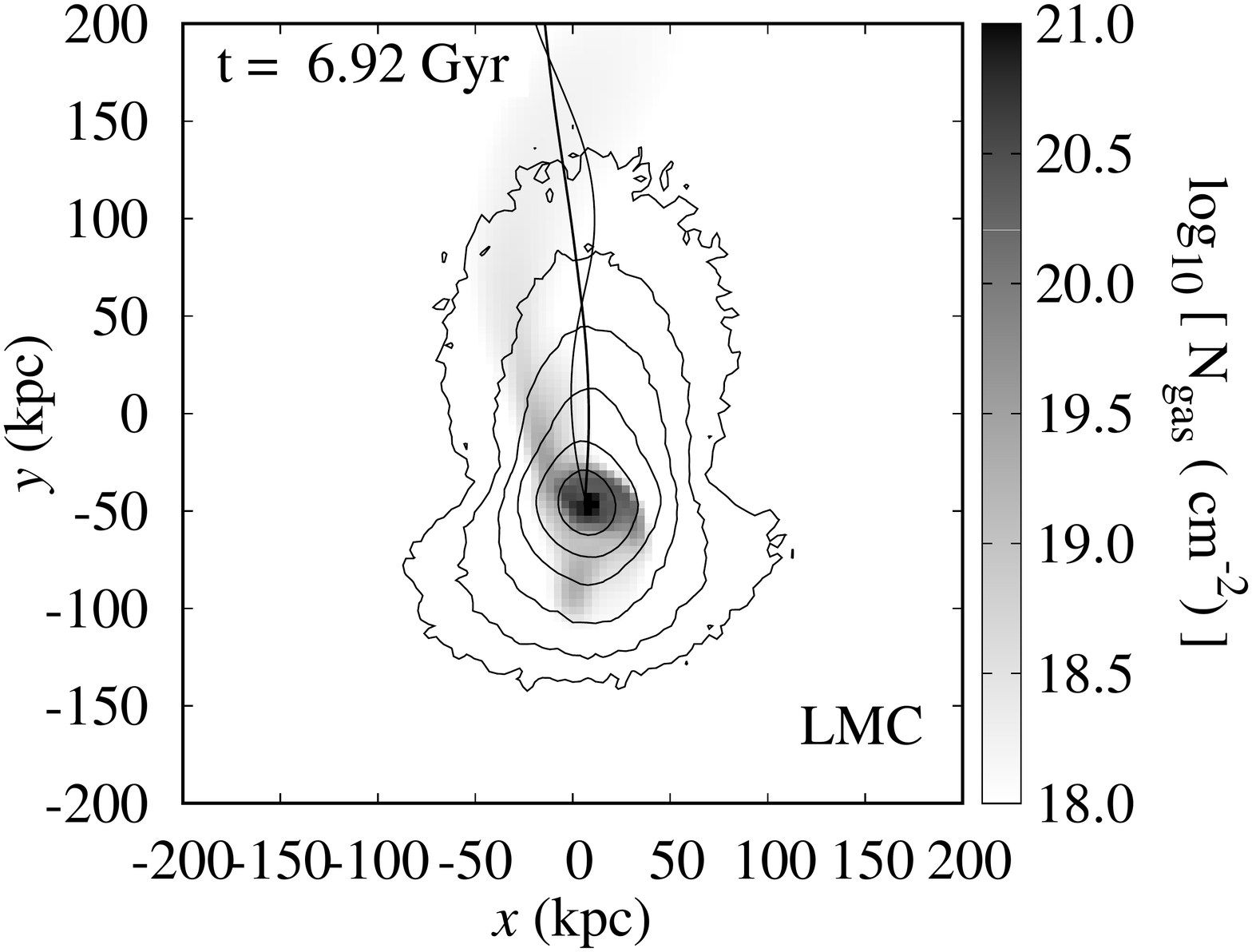}
\includegraphics[width=0.24\textwidth]{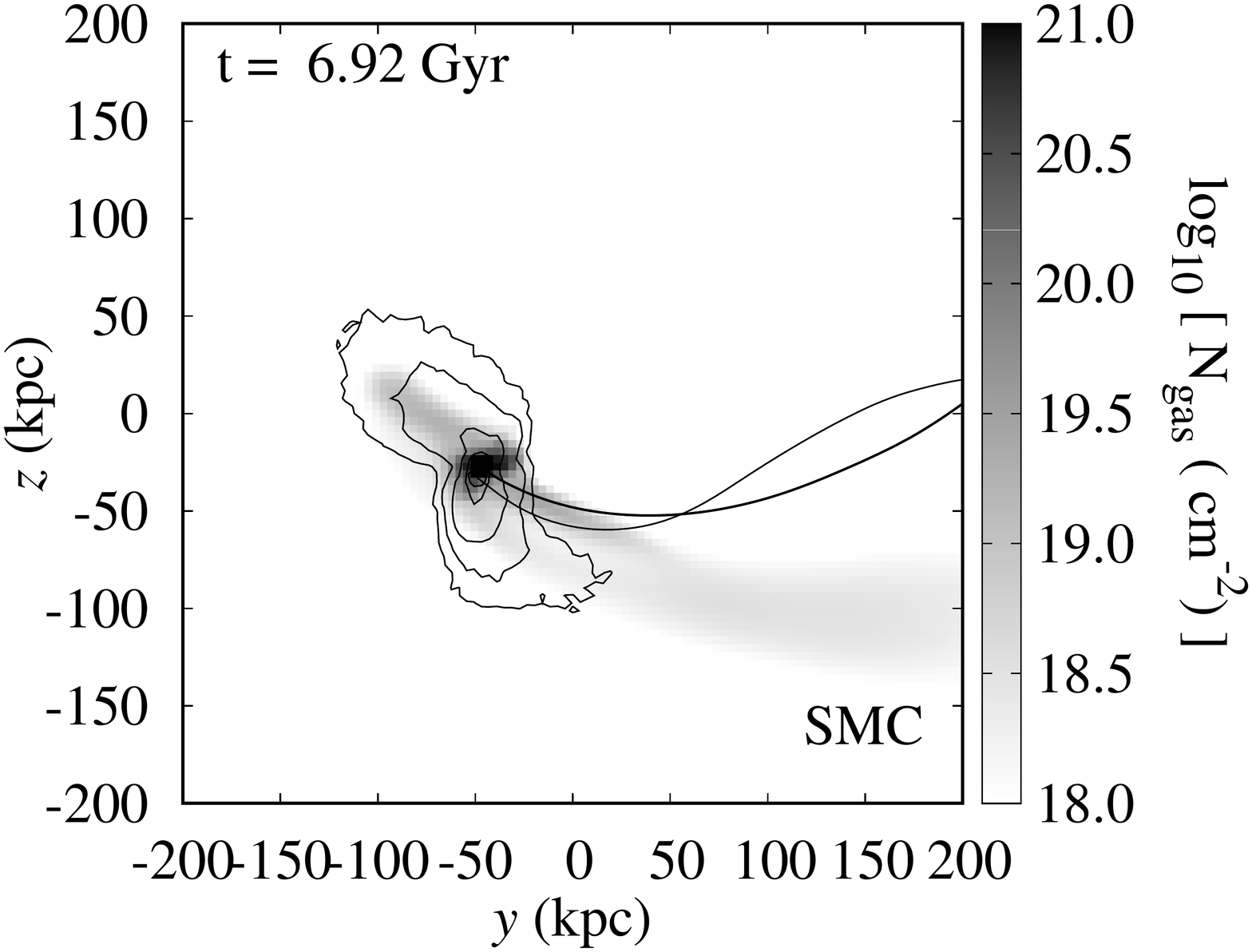}
\includegraphics[width=0.24\textwidth]{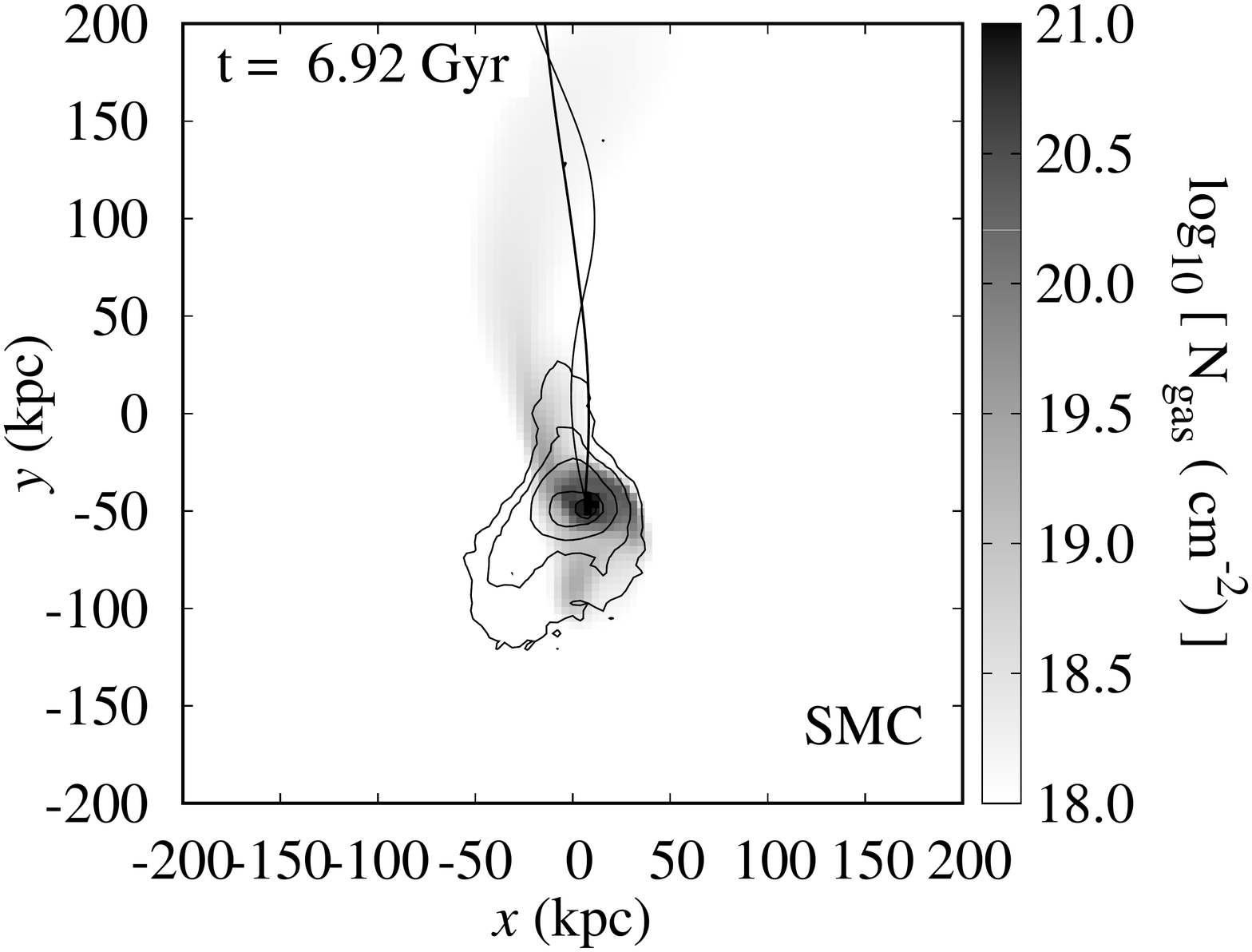}\\
\includegraphics[width=0.24\textwidth]{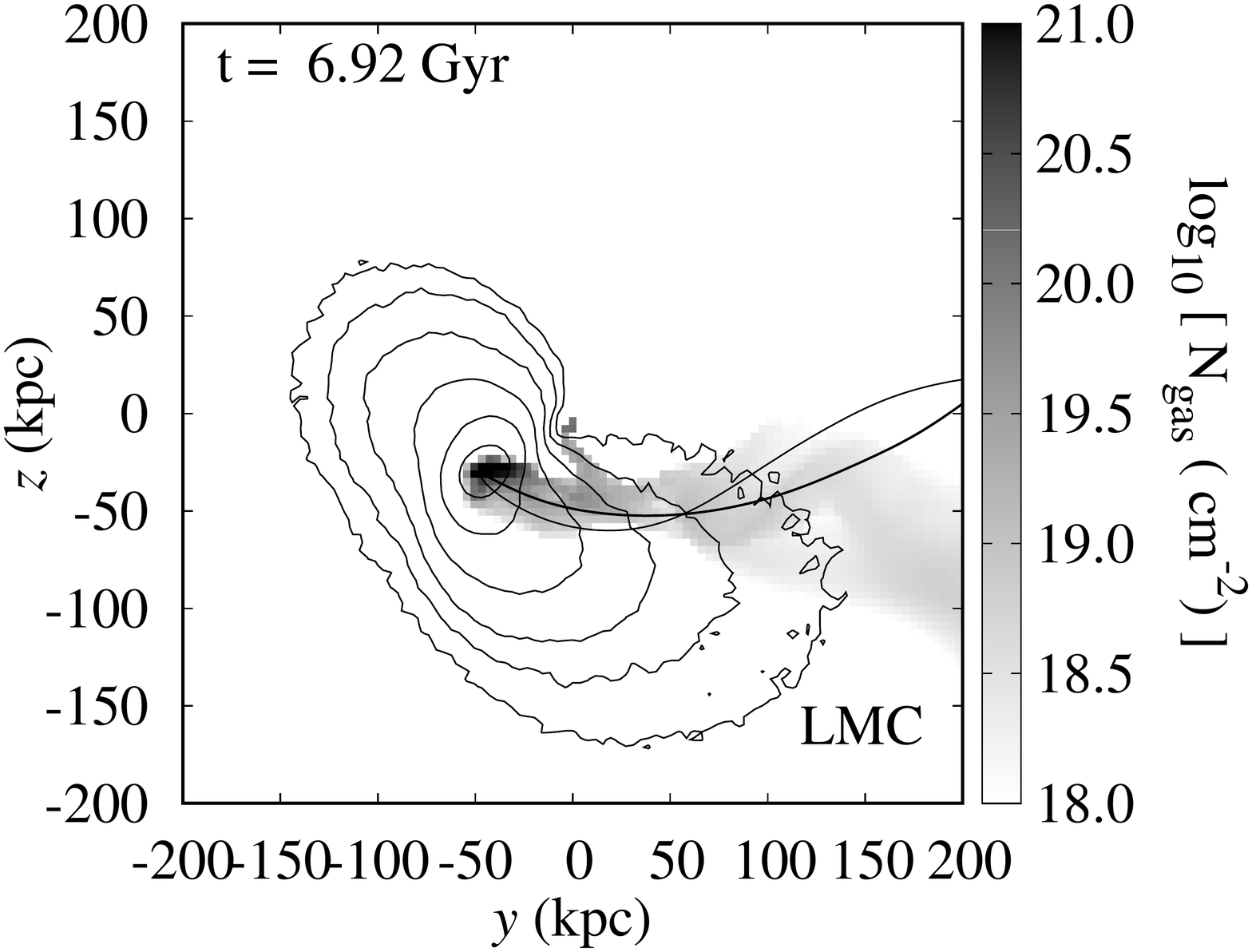}
\includegraphics[width=0.24\textwidth]{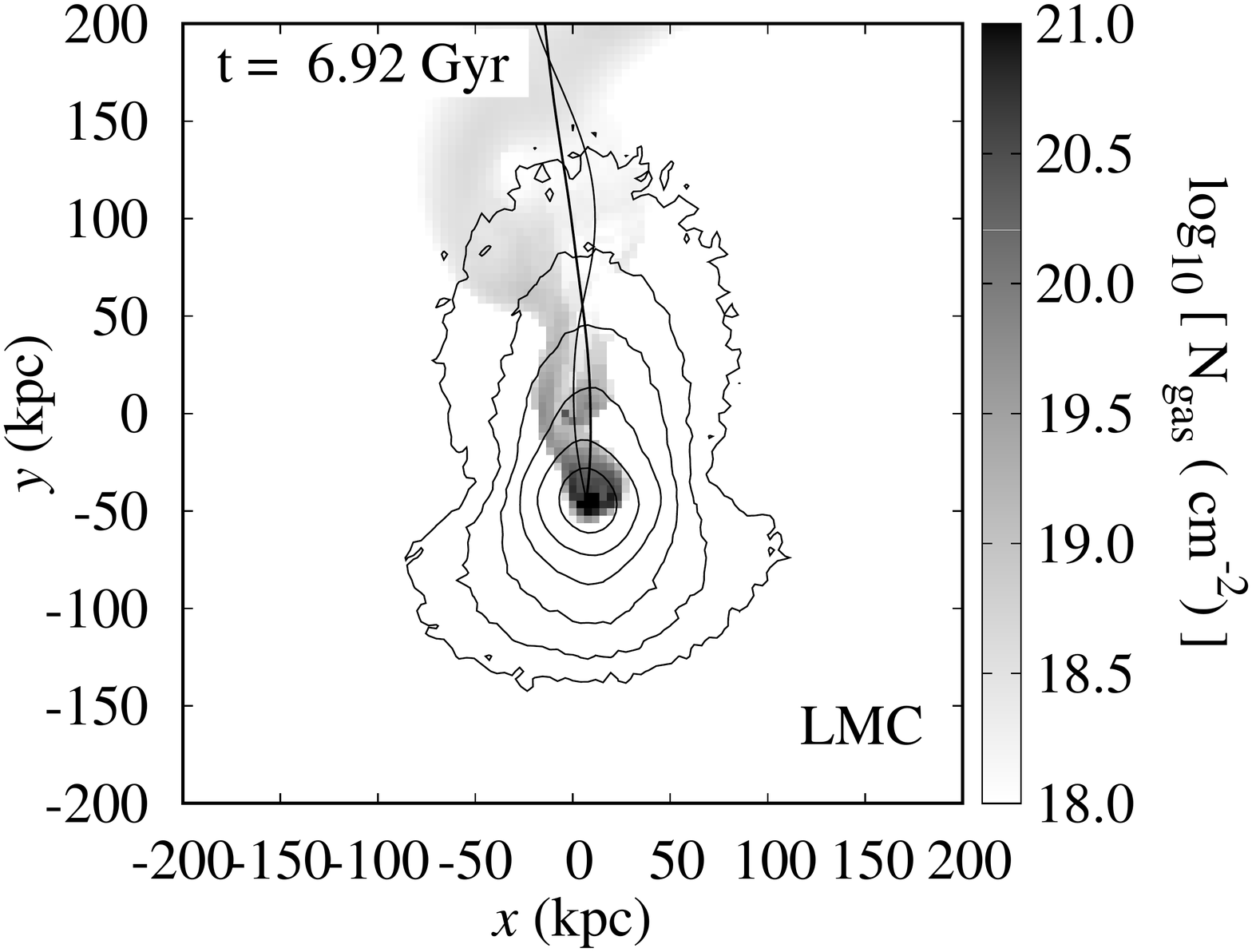}
\includegraphics[width=0.24\textwidth]{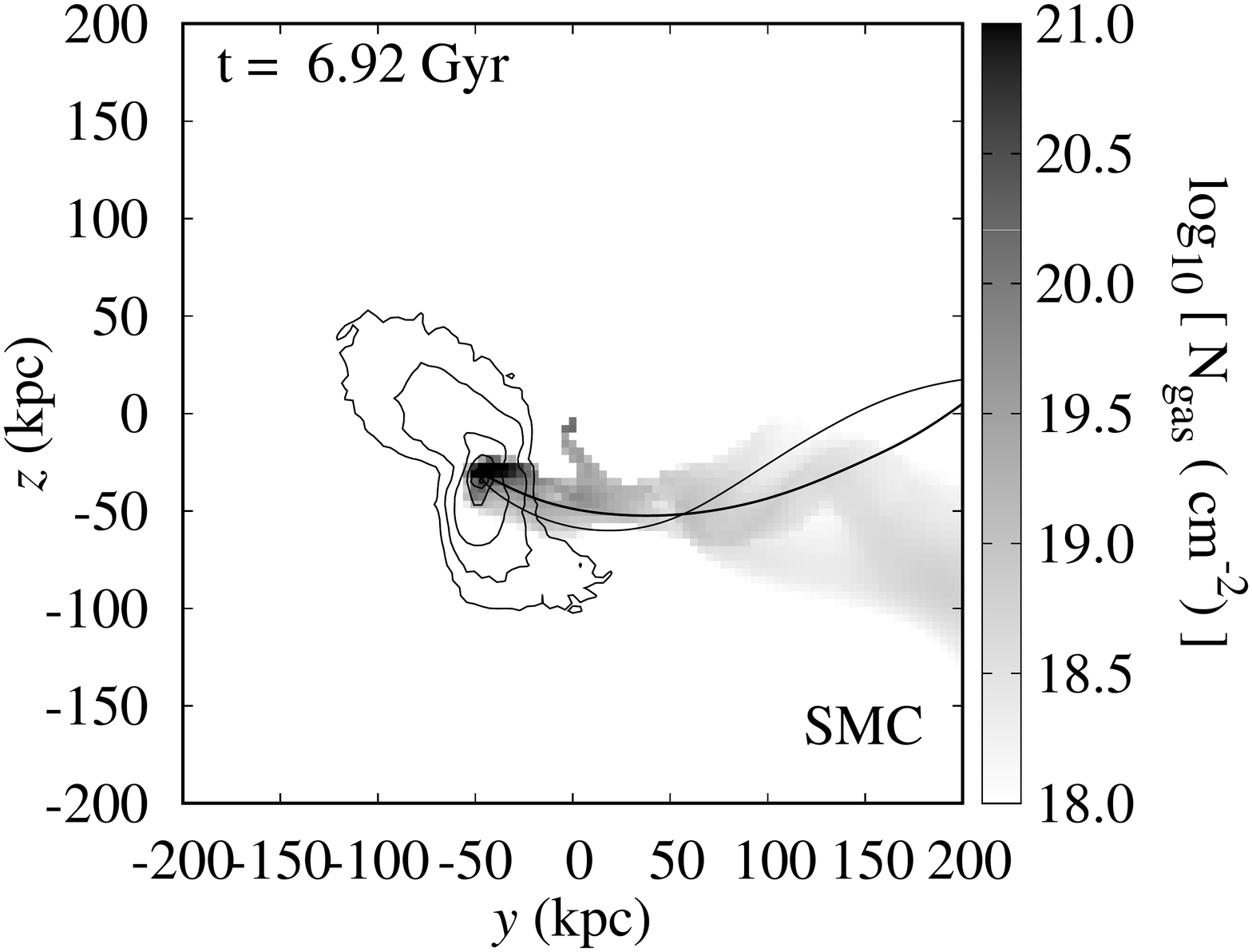}
\includegraphics[width=0.24\textwidth]{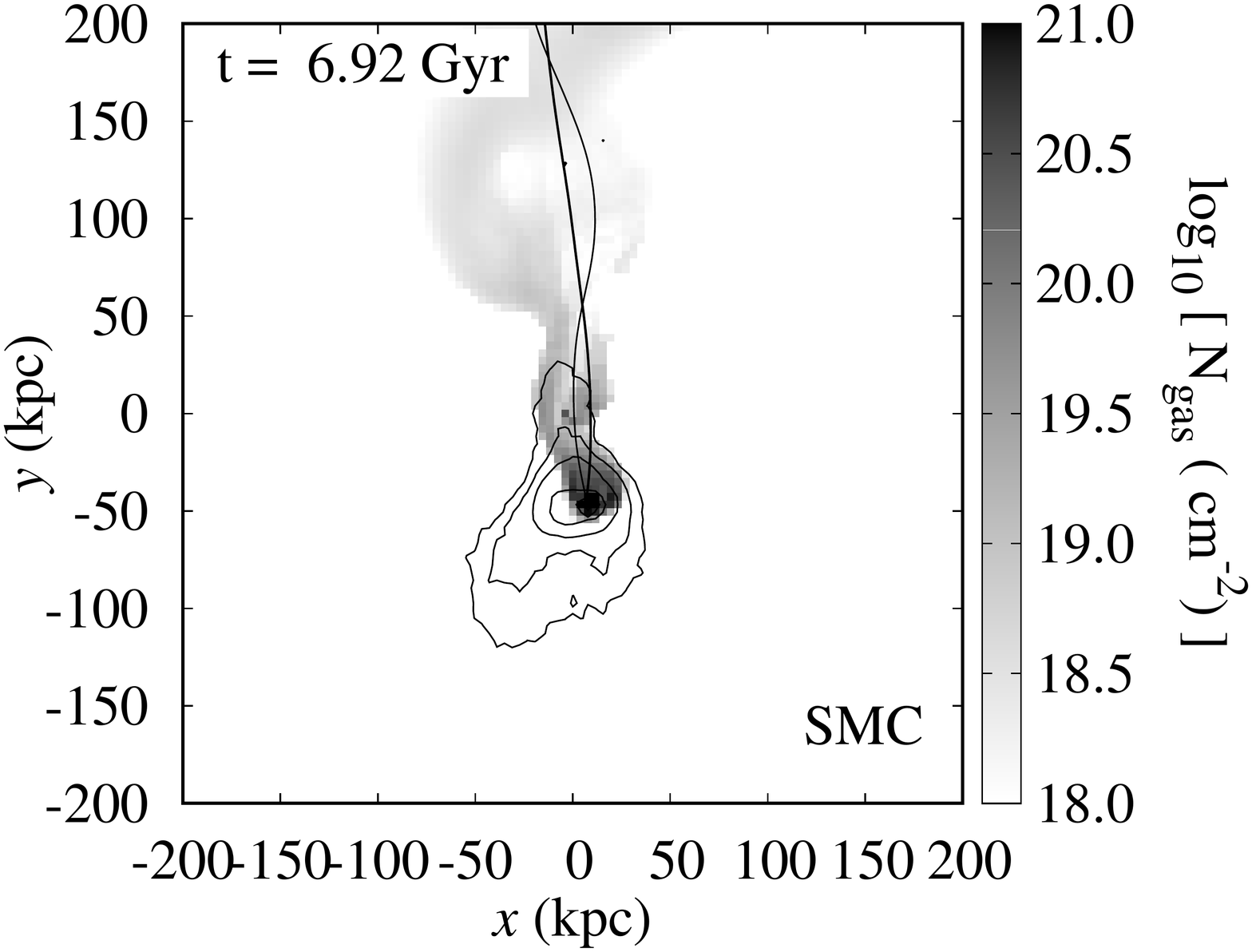}
\caption[ ]{ DM confinement. The panels display the total gas column density in physical space on an edge-on (columns 1 and 3) and face-on (columns 2 and 4) projection with respect to the Galactic plane, for model `Live DM Halo' (top row) and `Non-rotating corona' (bottom row). The contours indicate to the projected DM density ($\Sigma_{\sc \rm DM}$) of the halo initially around the LMC (columns 1 and 2) and the SMC (columns 3 and 4) in each corresponding model in the range $\log_{10} \left[ \Sigma_{\sc \rm DM} (\Msun\ \kpc^{-2}) \right] = (4.5,7)$ in steps of 0.5 dex. The space orbits of the MCs are indicated by the thin / thick solid curves in all panels. The total gas column densities and MCs orbits in each panel are identical to their counterparts in Figures \ref{fig:ms} and \ref{fig:ms3}.  Note that the result for the other corona models, in particular the `Magnetised corona', are comparable to the result shown in the bottom row.  }
\label{fig:dmconf}
\end{figure*}

In the absence of a Galactic corona (left column), the gas the MCs have stripped from one another while evolving in isolation, and which is surrounded {\em by design} by a gas structure consisting of a leading and a trailing components, simply becomes tidally stretched as the system falls into the Galaxy. In the presence of a Galactic corona (central and right columns), the infalling gas is affected early on by the corona's drag. The build-up of ram pressure ahead of the MCs and the resulting gradient along the gas as it falls in is apparent. Moving at roughly 200 \kms\ through a stationary corona with a density ${\rm n}_{\rm c} \sim 10^{-5}$ \pcc\  (at $\sim 200$ kpc), the leading edge of the gas experiences a ram pressure a few times $10$ \pcc\ K. The ram pressure is significantly higher at smaller distances because of the increasing ambient density. It appears somewhat lower if the corona is rotating, which is a consequence of two factors: 1) the reduced velocity difference (slightly prograde orbit); ii) the flattened density profile, which features a lower density at a given distance compared to a non-rotating corona (see Fig. \ref{fig:dens2}).

At infall, the leading gas stream that otherwise gives rise to the Leading Arm in corona-free models is pushed against itself by drag forces and `sweeps' past the denser gas around the MCs. The formation of the Leading Arm is hindered early on. The effect of the gas being compressed along its leading edge and ablated is consistent with the shock cascade found in our previous constrained simulations \citep[][]{bla07a,tep15a}. In consequence, the total mass in the trailing gas stream is slightly higher in these models compared to models that lack a corona.

It could be argued that the gas structures around the MCs that lead to the formation of the leading gas stream in corona-free models may be negatively impacted when the evolved isolated binary system is placed into the model Galaxy that features a corona. To test our results against a potential numerical artefact, we have run a control model similar to the `Non-rotating corona' model, but with the initial extension of the Galactic corona limited to 150 kpc (rather than $\sim 250$ kpc). In this model, the MCs and their associated gas do not interact with the corona until their reach a Galactocentric distance of $\sim 150$ kpc. Reassuringly, we find no significant difference in the final configuration in these control models compared to our models presented here. In particular, none of the control models displays a leading gas stream (not shown).

We note finally that, because the drag depends on the mass density, rather than the total particle density, of the ambient medium, models that take into account the ionisation state of the gas (which we have ignored in our models) should not yield significantly different results.

\subsection{Confinement by an ambient magnetic field} \label{sec:mag}

The Galactic corona and disc gas have long been known to be magnetised \citep[][]{bec96a}, with a complex structure and a strength that generally declines rapidly away from the Galactic plane \citep[][]{jan12a}. Moreover, there is evidence for the presence of an enhanced, coherent magnetic field around the Leading Arm\footnote{In the region identified as LA~II north of the Galactic plane \citep[][]{ven12a}.} with a magnitude along the line of sight $\gtrsim 6$ \mG\ \citep[][]{mcc10b}.

Early studies based on two-dimensional (2D) simulations of the interaction between gas clouds and magnetised media \citep[][]{mac94a,jon96a,kon02a} found that the presence of even a weak ambient magnetic field can extend a cloud's lifetime by draping around the gas and dampening the onset of boundary instabilities that eventually lead to the destruction of the cloud. This result has been confirmed by recent, more advanced 3D numerical experiments \citep[][]{mcc15b,ban18a}, which also have shown that the draped field around the infalling gas can be amplified by factors of $\sim 100$ relative to the background \citep[][]{gro17a}.

But magnetic fields may have a negative impact on infalling clouds as well. Sophisticated 3D simulations \citep[e.g.][]{kwa09a,gro18a} demonstrate that clouds moving through a magnetised medium experience a somewhat enhanced drag, compared to initially identical clouds moving in a non-magnetised medium. In other words, infalling gas clouds may be significantly delayed from accreting onto the Galaxy as a consequence of the  Galactic global magnetic field \citep[see also][]{bir09a}. Thus, whether and on which timescale the infalling gas reaches the disc will depend on the competition between two opposing effects: An enhanced drag force on the one hand, and an extended cloud lifetime on the other.

Our model dubbed `Magnetised corona' is designed to investigate the outcome of this competition. As detailed in Section \ref{sec:mod2}, we set up a model where the Galactic corona is initially identical to the `Non-rotating corona' model, but threaded by an initially uniform magnetic field at a level of 0.1 \mG, and use it to recompute the evolution of the Magellanic System.\footnote{Given the absence of any significant dependence on the kinematic properties of the corona discussed in Section \ref{sec:gas}, for simplicity we do not consider here a spinning, magnetised corona model.} The result is shown in Figure \ref{fig:ms3}. In agreement with previous studies, we find that the motion of the infalling gas trough the magnetised corona sweeps-up the ambient magnetic field, yielding a draped, enhanced field with a magnitude $\sim 0.1 - 10$ \mG\ all around the infalling gas (see bottom panels), a range of values consistent with observations. Although we do not see evidence for an enhanced drag compared to the results of our previous models, the leading gas stream apparently does not survive in this model either.

It is fair to ask whether this result is sensitive to initial configuration and magnitude of the magnetic field we have chosen, both of which are somewhat arbitrary. As by shown previous work \citep[][]{kwa09a,gro18a}, an initially uniform ambient magnetic field transverse to a gas cloud's motion enhances its survival. Also, the shielding effect increases with the field strength. Therefore, our adopted initial field configuration, with an strength far away from the Galaxy that is high relative to the strength of the Galactic magnetic field at the same location, and a significant transverse component along the orbit of the MCs, should favour the survival of the infalling gas.

We thus conclude that the presence of an ambient magnetic field does not help the leading gas stream survive.

\subsection{Confinement by dark matter} \label{sec:dm}

Finally, we explore whether the survival of the Leading Arm can be attributed to the confinement of the infalling gas by DM. As suggested by previous work \citep[][]{qui01a,nic14b,plo12a,tep18a}, a DM halo around the infalling gas may reduce the stripping effect of ram pressure by confining the baryons within its potential well. In this scenario, DM tidally stripped from the MCs ahead of their orbit could have provided the required confinement to the Leading Arm, through the Galactic corona (and eventually across the Galactic gas disc).

In order to assess whether DM confinement may have played a role in the survival of the Leading Arm, we analyse the distribution of DM associated to the MCs at the present epoch in two of our models: `Live DM halo' and `Non-rotating corona'. The results are shown, respectively, in the top and bottom rows of Figure \ref{fig:dmconf}. There, we show for each of these models the gas distribution in physical space along two orthogonal projections. Overlaid we show the projected DM density initially associated to either the LMC or the SMC. The result for the other corona models, in particular the `Magnetised corona', are comparable to the result of the `Non-rotating corona' model, and are therefore not discussed here.

Clearly visible are, for each dwarf, a leading and a trailing DM stream. Since the LMC is initially larger, the streams associated to its DM subhalo are more extended. It is worth emphasising that the DM distribution of either dwarf galaxy is similar in both models. This is expected since DM, being collisionless by assumption, should be sensitive only to tidal forces. Since the potential is dominated globally by the Galactic DM halo and locally by the DM subhalos of the dwarfs, both of which are initially identical in both models, there is no reason for the DM distribution to be too dissimilar between models.

The DM streams around the MCs in the `Live DM halo' model (top row) appear to be confining the leading gas stream, and also -- although to a lesser degree -- part of the trailing gas stream. However, we just mentioned that the DM streams are similar in both models. Therefore, if DM confinement were at play here, the same effect should be observed in the `Magnetised corona' model (bottom row). Clearly, this is not the case.

\section{Discussion} \label{sec:dis}

\subsection{The need for ram-pressure models} \label{sec:rpm}

The difficulties arising in `ram pressure models' when attempting to reproduce a leading gas stream, with either an analytic description of the gas drag \citep[][]{meu85a,hel94a,moo94a} or with a self-consistent hydrodynamic treatment \citep[][]{mas05a}, were recognised early on. This circumstance led to the view that the competing `tidal models' \citep[e.g.][]{fuj76a,mur80a,gar94a} appeared more successful in explaining the formation of the Magellanic System \citep[e.g.][]{put98a}. But new, accurate proper motion measurements of the MCs \citep[][]{kal06a} would lead to a radical change of view. Specifically, these measurements implied that the MCs are likely on their first approach to the Galaxy \citep[`first-infall scenario';][]{bes07a}, as had been suggested much earlier \citep[q.v.][]{lin95a}. This implication was only reinforced by the latest measurements of the MCs' proper motions \citep[][]{kal13a}.

If the MCs are truly on their first approach to the Galaxy, there has not been enough time for the tidal interaction with the Galaxy to allow for the formation of the Magellanic System. Thus it was argued \citep[][]{bes10a} that the key to explaining the origin of the Magellanic System lies in the interaction between the LMC and the SMC prior to being accreted by the Galaxy. \citet[][]{bes12a} showed that many (but not all) of the properties of the Magellanic System can be reproduced by a model constructed based on this paradigm. Nevertheless, the result that the measurements of the MCs' proper motions favour a first approach of the MCs onto the Galaxy over alternative orbital histories depends strongly on the Galactic potential, and it is also subject to the condition that the MCs have been a bound binary pair for a significant fraction of the Hubble time; a strong assumption indeed. In fact, others \citep[][]{dia12a} have shown that the long-term connection of the MCs and their first approach to the Galaxy are unwanted assumptions in order to reproduce some key properties of the Magellanic System {\em not} reproduced by first-infall models, such as the observed spatial and kinematic bifurcation of the Stream \citep[][]{put03a,nid08a}.

On the other hand, circumstantial evidence for the long-term association of the LMC and the SMC is provided by a study of the global star-formation history of the LMC, which suggests that the LMC and SMC may have had coupled episodes of star formation, one occurring roughly 2 Gyr ago and the other some 500 Myr ago \citep[][]{har09a}, indicating that the MCs have been a strongly interacting pair in the past \citep[but do not need to be bound; see ][]{gug14a}. Moreover, the existence of a large number ($\sim 70$) of Galactic satellites associated with (and moving ahead of) the LMC as suggested by data of the Dark Energy Survey  provides strong and independent evidence for a first passage of the LMC around the Galaxy \citep[][]{dea15a,jet16a}. We thus consider it highly likely that the MCs are on their first approach as a binary pair.

A noteworthy caveat of \citet[][]{bes12a}'s model is that it ignores the presence of the Galactic corona. First predicted by \citet[][]{spi56a} to explain the long-term survival of \HI\ high-velocity clouds in the halo of the Galaxy, the presence of a diffuse (${\rm n} \sim 10^{-5} - 10^{-2} ~\pcc$), hot ($\sim 10^6$ K), gaseous component encompassing the Galaxy is now well established \citep[q.v. \citealt{bre18a};][]{sor18a}. The impact of a diffuse background on the evolution of infalling gas clouds has been demonstrated repeatedly by a number of independent studies (see below). Therefore, ram pressure models are no longer a matter of choice; they are a physical necessity.

We have shown here that a new ram pressure model based on a realistic description of the Galactic corona, coupled to the first-infall scenario, does {\em not} allow for the formation of a leading gas stream like the Leading Arm. 

\subsection{Potential limitations} \label{sec:lim}

The results discussed above are robust to variations in the kinematic properties of the corona, or to the absence~/~presence of an ambient magnetic field. 
However, one may wonder whether our main result -- the absence of a leading gas stream in models that feature a Galactic corona -- is affected by a lack of adequate numerical resolution.

It has long been known that the hydrodynamic interactions governing the evolution of gas clouds moving through a diffuse medium depend on resolution. Recently, \citet[][]{hum18a} have shown that the presence and distribution of cool gas in the circumgalactic medium of galaxies can be dramatically affected by resolution. At a more fundamental level, a progressively increasing resolution has been shown to reduce the efficiency of cloud acceleration \citep[i.e. drag;][]{sch17a}. However, these authors also point out that a cloud's survival is appreciably reduced when the resolution is increased. In other words, at a higher resolution the drag experienced by a cloud may be less efficient, but at the same time the cloud is destroyed more rapidly.

A number of studies based on wind-cloud simulations show that infalling clouds with HVC-like densities typically break apart within a few 100 Myr, regardless of the presence of magnetic fields \citep[e.g.][]{gre00a,hei09b,kwa11a,ban16a}. Hence, while the typical survival timescale is much shorter than the infall timescale of the MCs onto the Galaxy in our models ($\sim 1$ Gyr), it is not obvious that our adopted resolution is adequate to properly account for the hydrodynamic interaction between the infalling gas and the hot corona.

In order to address this potential shortcoming, we have run two additional simulations, which are in essence identical to our `Slow corona' model, but each at a progressively increasing resolution. The details and result of this experiment are discussed in Appendix \ref{sec:conv}. In brief, we find no appreciable difference between these runs; i.e. in none of them does a leading gas stream survive. This experiment strongly suggests that, while gas drag is less efficient at higher resolution, the leading gas clouds are destroyed early on regardless.

Based on these tests, we conclude that our results -- in particular the destruction of the leading gas stream -- have not been affected in any significant way by our adopted standard limiting resolution. Our belief is strongly supported by similar results found by previous,  albeit less sophisticated ram pressure models of the Magellanic System (see above).

Similarly, we do not believe that our main result is overly sensitive to the physical processes  -- notably star formation and its associated feedback -- we have ignored in our models. We do so even in spite of the early suggestion that the Leading Arm may be a feature associated with stellar feedback within the LMC \citep[the `blowout hypothesis';][see also \citealt{ven12a}]{nid08a}. According to this scenario, the Leading Arm was created out of gas outflows from a strong star-forming region in the LMC. However, recent numerical work \citep[][]{sal15a,bus18a} shows that outflows coupled to the effect of ram pressure may indeed aid to the formation of a {\em trailing} gas stream, but they do not contribute in any significant way to the formation of a {\em leading} gas stream.
We stress that if we insisted in including stellar feedback in our models, a fully self-consistent treatment would require us to take into account the feedback from the Galaxy (i.e. Galactic winds), which may only further contribute to the destruction of the infalling gas \citep[e.g.][]{coo09a}.\\

So, taking the results presented here at face value, how can we understand the existence of the gas referred to as `Leading Arm'? In our view, there are only two plausible interpretations, both with far-reaching consequences and issues, which we discuss in the following.\\

\subsection{An alternative Galactic corona model} \label{sec:alt1}

\begin{figure*}
\centering
\includegraphics[width=0.33\textwidth]{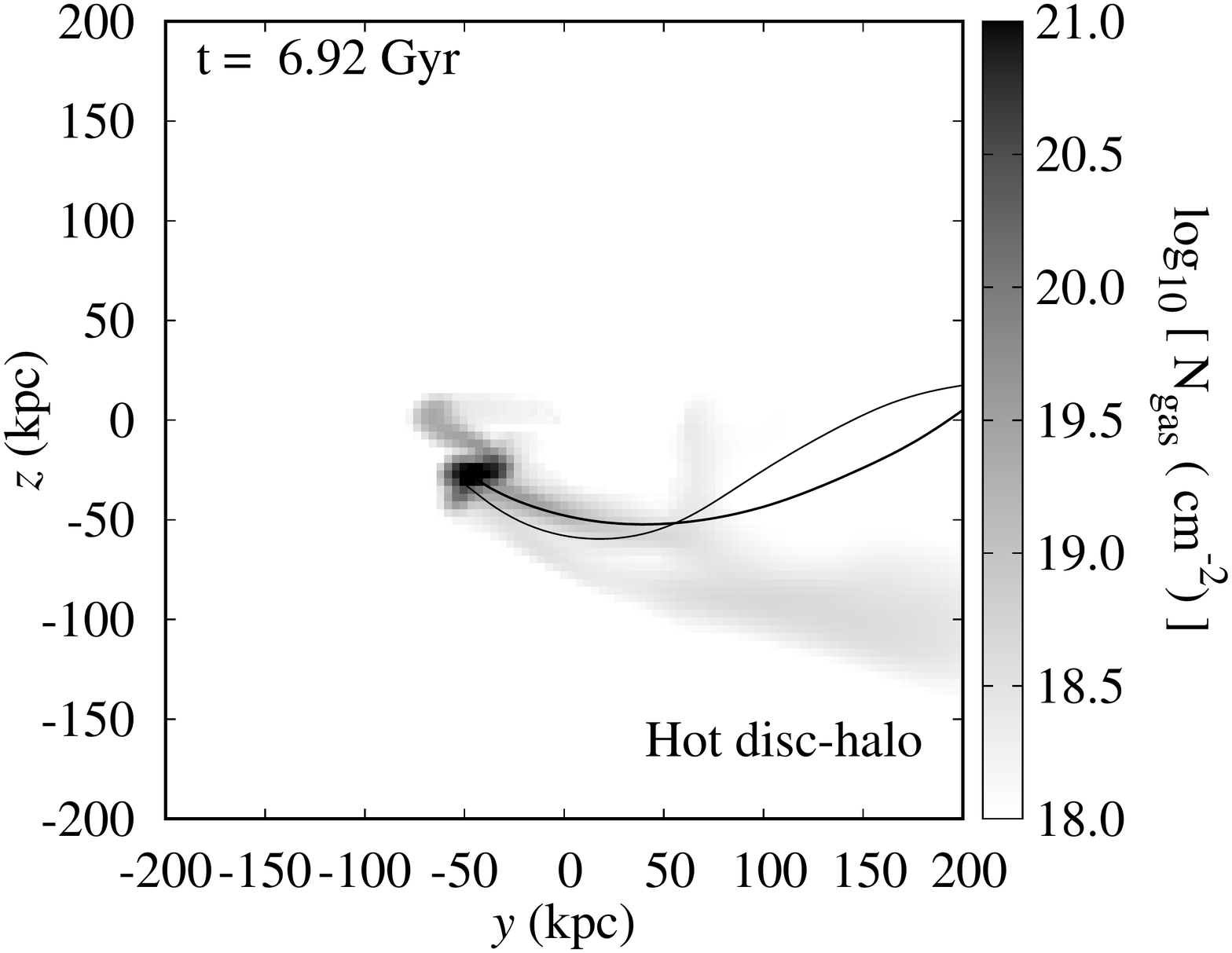}
\includegraphics[width=0.33\textwidth]{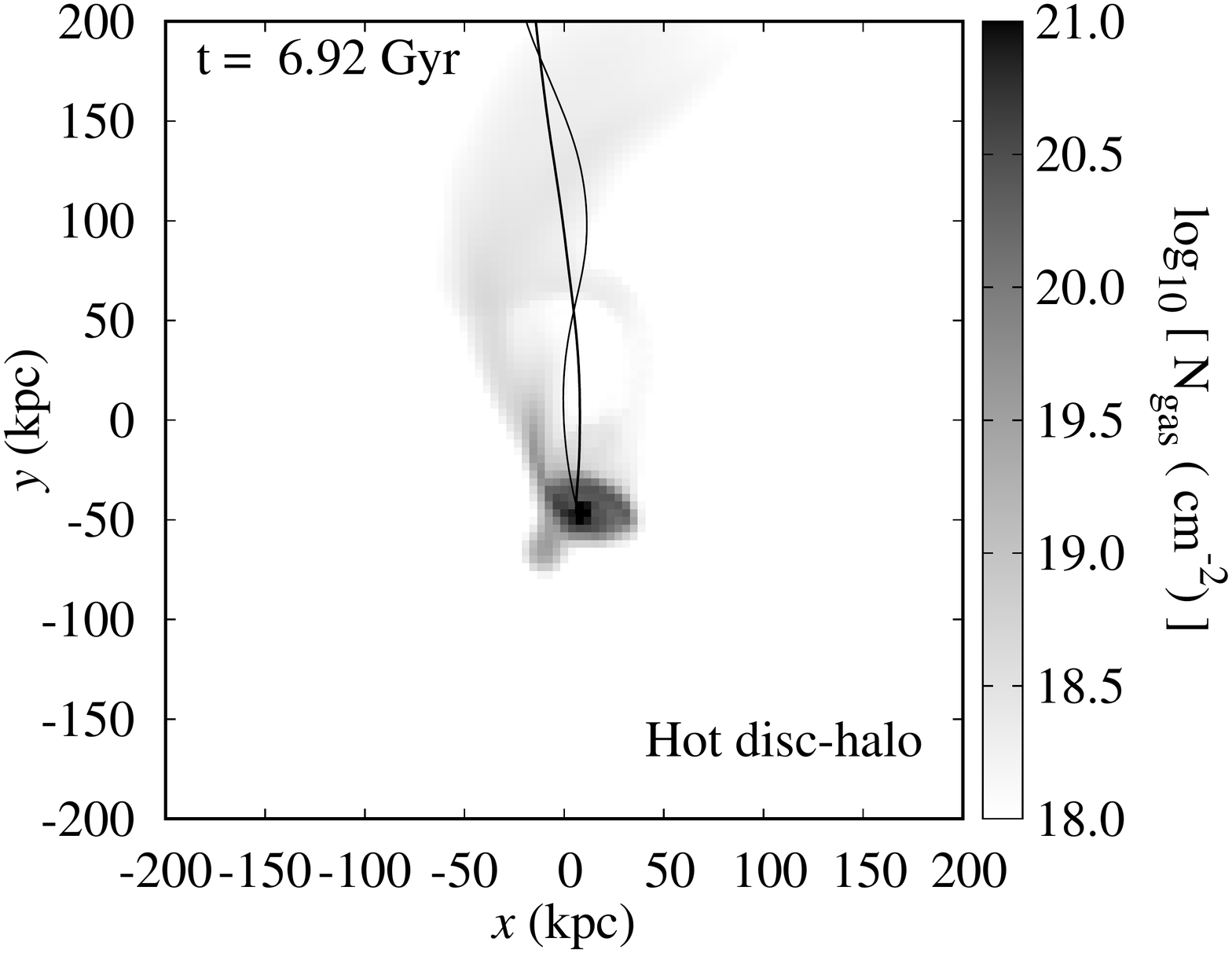}
\includegraphics[width=0.33\textwidth]{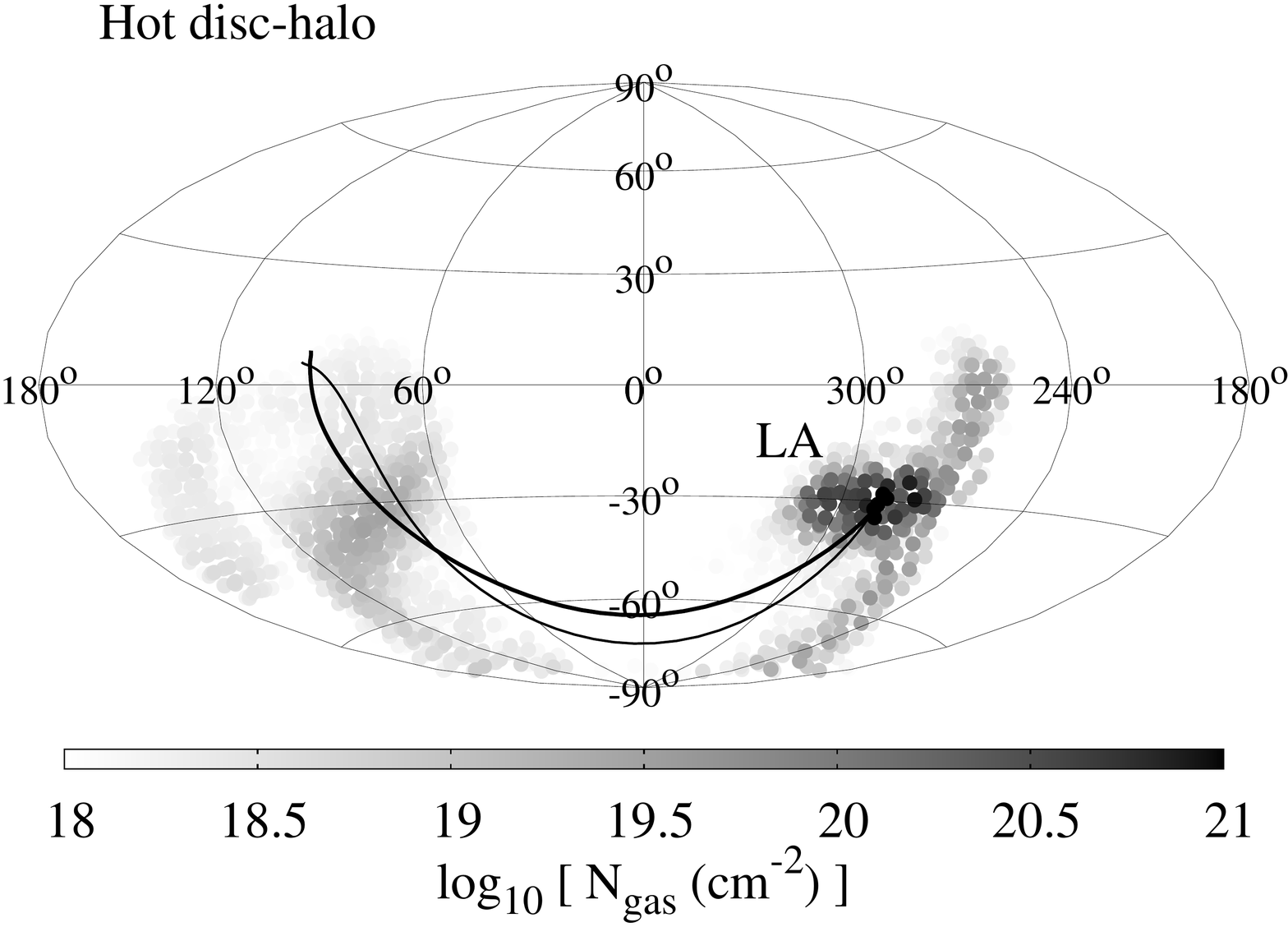}
\caption[  ]{ Distribution of the gas in the Magellanic System roughly a Gyr after infall into the Galaxy in the presence of {\em hot disc and rarified hot halo} around the Galaxy. See text and caption of Figure \ref{fig:ms} for details.  }
\label{fig:rare}
\end{figure*}

Although the presence of diffuse, hot gas around the Galaxy seems secure, its structure is still fiercely debated. Two competing models are generally invoked to explain the local x-ray observations: i) a disc-like structure with a scalelength and a scaleheight both of a few kpc \citep[e.g.][]{yao09a}; {\em or} ii) an extended ($\gtrsim 100$ kpc), spheroidal gaseous halo \citep[e.g.][]{mil13a}. The models of the evolution of the Magellanic System we have presented here so far are based on the latter, more conventional interpretation. A recent analysis of {\em Suzaku} archival x-ray data suggests, however, that neither of these interpretations may be appropriate \citep[][]{nak18a}.

The latter study shows that the x-ray coronal emission can be explained with a hybrid of the two competing models introduced above, i.e. a  model that consists of both a dense, disc-like component that dominates the x-ray emission; and a diffuse, spheroidal, extended component that dominates the mass of the corona, with a mean temperature of $\sim 3 \times 10^6$ K. As noted by \citet[][]{nak18a}, the x-ray data is highly anisotropic, with large scatter across the sky, making simple spherically symmetric models rather unlikely. Given this result, it is interesting to explore whether a model of the evolution of the Magellanic System within a Galaxy that includes a `corona' based on this hybrid interpretation could allow the formation and survival of the Leading Arm .

To this end, we construct a model of a hot gas distribution around the Galaxy following \citet[][]{nak18a}'s parametrisation, consisting of an axisymmetric disc described by
\begin{equation} \label{eq:nak1}
	n_{\rm d}(R,z) = n_{\rm d,0} \exp\left[ - { R }/{ R_{\rm d} } \right]  \exp\left[  - { z }/{ z_{\rm d} } \right] \, ,
\end{equation}
with an extension of 30 kpc in both directions, and total mass of $\sim 5 \times 10^7$ \Msun; and a spherically symmetric halo described by
\begin{equation} \label{eq:nak2}
	n_c(r) = n_{\rm c,0} \left[ { r }/{ r_c } \right]^{-3 \beta} \, ,
\end{equation}
with a mass of $\sim 2 \times 10^9$ \Msun\ enclosed within 250 kpc. The corresponding parameter values are $n_{\rm d,0} \approx 4 \times 10^{-3}$ \pcc, $R_{\rm d} = 7.0$ kpc, $z_{\rm d} = 2.0$ kpc, and $n_{\rm d,0} \approx 10^{-3}$ \pcc, $r_{\rm c} = 2.4$ kpc, $\beta = 0.51$.\footnote{We neglect the metallicity dependence introduced by \citet[][]{nak18a} in their calculations of each component's mass. } The composite radial density profile corresponding to this model is compared to our `Non-rotating corona' model in Figure \ref{fig:dens}. We refer to this as the `Hot disc-halo' model. It is worth emphasising that at $R \gtrsim 10$ kpc the density profile resulting from this model roughly corresponds to the density profile of our `Non-rotating corona' model but scaled by a factor $\sim 1/10$.

Using the `Hot disc-halo' model, we calculate the evolution of the Magellanic System as we did before with our conventional corona models. The result of this exercise is shown in Figure \ref{fig:rare}. Clearly, a leading gas stream survives in this model up to the present-day. Note that the leading stream is, however, less extended compared to the leading gas stream in models that lack a hot corona. These results are not fully unexpected in view of the role ram pressure plays in the survival of a leading gas stream discussed earlier and the lower density of the hot gas in this model compared to our standard corona models.

Our main result is in line with \citet[][]{mur00a} who found using pure analytic methods that an ambient density $< 10^{-5}$ \pcc\ at 50 kpc, i.e. an order of magnitude {\em lower} than our standard coronal model, is required for the leading gas stream of the Magellanic System to survive. The density profile of our `Hot disc-halo' model is consistent with this upper limit (Figure \ref{fig:dens}). Thus we conclude that any coronal model with a mean density lower by a factor of $\gtrsim10$ compared to our standard coronal model allows the leading gas stream to survive.

\subsubsection{Spheroidal vs. flattened coronae}  \label{sec:comp}

The survival of the leading gas stream in our `Hot disc-halo' model is an argument in favour of flattened over spheroidal coronal models. Here we present additional evidence in support of flattened hot halo models.

Diffuse, hot coronae around massive ($\gtrsim 10^{12}$ \Msun) spiral galaxies extending out the virial radius of their host halo in near hydrostatic equilibrium are expected in theories of galaxy formation \citep[][]{fal80a}. X-ray images of quiescent spiral galaxies \citep[e.g.][]{bog13a,bog13b} seem to support this picture, although no x-ray haloes extending beyond $\sim 70$ kpc have been observed. Nonetheless, extended, spheroidal coronae are still seriously considered in the context of the so-called `missing baryons' problem \citep[][]{fuk98a}. Indeed, some claim \citep[q.v.][]{fae17a} that the Galactic corona may be massive enough to bring the baryon fraction of the Galaxy close to the expected cosmic value.

Simulations of galaxy formation within a cosmological framework \citep[][]{cra10a,nuz14a,nel16a} self-consistently lead to the formation of extended (i.e. beyond the optical disc) structures of hot gas around spiral galaxies; however their shapes are rarely spheroidal, as some models of the Galactic corona assume \citep[e.g.][]{mil15a}. If the Galactic corona is spinning fast as inferred from x-ray observations of the Galaxy \citep[][]{hod16a}, then it cannot be spherical, i.e. an initially spheroidal spinning corona will eventually flatten as it spins up \citep[][]{pez17a,sor18a}. Conversely, a non-rotating corona will eventually acquire enough angular momentum as a result of the momentum transfer from the gas disc ejected in the form of a Galactic fountain \citep[][]{fra08a,mar11a}. In other words, both a significant rotation speed and a spherical shape -- both of which are inferred from the same x-ray data, it should be noted--, are mutually exclusive properties of the Galactic corona.

Therefore, we believe that the hot gas around the Galaxy is likely to be a moderately flattened, spinning structure although the degree of flattening is an open question. An equilibrium, baroclinic model of the Galactic corona that is consistent with a number of observational constraints (but that ignores x-ray observations) also shares these properties \citep[][]{sor18a}. We note with interest that some evidence for a flattened hot gaseous structure around the system NGC~891 has been recently presented \citep[][]{hod18a}. This galaxy is widely considered to be the best edge-on surrogate or analogue to the Milky Way.

The specific model of this kind considered here implies a total mass of hot gas ($< 10^{10}$) \Msun\ that is low compared to conventional models, and clearly cannot account for the missing baryons in the Galaxy, which are expected to have a total mass $\sim 10^{11} \Msun$. This is a consequence of the density profile of this model, which is low compared to {\em some} of the available estimates of the gas density at $r \gtrsim 50$ kpc all around the Galaxy (Figure \ref{fig:dens}). However, it is consistent with the upper limits inferred by \citet[][]{sta02a}, and \citet[][]{mur00a}, as mentioned above.

In summary, taking both the observational evidence and theoretical arguments into account, the distribution of hot gas around the Galaxy likely follows a flattened distribution as a result of its relatively high rotation speed. Such a distribution may permit formation of the Leading Arm within the framework of a ram-pressure model coupled to the first-infall scenario. The caveat is that such a distribution does not leave room to accommodate the large reservoir of baryons required to raise the total baryonic mass fraction of the Galaxy to the cosmic value, as argued by others \citep[][]{fae17a}.

\subsection{An alternative view on the nature of the Leading Arm } \label{sec:alt2}

\begin{figure*}
\includegraphics[width=0.49\textwidth]{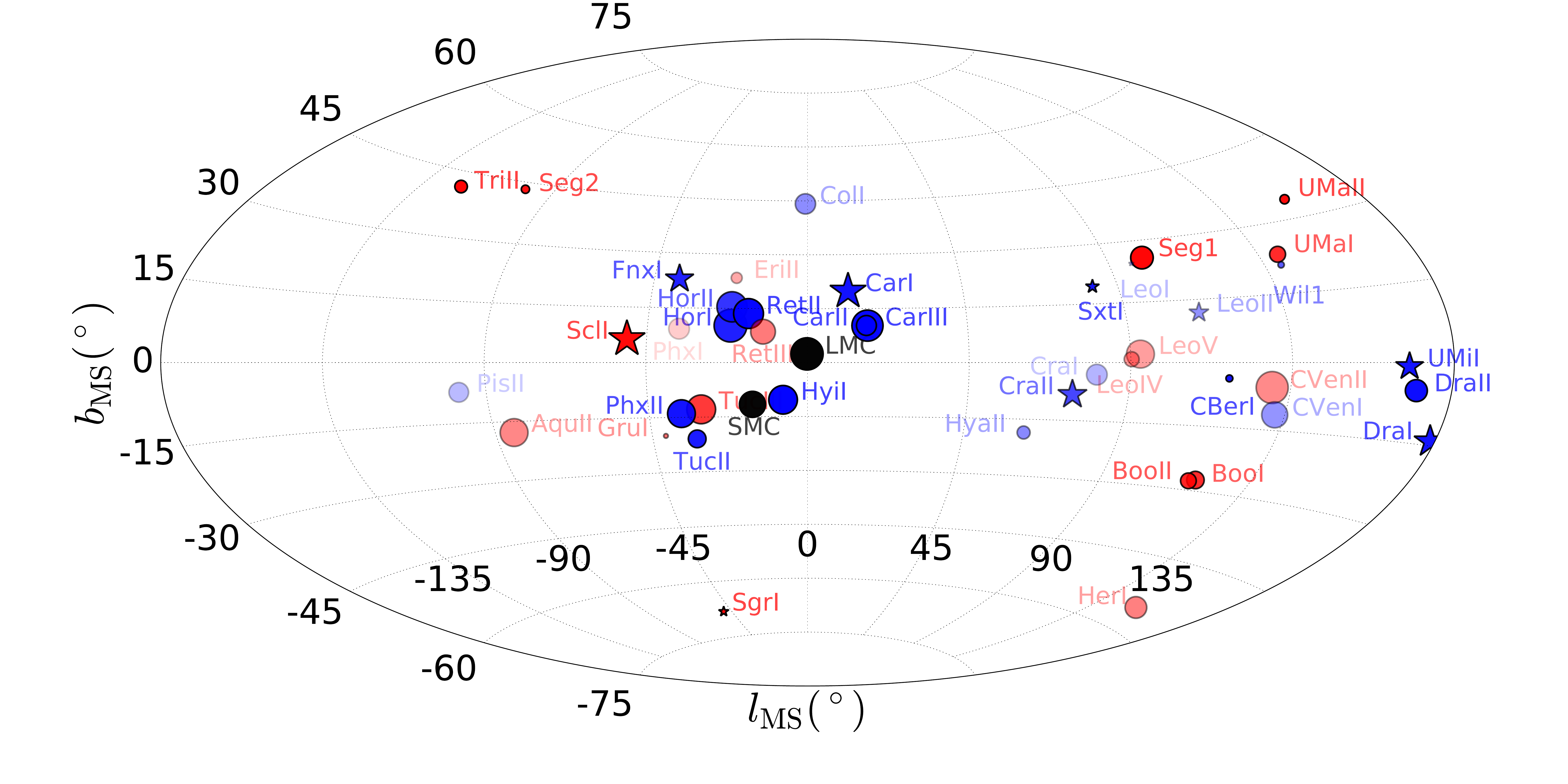}
\includegraphics[width=0.49\textwidth]{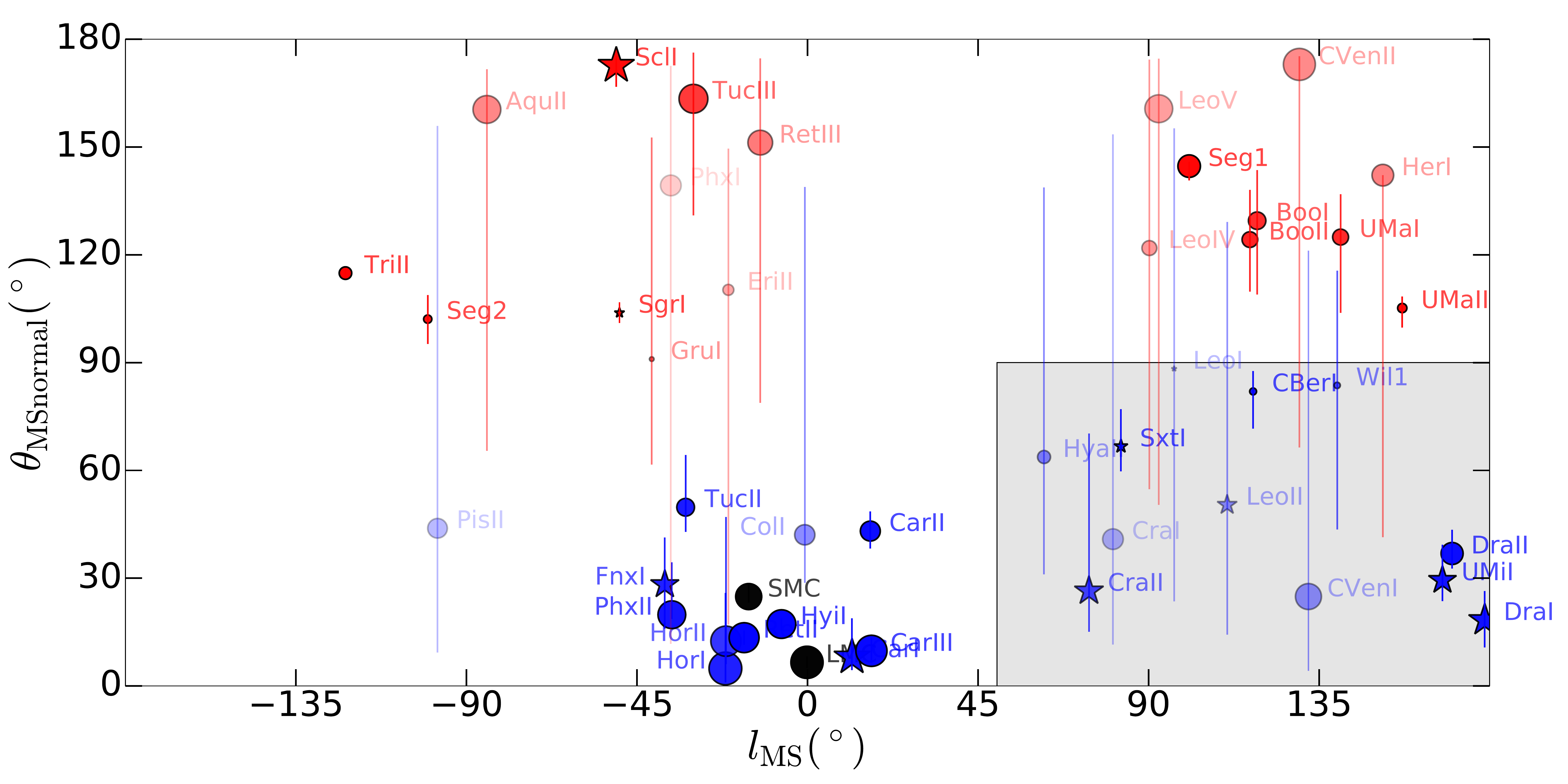}
\caption[ ]{Left: Positions of the satellites in Magellanic Stream (MS) coordinates, as defined in \citet[][]{nid08a}. The LMC is at $l_{\rm MS} = 0$\deg\ and $b_{\rm MS} = 0$\deg\ and moves towards positive $l_{\rm MS}$ (i.e. to the right in the plot). The LMC and the SMC are shown in black. Blue (red) symbols indicate prograde (retrograde) motion relative to the MS normal $\hat{n}_{\mathrm{MS}}$, larger symbols correspond to orbital poles that are more closely aligned with the MS normal. Satellites with less certain orbital pole direction are flagged by fainter colours. The  $l_{\rm MS}$ contours run from -180\deg to +180\deg from left to right in steps of 45\deg. Right: Angle between a satellite's orbital pole and the MS normal.  Vertical lines indicate the 90\% confidence interval on the angle's value. The shaded area flags the region of interest to search for frontrunner candidates. Symbol colours, size, and brightness as in the left panel. }
\label{fig:sats}
\end{figure*}

The physical need for the Galactic corona, as well as our current ignorance about its detailed structure, and the difficulty in the formation of a leading gas stream in its presence, leads us to consider an alternative scenario: that the Magellanic Leading Arm is not in fact directly associated to either the LMC or the SMC.

The interpretation of the Leading Arm as the tidal counterpart of the Magellanic Stream is based on two observations: i) The apparent continuity in the kinematic properties of the \HI\ gas from the tip of the Leading Arm (regions LA~II and LA~IV)\footnote{See \citet[][]{ven12a} for a definition of the different components associated with the Leading Arm. } across the Galactic plane to LA~I and to the MCs \citep[][]{put98a}; and ii) the chemical signature of the \HI\ gas across the Leading Arm region, which is comparable to the chemical signature of the MCs gas \citep[e.g.][]{sem01a}.

However, while recent kinematic \HI\ data suggests an LMC origin \citep[][]{nid08a,ven12a}, accurate measurements of the chemical composition of the Leading Arm clouds favour a SMC origin \citep[][]{fox18a,ric18a}. There is, however, an inherent danger in the use of the metallicity as a constraint on the origin of the gas, in that the interaction between two gas phases results in mixing that inevitably washes out the original chemical signature \citep[][see also \citealt{tep18a}]{hen17a}.

But there is an additional issue specific to the components LA~II and LA~III north of the Galactic plane ($b > 0$\deg), henceforth collectively referred to as {\em LA-North}. A number of models of gas cloud-gas disc interactions indicate that gas clouds will not survive a collision with the Galactic gas disc \citep[][]{ten86a}, regardless of the presence of an ambient magnetic field \citep[][]{san99a,gal16a}, unless their density is high enough compared to the disc or surrounded by a DM subhalo \citep[][]{nic14b,tep18a}. Assuming the Galactic gas disc follows the radially declining \HI\ surface density profile as estimated by \citet[][]{kal08a} at the time of the interaction, the total gas column density of the Leading Arm prior to collision had to be $\gtrsim 10^{21}$ \psc, substantially higher than total gas column density of LA~I (which has not yet crossed the disc) as inferred from observations \citep[][]{fox18a}.

We conclude that the evidence so far for the connection of the Leading Arm with the MCs is somewhat inconclusive. This opens the door to, or perhaps even demands, an alternative interpretation about the nature of the Leading Arm.

\citet[][]{mas05a} showed that the Leading Arm could be explained as a {\em trailing} stream that arises from the LMC as it encircles the Galaxy. This interpretation requires the LMC to have orbited the Galaxy once already. Although unlikely, it should be noted that such an orbit is not entirely ruled out by recent measurements of the MCs' proper motion \citep[][]{kal13a}. Nonetheless, it would appear difficult for this model to explain the significant spread in elemental abundances observed among the different Leading Arm components \citep[$\sim0.1 - 0.3$ \Zsun;][]{fox13a,ric18a}.

A different paradigm has recently been laid out by \citet[][]{ham15a}, based on an scenario originally proposed by \citet[][]{saw05a}, according to which the Magellanic gas structures are associated with a large number of baryon-dominated tidal dwarf systems, whose largest members are the MCs \citep[see also][]{paw14a}. In this picture, a group of essentially DM-free, tidal dwarf galaxies lost their gas along their orbit all the way from the system M31 to the Galaxy as a result of ram pressure by the intergalactic medium and the Galactic corona.\footnote{The idea that the MCs underwent a close encounter with M31 before being captured by the Galaxy dates back to \citet[][]{shu92a,byr94a}.} As for \citet[][]{mas05a}'s model, in \citet[][]{ham15a}'s model the Leading Arm is interpreted as a trailing gas stream of the tidal dwarfs moving ahead of the MCs. We consider this idea by appealing to newly discovered satellites (including streams) around the Galaxy as the possible origin of the Leading Arm \citep[e.g.][]{hel18a}.

\subsubsection{Is the Leading Arm linked to the infall of a satellite moving ahead of the MCs?} \label{sec:sat}

The paradigm we are proposing is that the Leading Arm components south of the Galactic plane (i.e. regions LA~I, LA~III at $b < 0$\deg, henceforth referred to as {\em LA-South}) can still be directly associated to the MCs, but LA-North most likely are not. Our conviction comes from the fact that no pure gas structure could have survived the interaction with the gas disc, as discussed above, unless confined by a deep DM potential, i.e. a dwarf subhalo possibly (but by no means necessarily) associated to the MCs, a system collectively referred to as the Magellanic Group \citep[][]{don08a,nic11b}.

Evidence for the existence of a Magellanic Group is widespread.\footnote{The idea that other Galactic satellites may be associated to the Magellanic System dates back to \citet[][]{lyn76a}.} Data taken as part of the {\em Dark Energy Survey} (DES) suggest that there are $\sim 70$  satellites associated to the LMC \citep[][]{jet16a}. A number of potential candidates of satellites associated to the LMC have also been identified in the {\em Gaia} DR2 \citep[][]{kal18a}. Simulations of structure formation within a cosmological context predict that LMC-SMC-like dwarf systems are surrounded by their own satellites at all epochs \citep[][]{whe15a}.

Our conjecture is that the LA-North could be a mixture of gas that was stripped from a `frontrunner', a satellite galaxy moving ahead of the MCs, and the interstellar medium (ISM) of the Galaxy at the crossing point. LA-South would then correspond to the gas stripped away from the dwarf by ram-pressure by the corona prior to the frontrunner's transit of the disc, much alike what has been observed in an external system \citep[][]{cra18a}.

We have illustrated these mechanisms in our work on the Smith Cloud \citep[][see also \citealt{gal16a}]{tep18a}. In essence, in that work we showed that a $\sim 10^9$ \Msun\ {\em gas-bearing} DM subhalo traveling at $\sim 350$ \kms\ is partially stripped of its gas while interacting with the Galactic corona and mostly stripped when transiting the Galactic gas disc, lifting up gas from the ISM, and leaving behind a gas streamer of mixed composition. Crucially, since the streamer moves around with the Galactic disc, it will in general not be aligned with the orbit of the infalling subhalo, unless it is very young. In other words, the stripped gas will not generally track its parent satellite orbit.

If the Leading Arm components are in fact fragments of a gas stream created by a gas-bearing satellite that has lost its gas via ram-pressure by the Galactic corona and the Galactic gas disc, the only way we can tightly constrain its progenitor is by having a handle on the distance and the age of the Leading Arm. A recently discovered star cluster in the Galactic halo, believed to be associated to the LA~II, would place it at a Galactocentric position of $(R,z) \approx (23, 15)$ kpc \citep[or an heliocentric distance of $D \approx 29$ kpc;][]{pri18a}. The LA~II component extends over roughly 30\deg\ on the sky. At an heliocentric distance of $\sim 29$ kpc, this corresponds to a physical extension of $\sim 16$ kpc. Based on our work on the Smith Cloud, we estimate an age on the order of 200 Myr for the LA~II, and a distance of the progenitor of $\lesssim 100$ kpc. It is worth noting that our age estimate is consistent with the age of the star cluster believed to be associated to the LA~II \citep[$\sim 130$ Myr; ][]{pri18a}.

To sum up, our key constraints on the frontrunner are:  i) moving ahead of the MCs on a prograde orbit with respect to theirs; ii) an orbital history that brings it within the optical radius of the Galaxy; iii) a present-day distance consistent with having travelled $\lesssim 100$ kpc along its orbit. It would be desirable to have evidence that the candidate may have lost all of its gas only until recently; and perhaps less important: an orbit roughly consistent with the orbit of the MCs. The reason for the last constraint to be a weak one is two-fold: First, the frontrunner need not be directly associated to the MCs. Secondly, its orbit may have been significantly affected by the interaction with the Galactic gas disc \citep[cf.][]{tep18a,tep18b}.

We consider plausible candidates from a subset of the Galactic satellites discovered so far. In the left panel of Figure \ref{fig:sats}, we show the positions of the satellites in Stream (MS) coordinates \citep[][]{nid08a}, where the LMC is at $l_{\rm MS} = 0$\deg\ and $b_{\rm MS} = 0$\deg, by definition. The pole of this MS coordinate system is at $(l; b) = \left(188.5^\circ;  7.5^\circ\right)$, the MS normal $\hat{n}_{\mathrm{MS}}$. On the right of Figure \ref{fig:sats}, we show the angle between a satellite's orbital pole $\hat{n}_\mathrm{sat}$\ and the MS normal, $\theta_{\mathrm{MSnormal}} = \arccos\left(\hat{n}_\mathrm{sat} \cdot \hat{n}_{\mathrm{MS}}\right)$, as a function of $l_{\rm MS}$. The data has been taken from \citet{fri18b,fri18c}. Note that the gap at around $l_{\rm MS} = 30-50$\deg\ is due to obscuration by the Galactic disc. Systems with $\theta_{\rm MS} < 90$\deg ($\theta_{\rm MS} > 90$\deg) are moving on prograde (retrograde) orbits with respect to the MCs. The smaller $\theta$, the more consistent the orbit with that of the MCs. The region of interest here is thus identified by $l_{\rm MS} > 50$\deg and $\theta_{\rm MS} \ll 90$\deg.

Clearly, there is already a large number of dwarf galaxies observed to be moving ahead of the Magellanic Clouds, with orbits that are roughly consistent with the orbit of the MCs. And in the impending era of the Large Synoptic Sky Survey (LSST), it is likely that many more are yet to be discovered.

Of all, the most interesting candidates are perhaps Draco I and Ursa Minor. These are bound to the Galaxy \citep[][]{mcc12a,kaf14a}. Both Draco I's \citep[][]{mas06a} and Ursa Minor's \citep[][]{pia05a} proper motion measurements are consistent with orbital histories that bring the dwarfs within the Galactic optical disc, with orbital periods of $\sim 1.5$ Gyr. Their orbits are well aligned with the orbit of the LMC \citep[Figure \ref{fig:sats}; see also ][]{paw15a}. Both Draco I and Ursa Minor have dynamical masses in excess of $10^7$ \Msun, large enough to lift up ISM gas during a disc crossing, but it is currently unknown whether they initially had gas that was recently lost \citep[][]{mcc12a}. Interestingly, \citet[][]{cap15a} model the gas loss in Ursa Minor and conclude that stellar feedback is not enough, but additional mechanism are required to fully explain its gas loss, perhaps ram pressure stripping by the Galactic gas disc. Nevertheless, based on our current knowledge of their orbits, it is unlikely that any of these transited the Galactic gas disc within our constraint of $\sim 200$ Myr. 

The next most interesting candidate is Crater II with an estimated dynamical mass of $\sim 5\times10^6$ \Msun\ \citep[][]{cal17a}, at a distance of $\sim 117$ kpc, an orbit well aligned with that of the MCs, and a pericentric distance within the optical radius of the Galaxy \citep[][]{fri18c}. However, the epoch of its last disc crossing is likely too long ago to be consistent with our time constraint.

Thus, neither of these systems represents really a satisfactory candidate for our sought frontrunner. In particular, none of them shows evidence for young stars. A suitable candidate may have yet to be unveiled. The prospect of doing so, however, seems low, given the expectation of a young age, characterised by bright A-type stars, making it unlikely that such a system has escaped detection so far.

\section{Summary} \label{sec:con}

Our new study has highlighted the difficulty of understanding the Leading Arm as a gas stream that is tidally stripped and runs ahead of the Magellanic Clouds. The latter has become the conventional interpretation since the seminal \HI\ study of \citet[][]{put98a}. Simple dynamical models support this interpretation assuming the Clouds are on first infall. But we have shown that realistic models for the Galaxy that include a pervasive, hot coronal halo challenge this basic assertion because gaseous `leading arms' do not form.

\smallskip
\noindent
In brief:

1. It may be possible to rescue the conventional `Leading Arm' interpretation if the hot halo is truncated well within the distance of the Clouds, e.g. through flattening by rapid rotation. The appeal of this scenario is that it is consistent with the recent claims based on x-ray data of the Galaxy, and with the recent observation of a flattened hot halo in an external disc galaxy. The disadvantage is that a truncated hot halo cannot accommodate the `missing baryons' claimed by many authors, both observers and simulators, to be locked up in the hot halo of the Galaxy.

2. A belief shared by some is that a population of \HI\ clouds in the Northern Galactic hemisphere appears to be a continuation of the Leading Arm and may be associated with it. We consider this connection to be highly unlikely because the Leading Arm would then need to survive its transit of the Galaxy's gaseous disc. It is difficult to conceive how the latter can be true given the results of numerous studies that strongly argue against it.

3. We consider an alternative explanation involving a gas-bearing dwarf galaxy running ahead of the Clouds (`frontrunner') that is being ram-pressure stripped of its gas. The advantage of this scenario is that it requires the existence of an extended hot halo in line with the conventional picture, and may simultaneously explain: i) the enhanced metallicity of the Leading Arm gas (relative to the trailing Stream); ii) recent claims of star formation in the Leading Arm as a result of ram-pressure acting upon the gas \citep[][]{cra18a,ali18a}; and iii) the survival of the gas as it transits the disc as a result of the shielding effect of the host potential. The disadvantage is that a suitable candidate dwarf has yet to be identified.\\

Overall, it seems difficult to escape the conclusion that there is something fundamental about the Magellanic System and the gaseous structure of the Galaxy that we do not yet understand.

\section*{Acknowledgments}

We thank S.~Pardy for providing the geometric transformations used to setup the initial positions and velocities of the MCs for our infall models.
We thank the anonymous referee for carefully reading our manuscript and providing insightful comments that helped improving the presentation of our results.
TTG acknowledges financial support from the Australian Research Council (ARC) through an Australian Laureate Fellowship awarded to JBH. MSP acknowledges funding by NASA through Hubble Fellowship grant \#HST-HF2-51379.001-A awarded by the Space Telescope Science Institute, which is operated by the Association of Universities for Research in Astronomy, Inc., for NASA, under contract NAS5-26555.
We acknowledge the facilities, and the scientific and technical assistance of the Sydney Informatics Hub (SIH) at the University of Sydney and, in particular, access to the high-performance computing facility Artemis and additional resources on the National Computational Infrastructure (NCI),  which is supported by the Australian Government, through the University of Sydney's Grand Challenge Program the {\em Astrophysics Grand Challenge: From Large to Small} (CIs: Geraint F.~Lewis and JBH). All figures created with {\sc gnuplot},\footnote{ {http://www.gnuplot.info} } originally written by Thomas Williams and Colin Kelley, and {\sc matplotlib} \citep[][]{hun07a}.
This research has made use of NASA's Astrophysics Data System (ADS) Bibliographic Services\footnote{ {http://adsabs.harvard.edu} }, and of Astropy\footnote{ {http://www.astropy.org} }, a community-developed core Python package for Astronomy \citepalias{ast13a}.

\bibliographystyle{mnras} 
\input{manuscript_mnras.bbl} 


\appendix

\begin{figure*}
\centering
\includegraphics[width=0.24\textwidth]{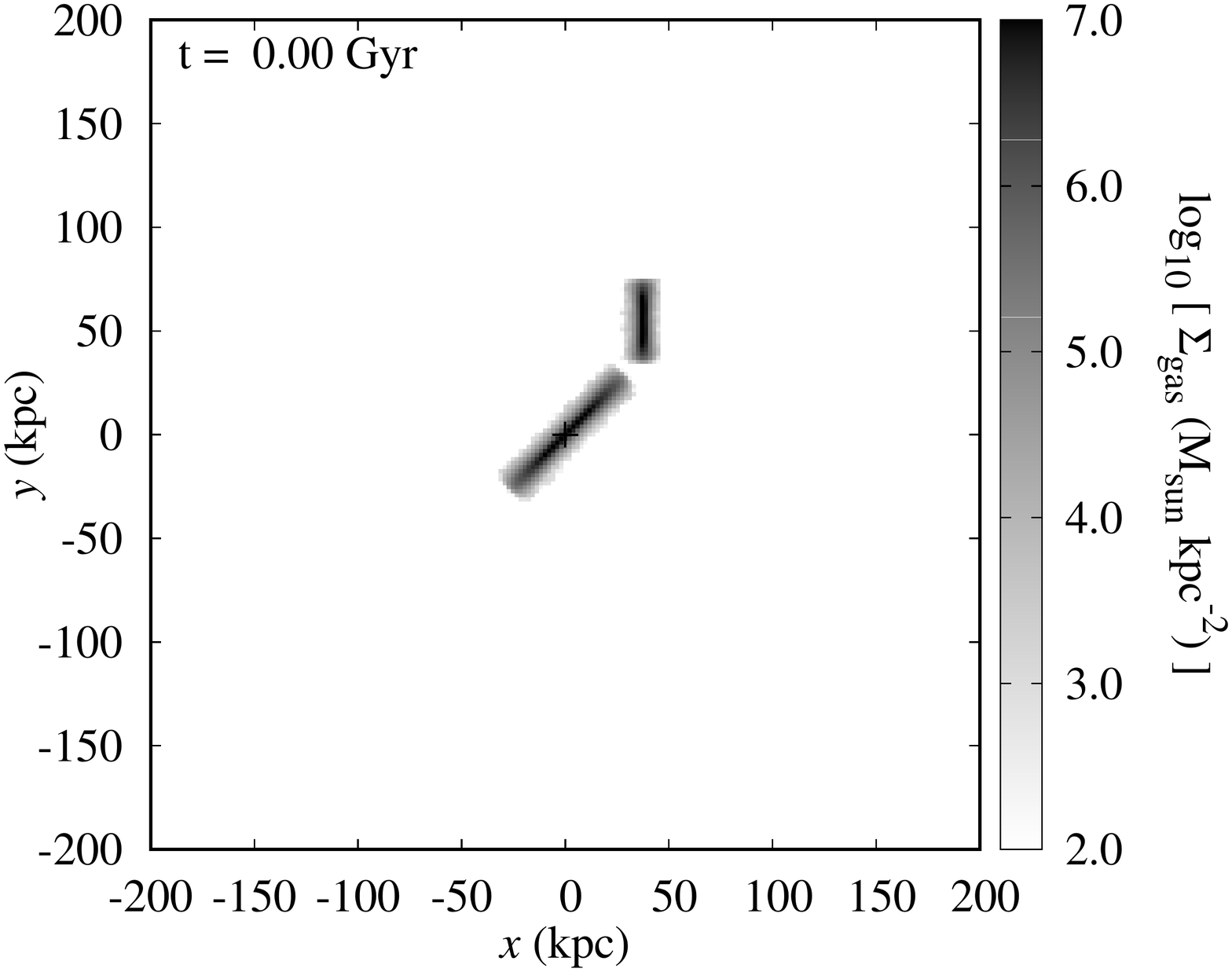}
\includegraphics[width=0.24\textwidth]{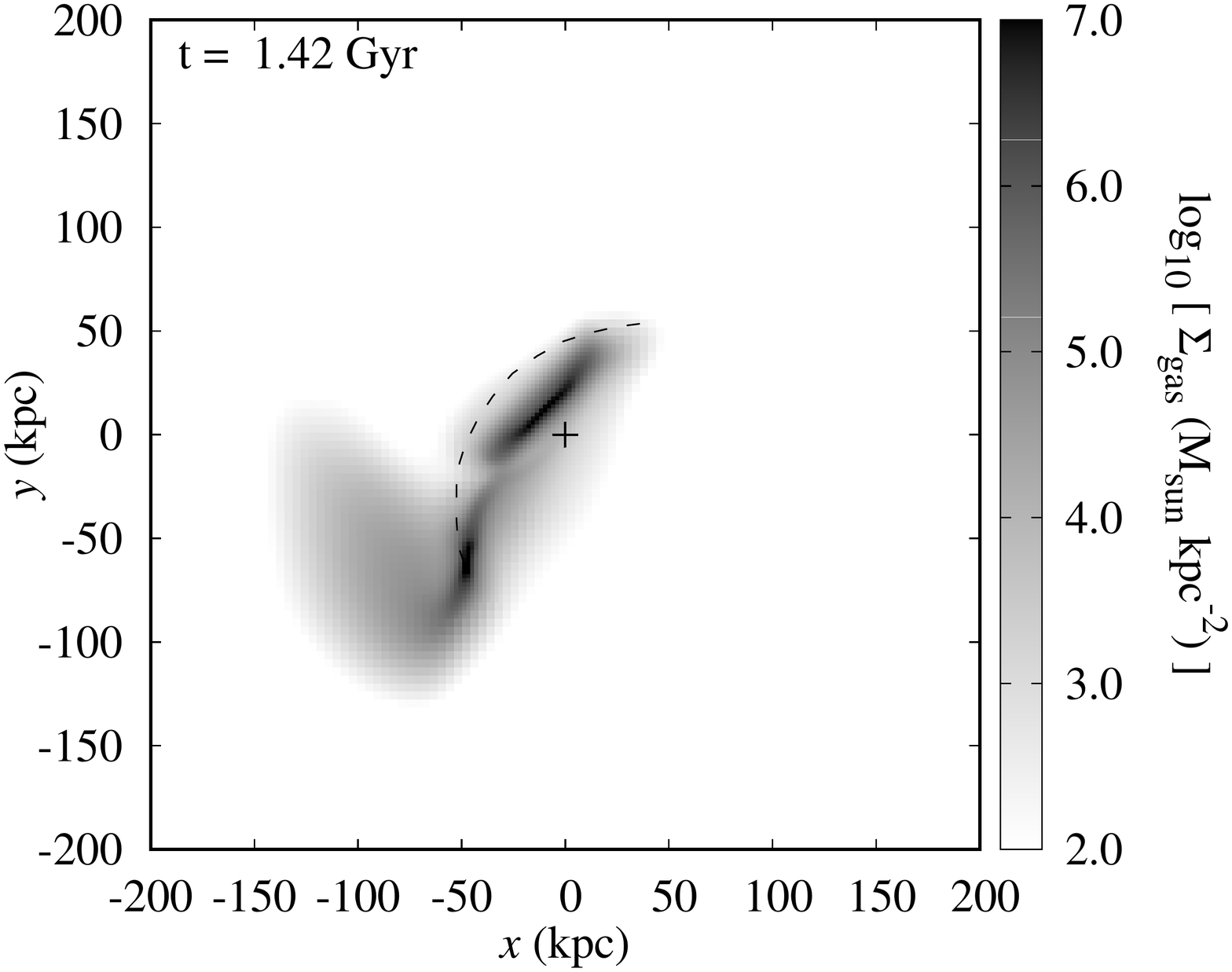}
\includegraphics[width=0.24\textwidth]{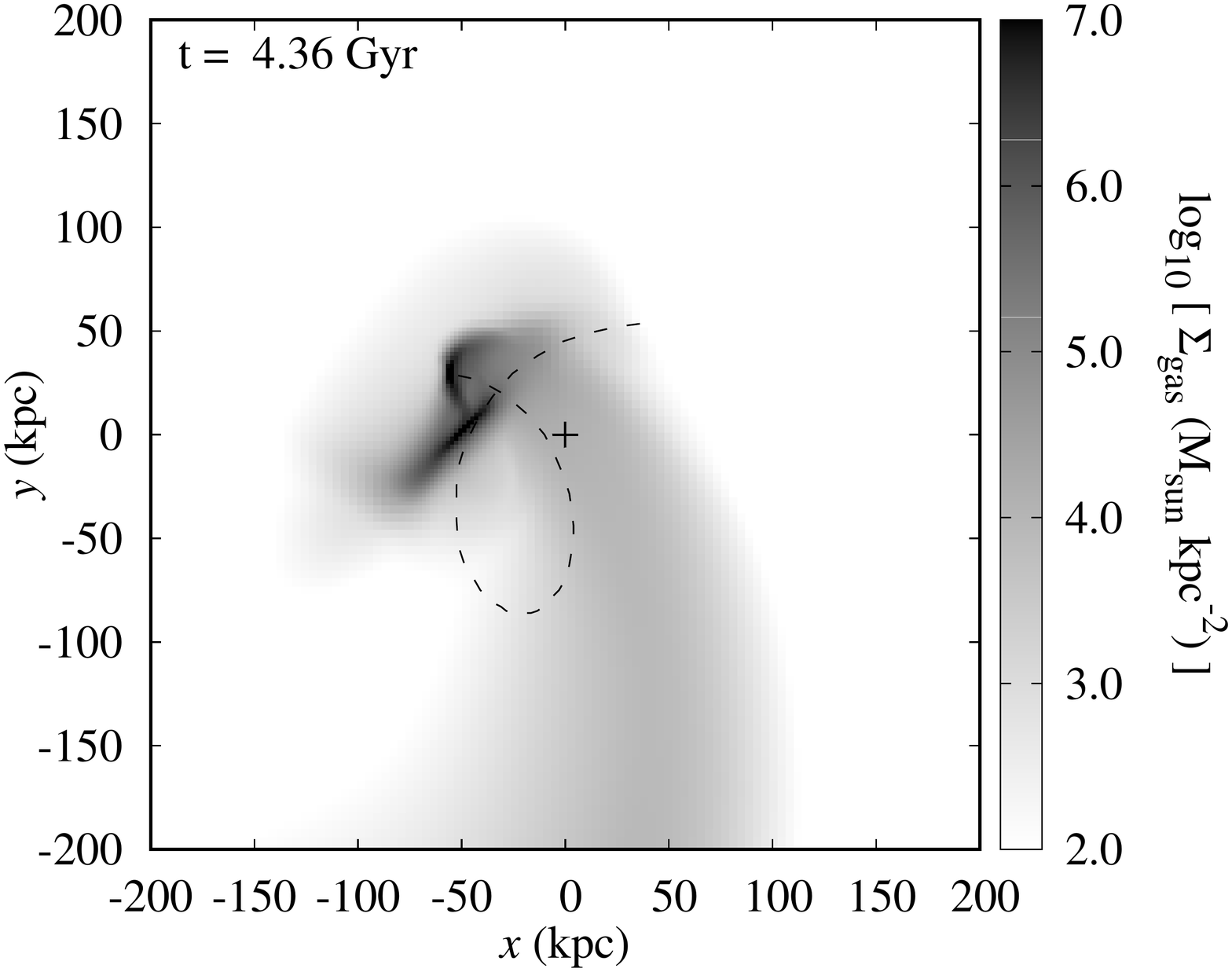}
\includegraphics[width=0.24\textwidth]{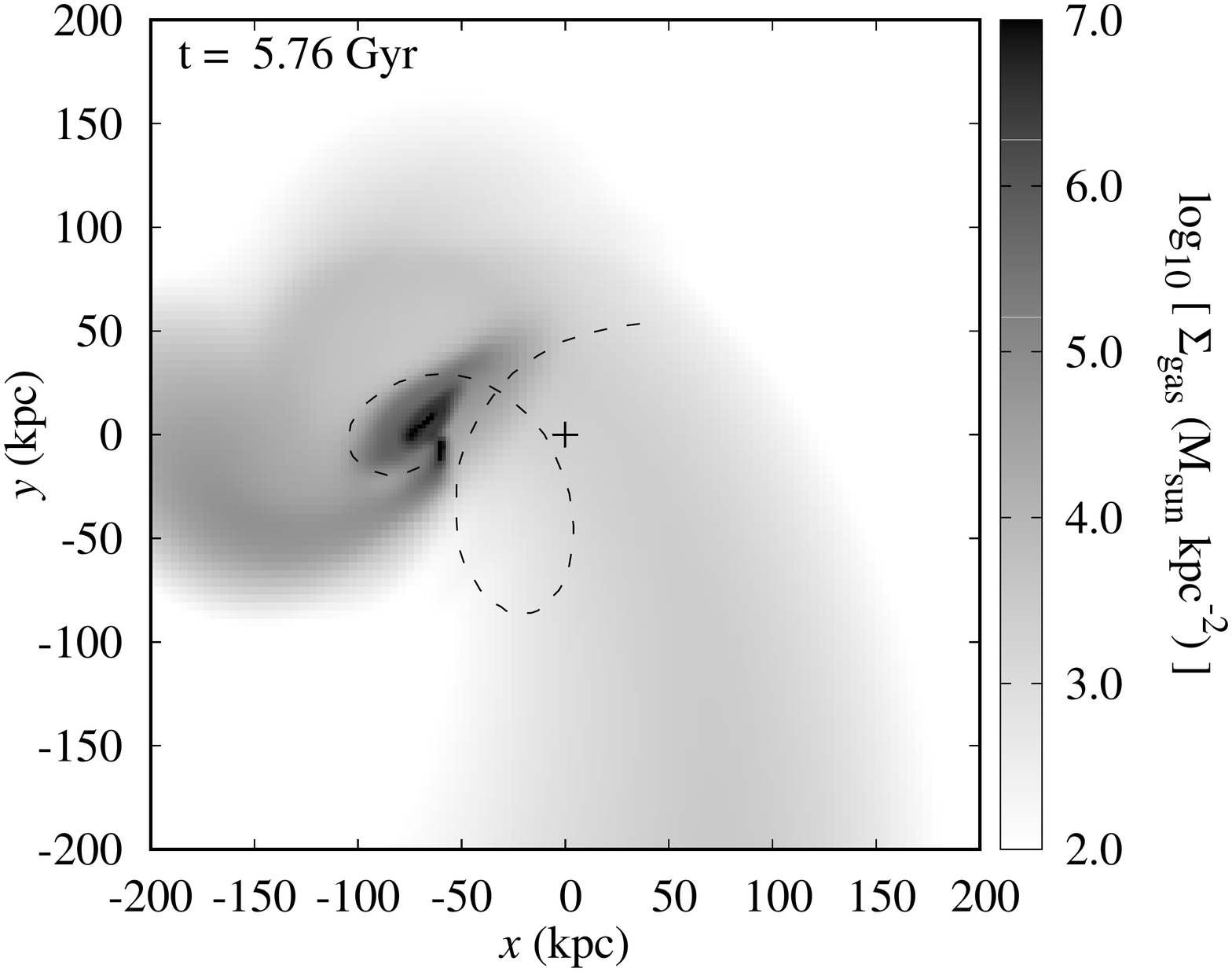}
\caption[  ]{ Replica of  \citet[][cf. their figure 2]{par18a}. Note that we have adopted less extended gas discs (but with the same mass) for the MCs, and therefore their gas discs do not overlap initially, as they do in \citeauthor{par18a}'s models. We do not shift the origin of the coordinate system to the initial barycentre of the LMC (indicated by the cross) for each snapshot as they do. }
\label{fig:ver1}
\end{figure*}

\section{Model verification} \label{sec:ver}

Here we present our results intended to replicate some of \citet[][]{par18a}'s results corresponding to their 9:1 model. This step is relevant to make sure that we are using the correct initial conditions for all our models. We stress that the agreement between their model and our replica is {\em non-trivial}, given the significant difference in codes (i.e. numerical techniques, and implementation) used by them and us to create and evolve the initial conditions.

\begin{figure*}
\centering
\includegraphics[width=0.33\textwidth]{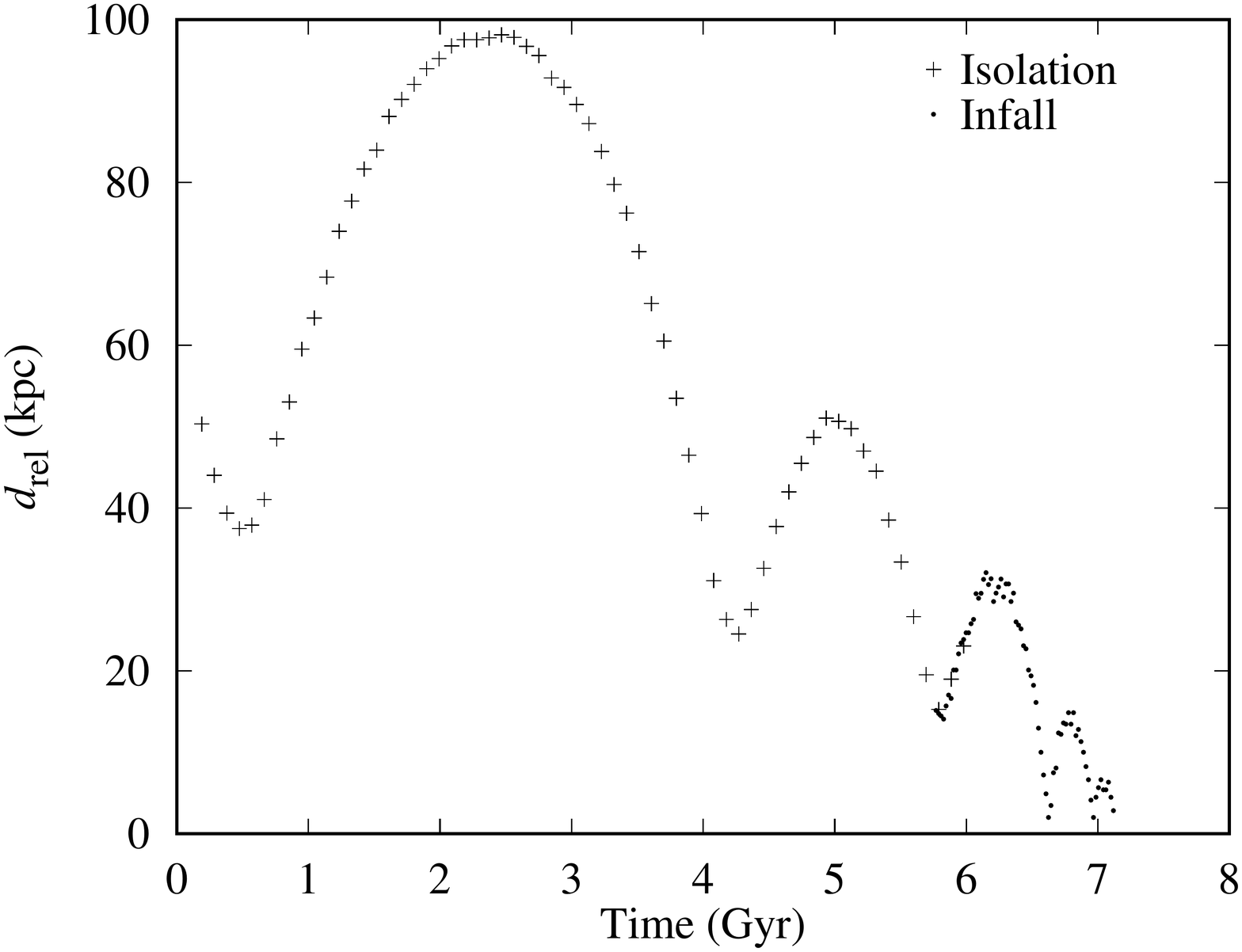}
\includegraphics[width=0.33\textwidth]{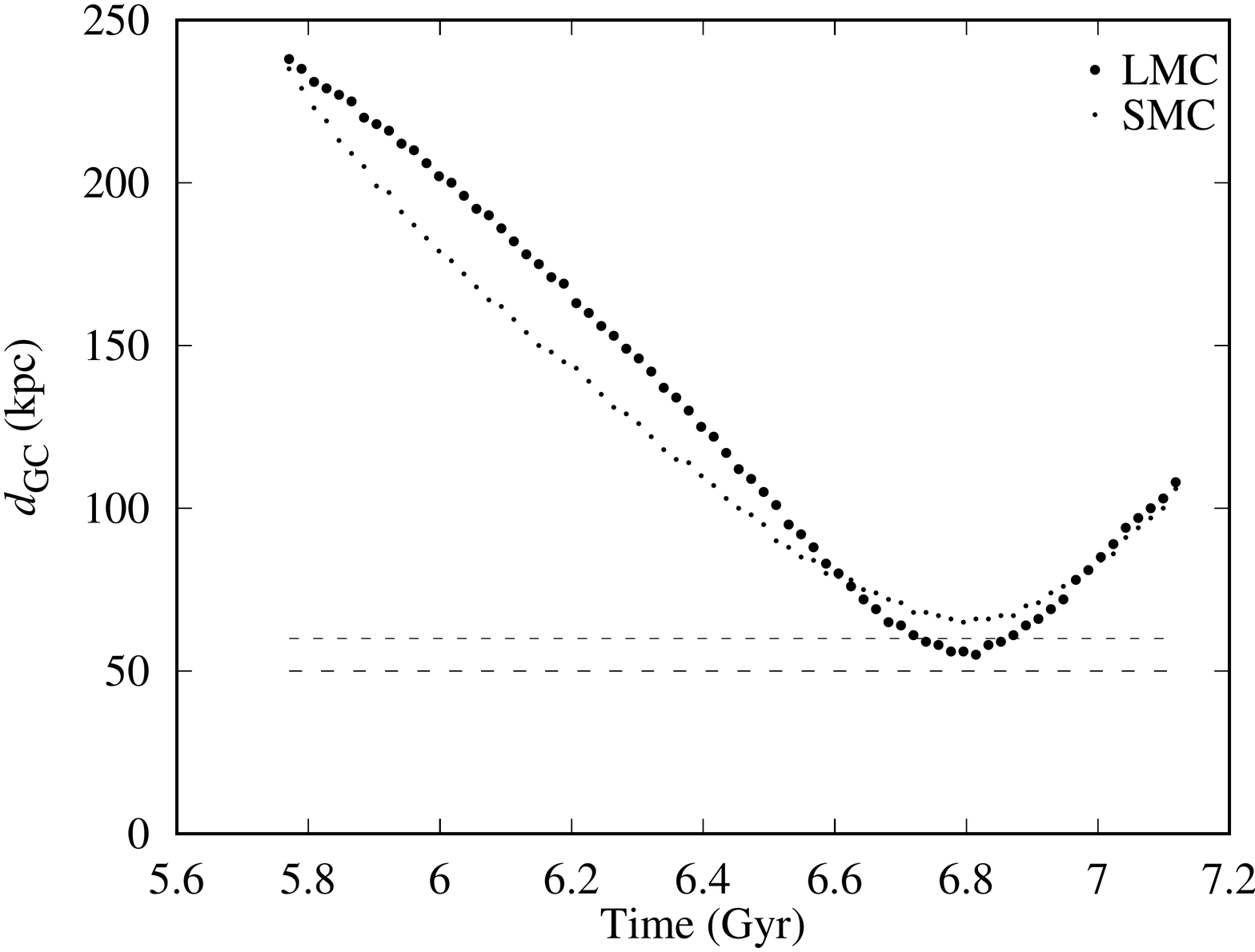}
\includegraphics[width=0.33\textwidth]{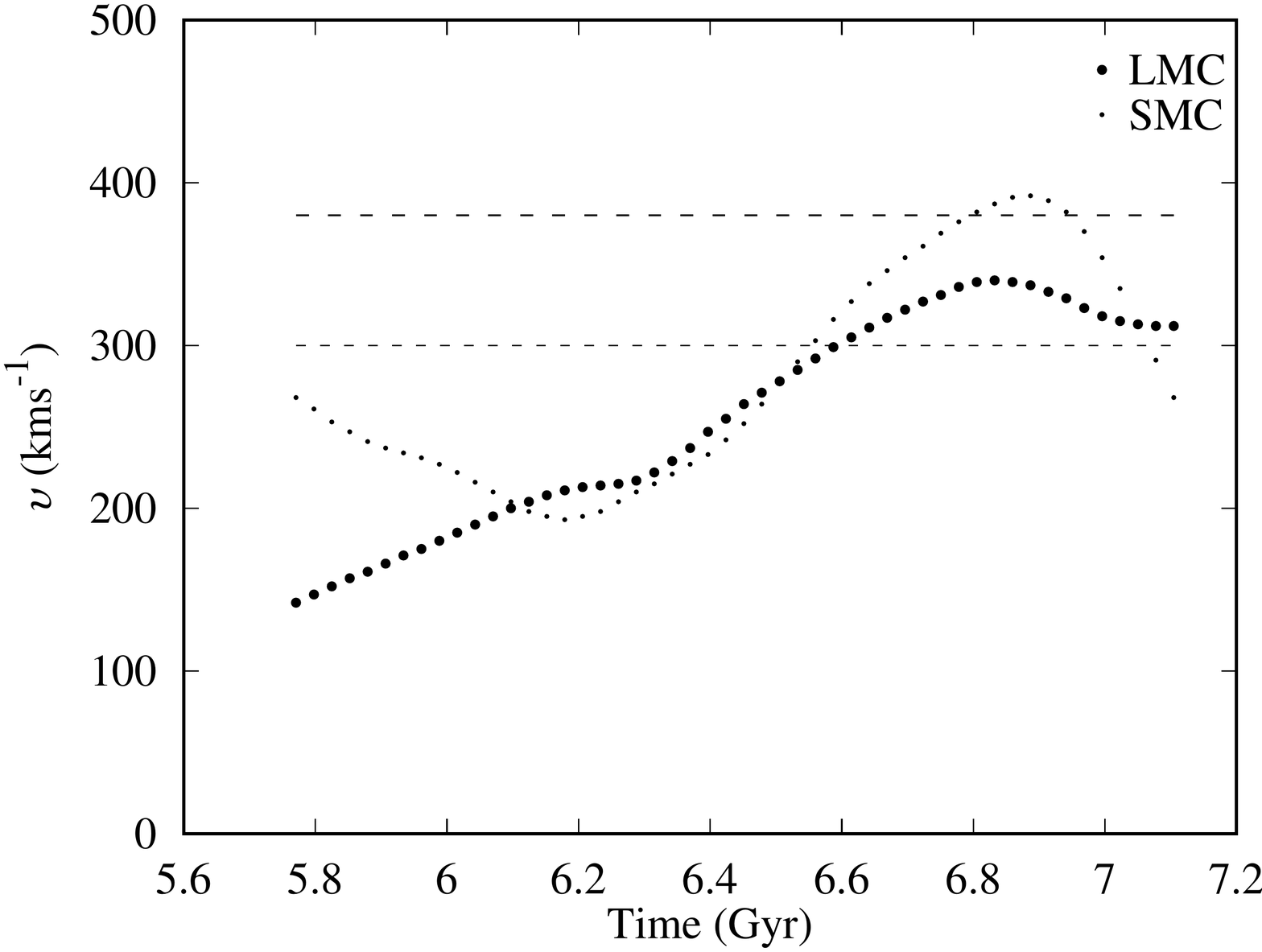}\\
\includegraphics[width=0.33\textwidth]{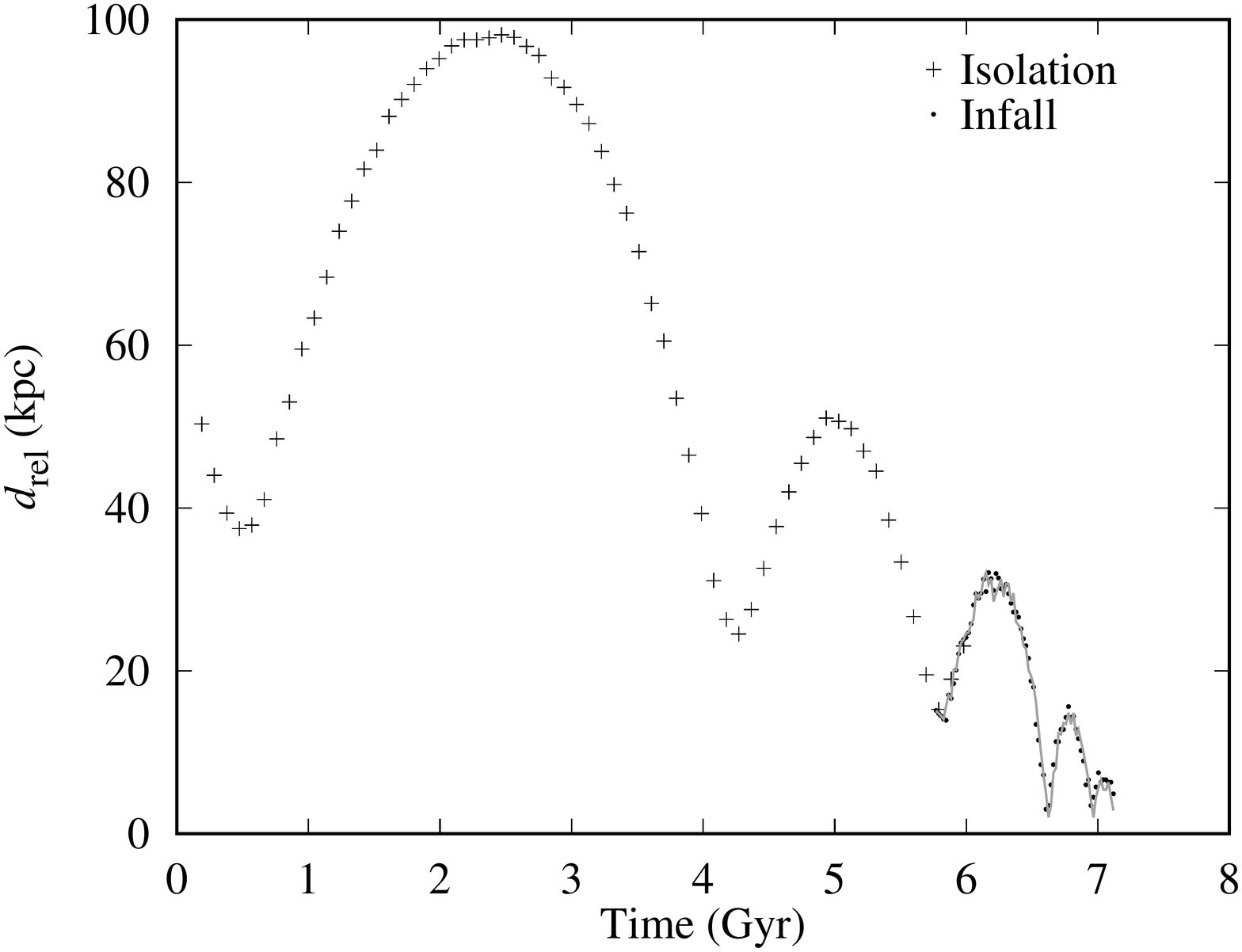}
\includegraphics[width=0.33\textwidth]{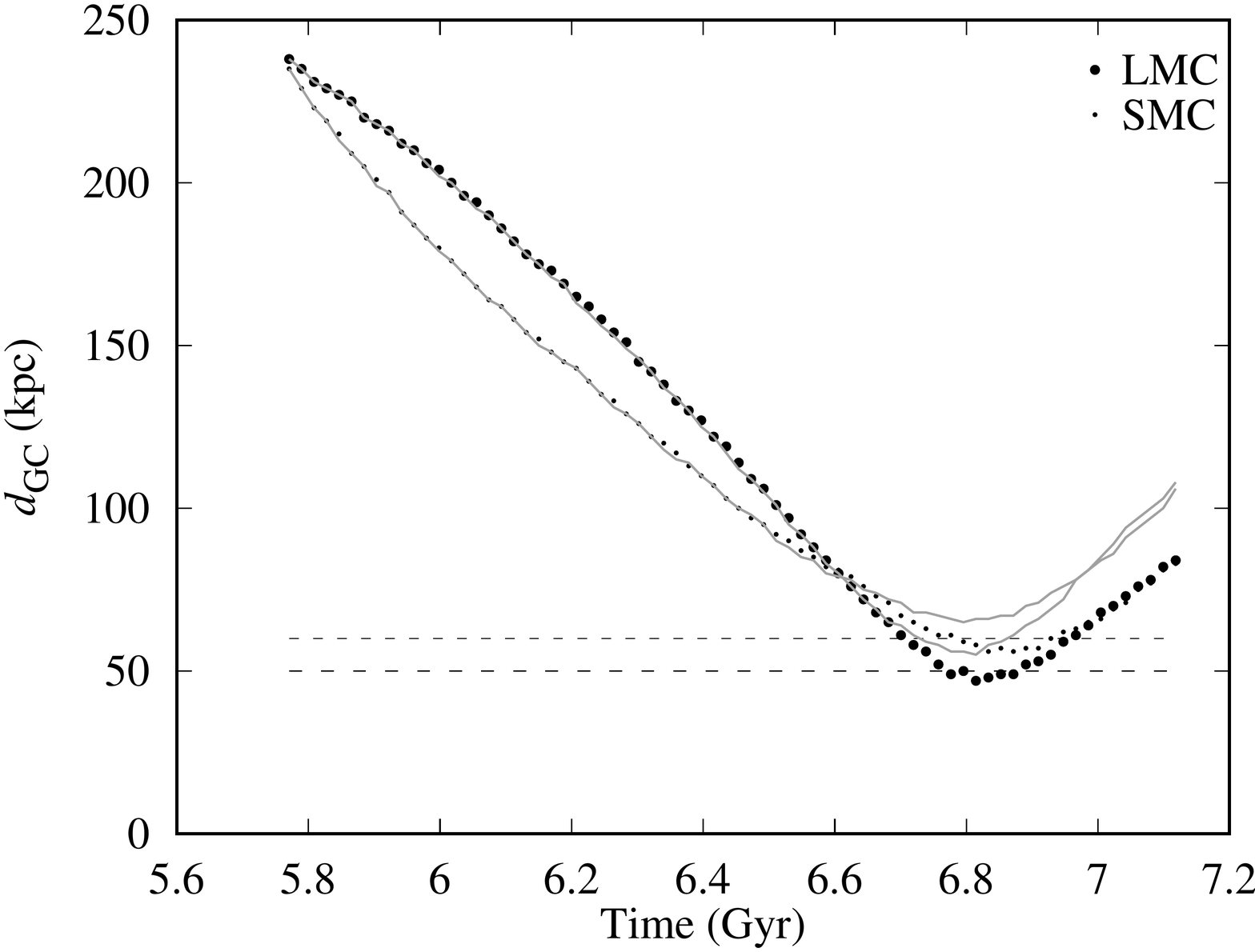}
\includegraphics[width=0.33\textwidth]{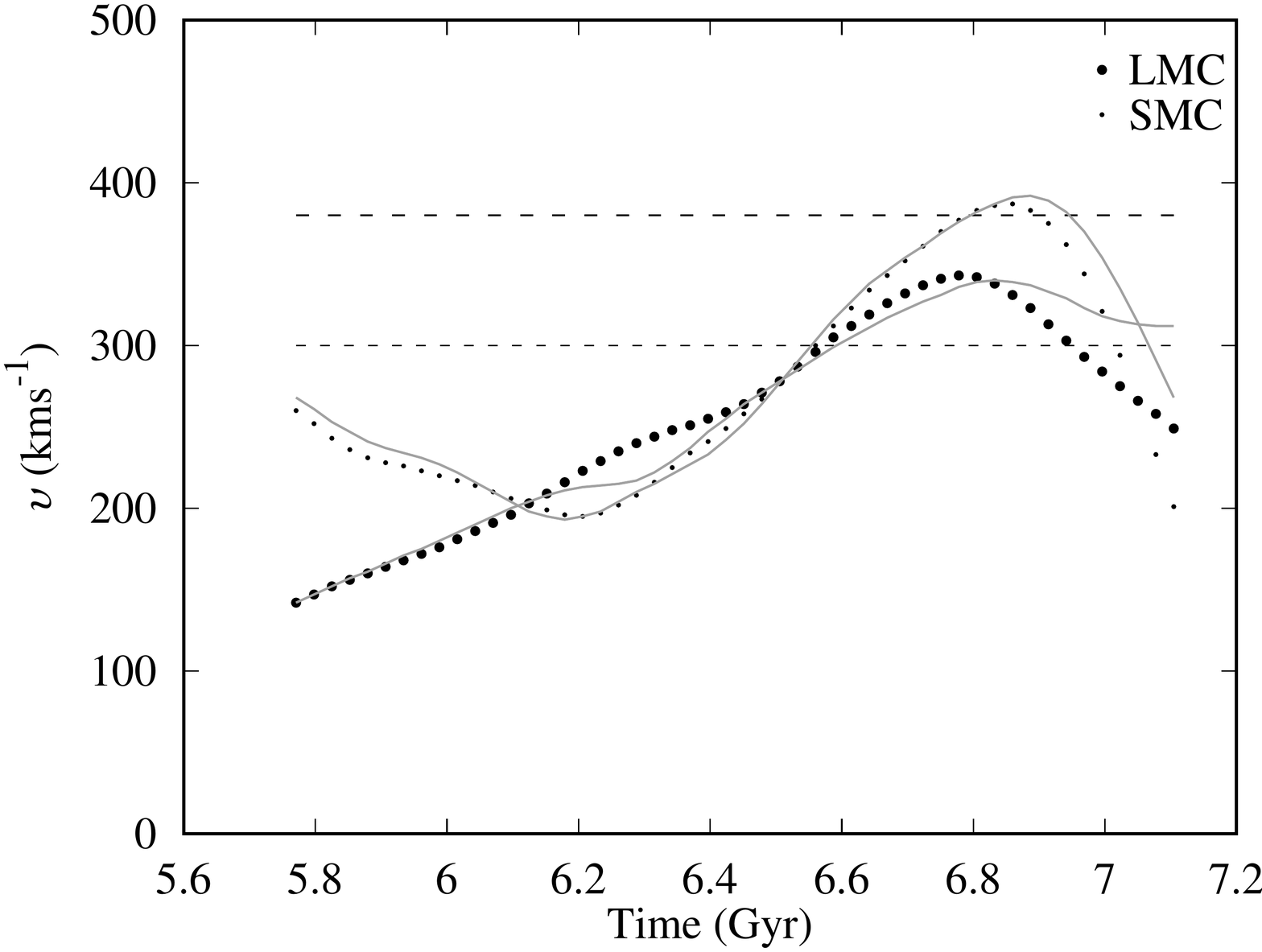}
\caption[  ]{ Orbital history of the MCs. Distance of the SMC around the LMC (left); Galactocentric distance (centre) and speed (right) of the LMC and the SMC. Top: `Static DM halo', our base model. This corresponds to \citet[][]{par18a}'s model (cf their figure 1), and \citet[][]{bes12a}'s model 2 (cf. their figures 2b, 4b and 4d). Bottom: `Live DM halo' model.  The results of our base model shown in the top panel are included in the bottom (grey thick curves) for comparison. Note the difference in the orbital history between these two models, demonstrating the impact of the presence of a live DM halo and the resulting dynamical friction which reduces both the relative distance and the relative speed of the MCs compared to a model with a static DM halo.}
\label{fig:ver2}
\end{figure*}

A glance at our Figure \ref{fig:ver2} and comparison to \citet[][their figure 2]{par18a} readily reveals the similarity in the evolution of the MCs in isolation in their model and our realisation. We note that the 9:1 model by \citet[][]{par18a} consists of extremely gas-rich dwarf galaxies with gas discs as extended as the Galactic gas disc.
We have chosen not to use gas disc as extended as theirs, but we do adopt the same mass.

We find the best match to the final configuration for the system evolved in isolation in terms of number of passages of the SMC around the LMC and its relative distance presented by \citet[][]{par18a} after 5.76 Gyr, a perfect match to their quoted time scale of 5.8 Gyr.

In order to model the infall of the MCs, the evolved binary system is mapped onto the Galactic reference frame. In brief, the MCs are placed around the Galaxy's centre by applying a series of geometric transformations to the position and velocities of the MCs' constituents. The following are the exact steps we took to set the initial Galactocentric positions and the Galactocentric velocities of the MCs in all of our models, {\em after} they have been evolved in isolation, kindly provided by S.~Pardy ({\em private communication}):\footnote{Note that the rotation as defined here follows the mathematical convention that a positive angle corresponds to a {\em counter-clockwise} rotation. }\\
\begin{enumerate}
	\item Calculate the centre of mass (CoM) and its velocity (CoV) of the LMC. We get $\vec{r}_{\textrm{\sc lmc}} \approx (-70, 3.5, 0.5)$ kpc and $\vec{v}_{\textrm{\sc lmc}} \approx (-19, -17, -0.2)$ \kms, respectively;
	\item Shift both the LMC and SMC to a coordinate frame where the CoM of the LMC is at $\vec{r} = (0, 0, 0)$ and its CoV at $\vec{v} = (0, 0, 0)$;
	\item Apply a rotation of:
	\begin{itemize}
		\item[-] 66.00 deg around the $x$ axis; then
		\item[-] 231.2 deg around the $z$ axis;
	\end{itemize}
	\item Apply a constant shift to the SMC and the LMC position and velocities of $\vec{r}_{\rm shift} = (48.0, 198, -85.0)$ kpc, and $\vec{v}_{\rm shift} = (-17.0, -160, -29.0)$ \kms, respectively;
	\item Apply a rotation of:
	\begin{itemize}
		\item[-] 31.9 deg around the $x$ axis; then
		\item[-] 18.5 deg around the $z$ axis.
	\end{itemize}
\end{enumerate}

The position and the velocity of the SMC we obtain after these transformations is $\vec{r} \approx (-31.4, 217,  20.8)$ kpc and $\vec{v} \approx (41.3 ,-274, -73.3)$ \kms, respectively, to be compared to $\vec{r} \approx (-29.9, 217,  22.0)$ kpc and $\vec{v} \approx (42.9 ,-284, -80.0)$ \kms from \citet[][see their table 3]{par18a}. Note that the position and the velocity of the LMC we get are identical to theirs, by construction.\\

The agreement between \citet[][]{par18a}'s and our results is reinforced by a comparison of the orbital history presented in Figure \ref{fig:ver1} and \citet[][their figure 1]{par18a}.\footnote{Compare also to \citet[][]{bes12a}'s model 2 (their figures 2b, 4b and 4d).} This demonstrates that not only our initial orbital parameters, but also the Galactic potential as well as the masses and structure of the MCs agree well with theirs.

\begin{figure*}
\centering
\includegraphics[width=0.33\textwidth]{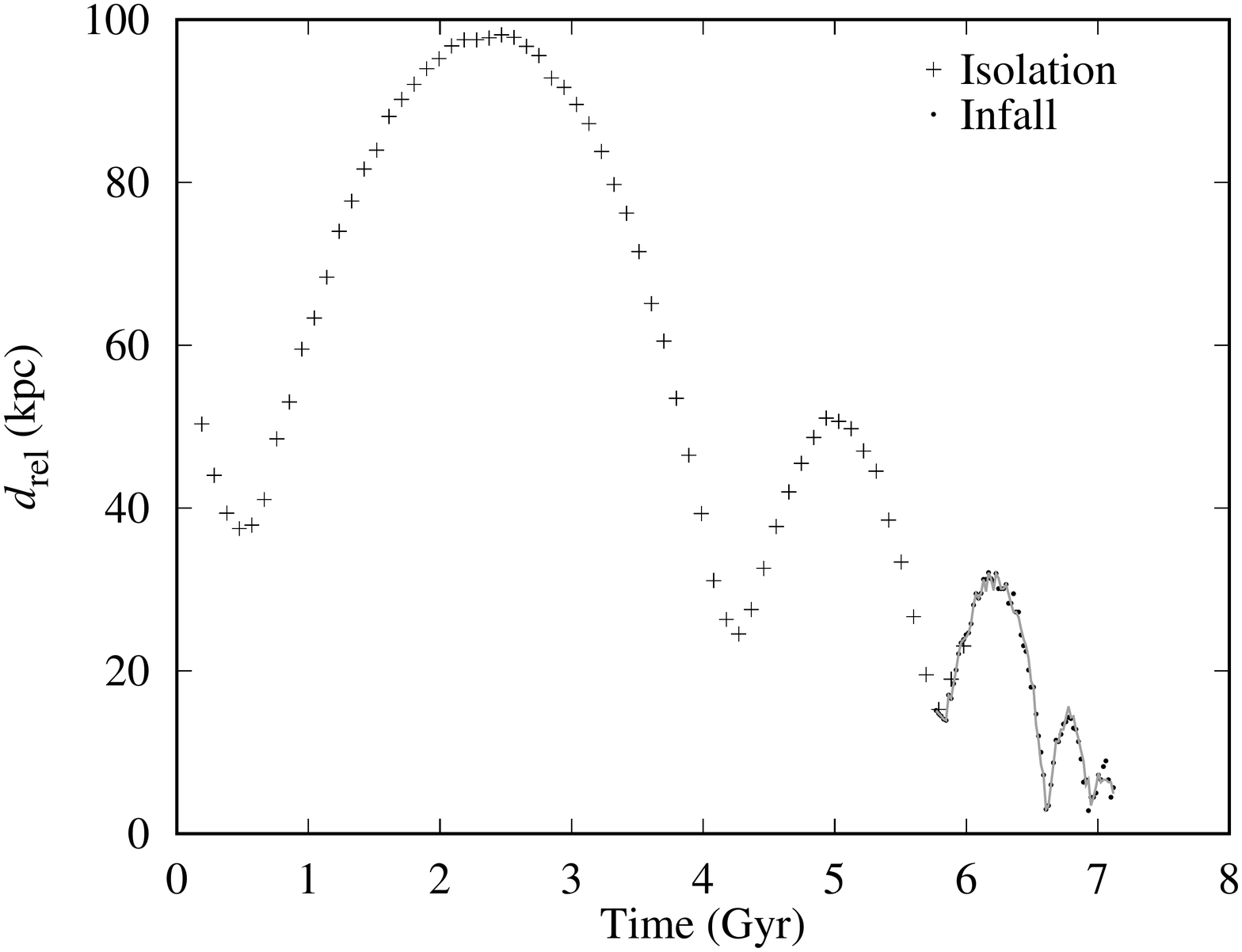}
\includegraphics[width=0.33\textwidth]{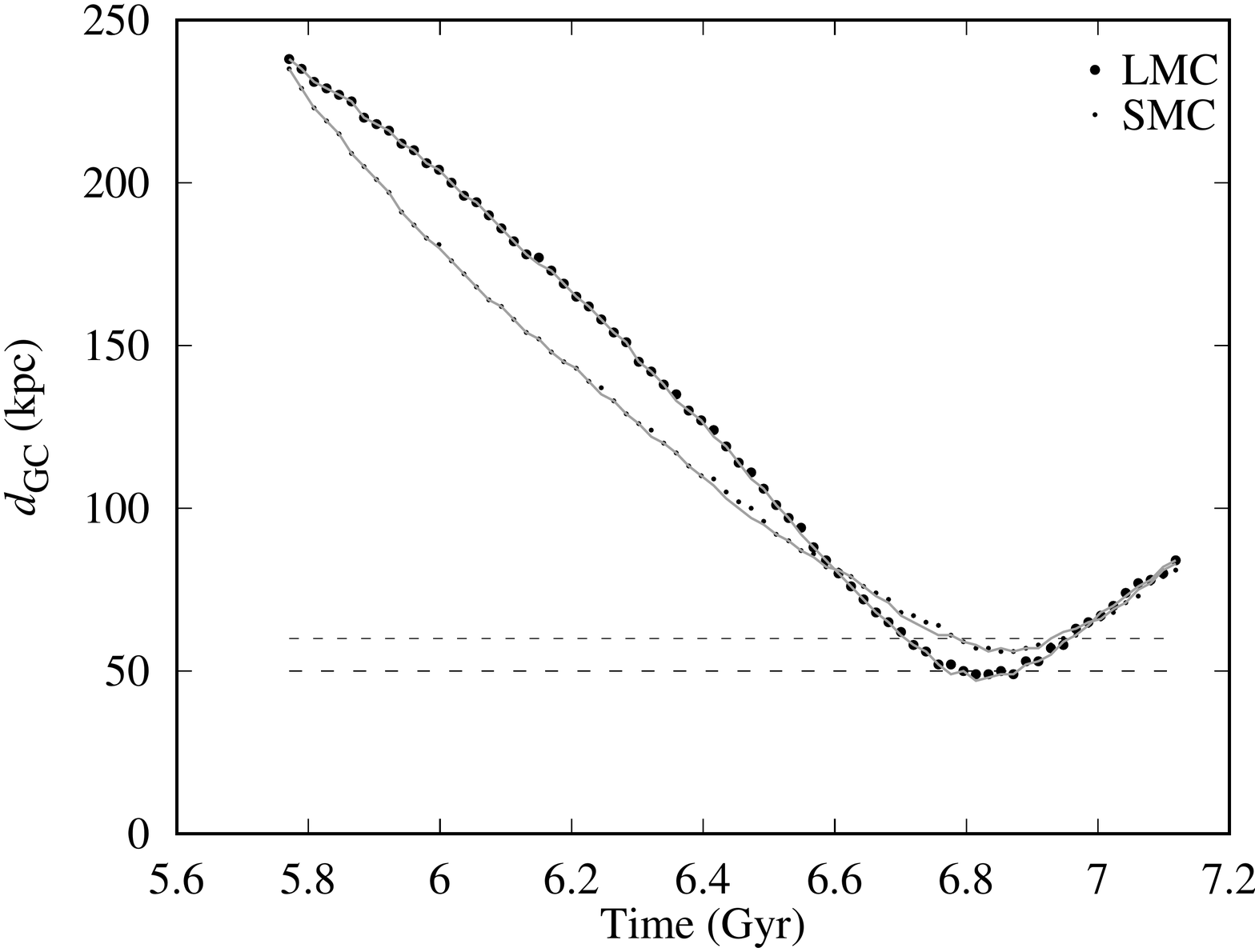}
\includegraphics[width=0.33\textwidth]{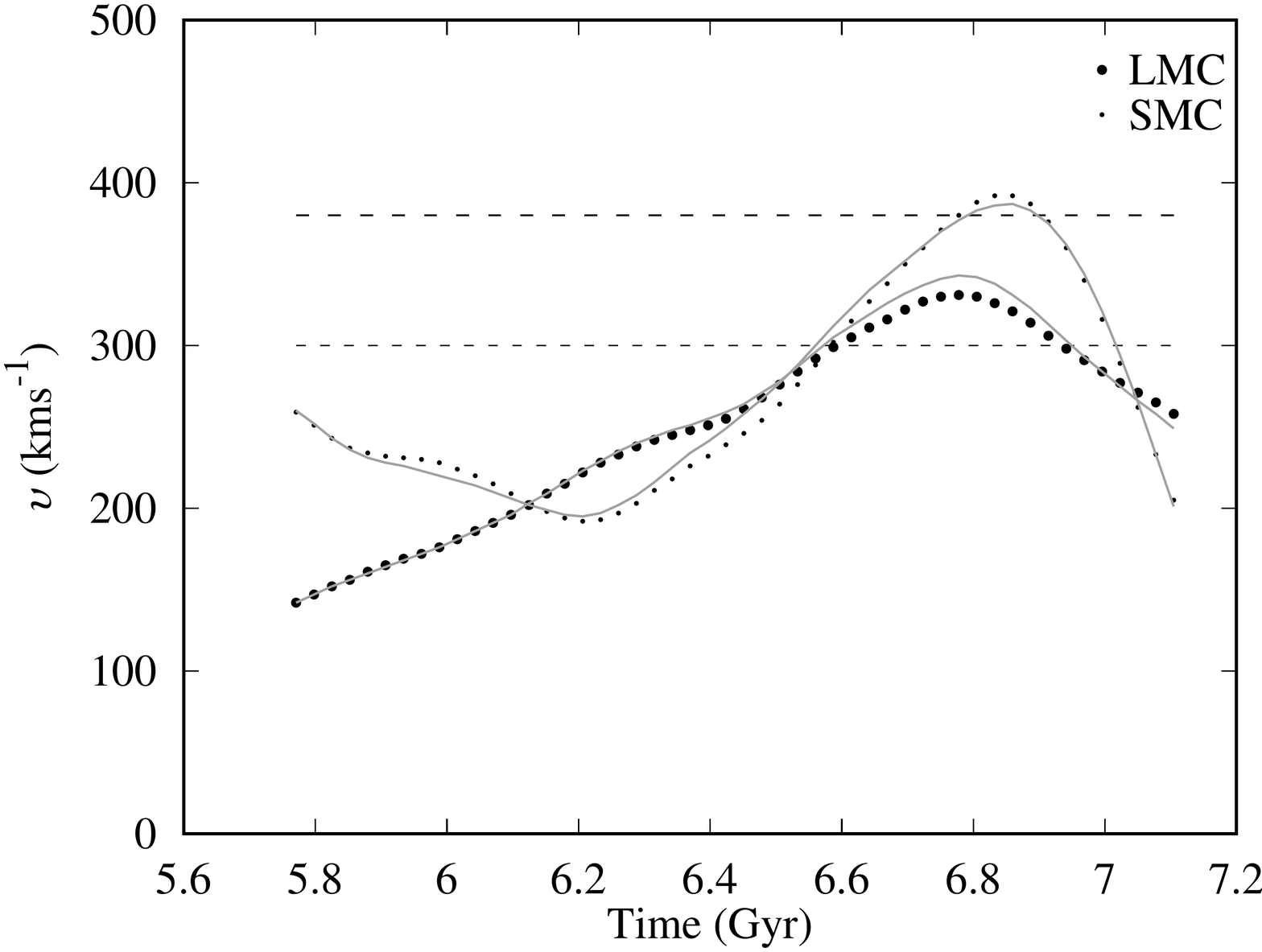}\\
\includegraphics[width=0.33\textwidth]{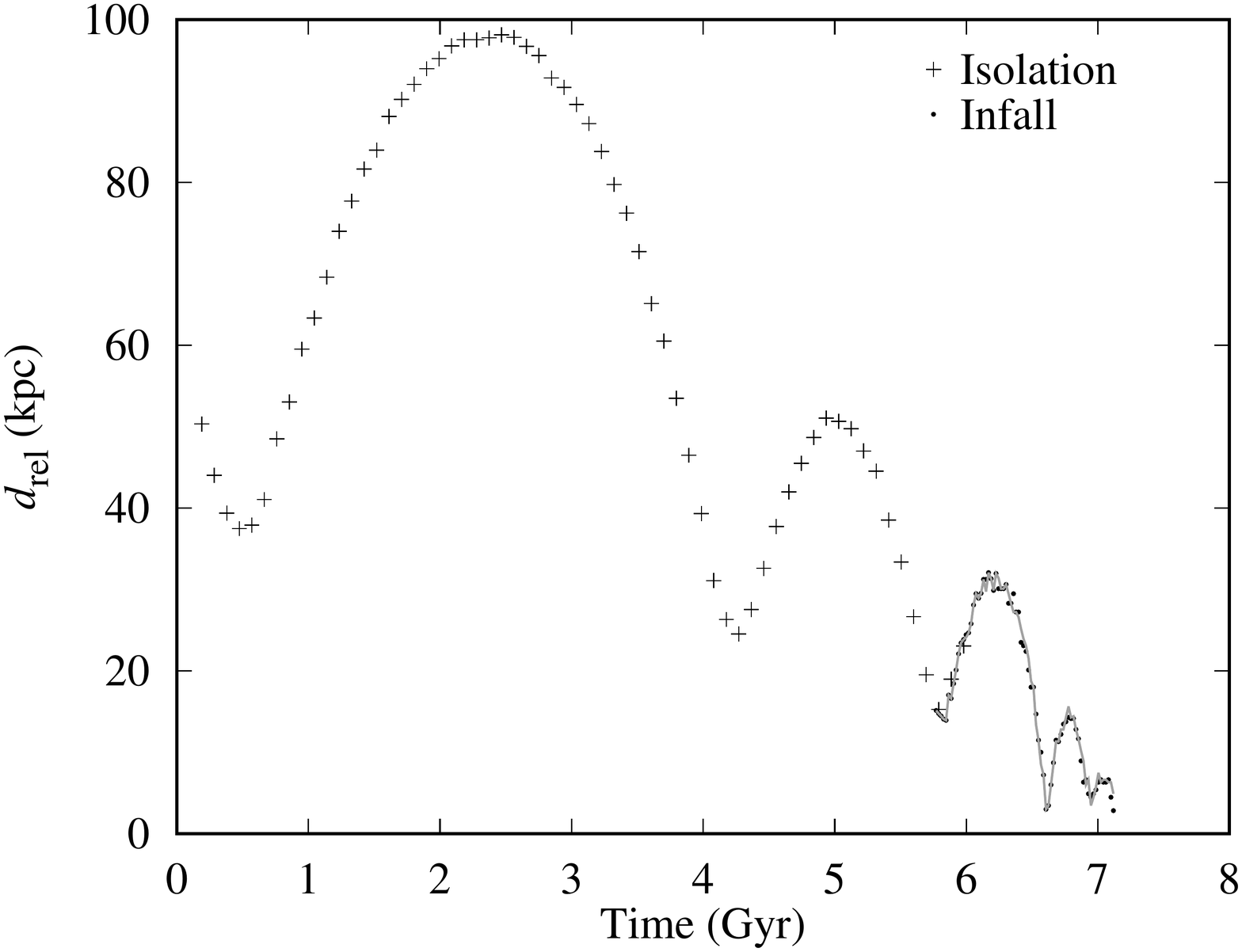}
\includegraphics[width=0.33\textwidth]{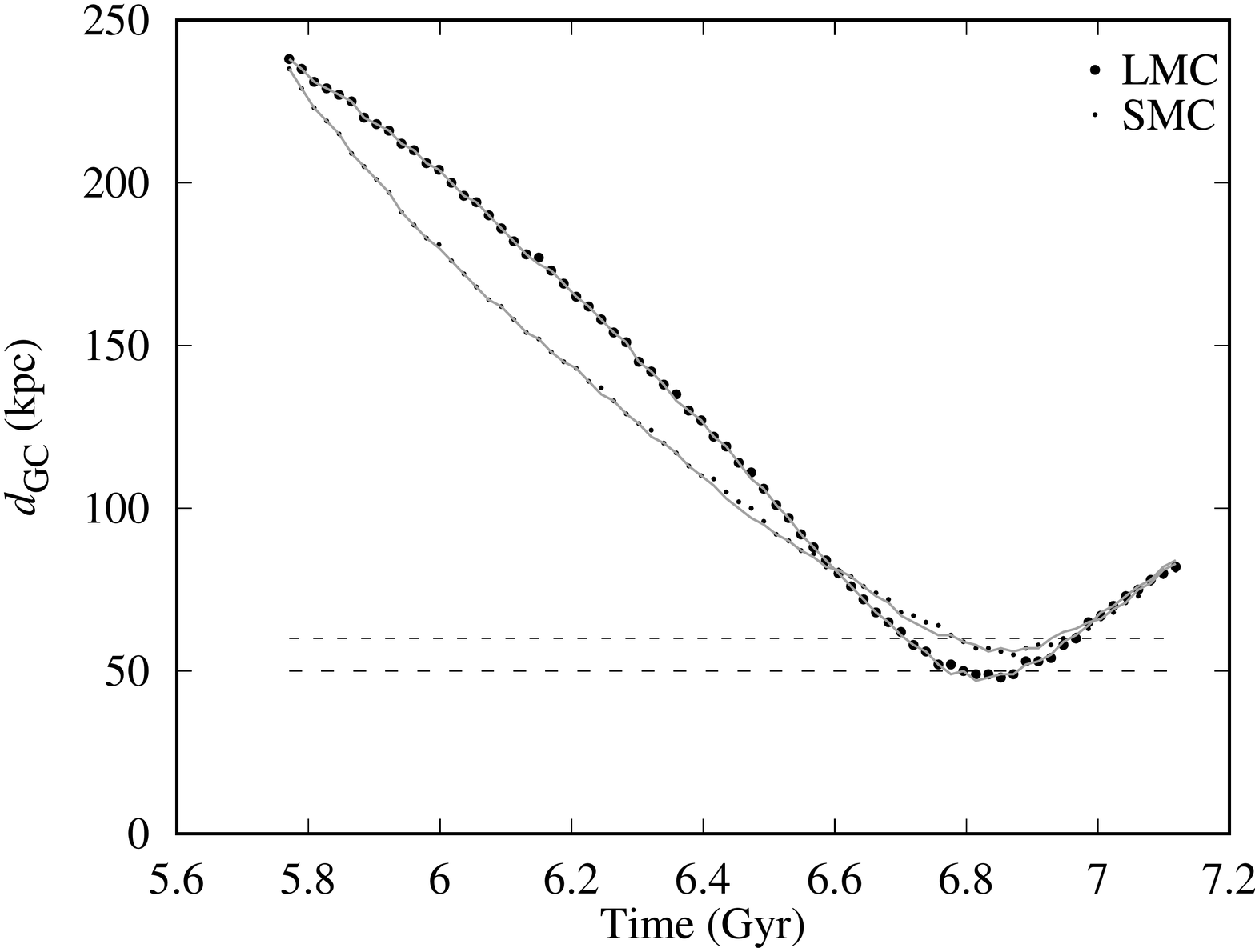}
\includegraphics[width=0.33\textwidth]{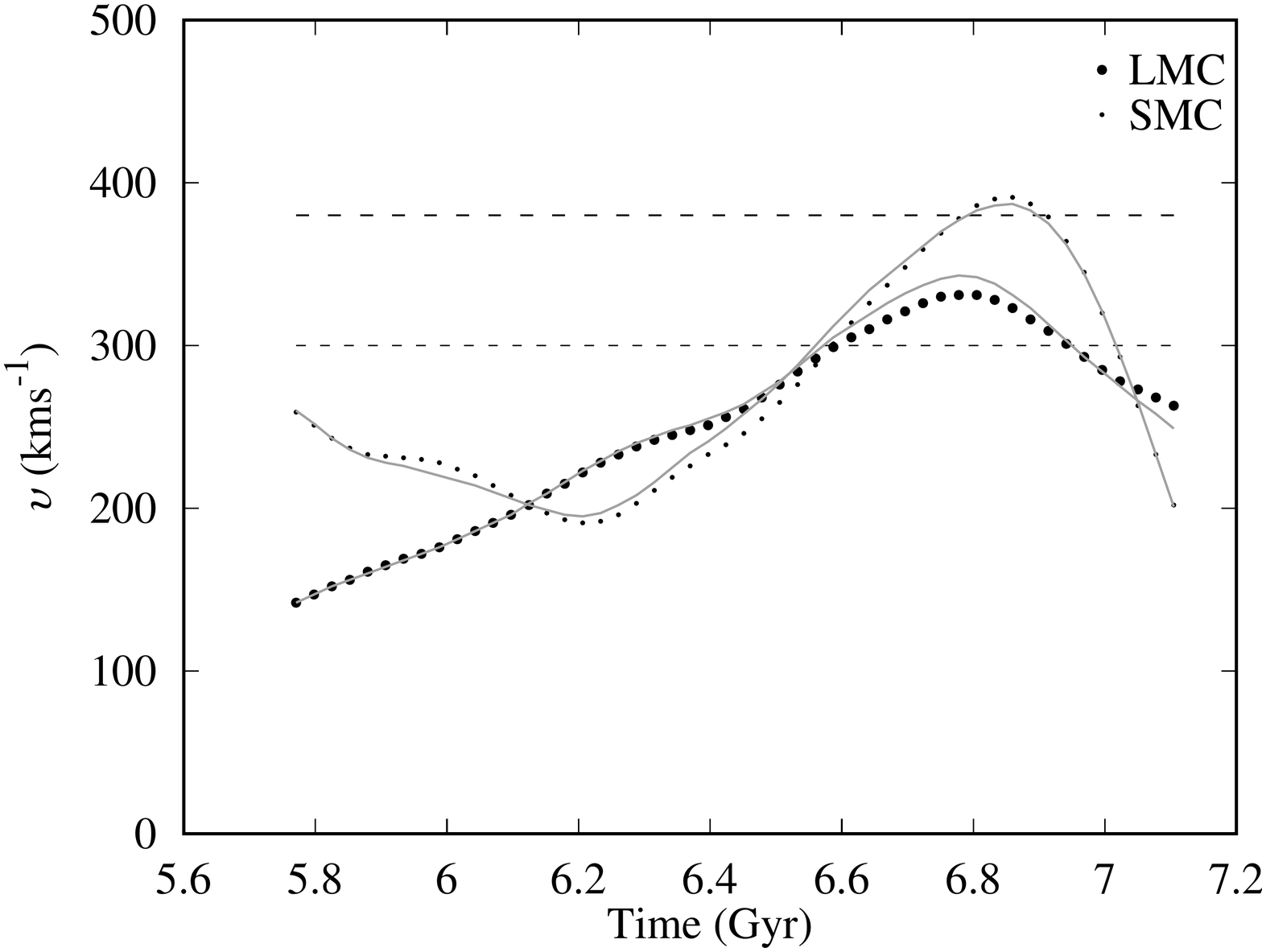}\\
\includegraphics[width=0.33\textwidth]{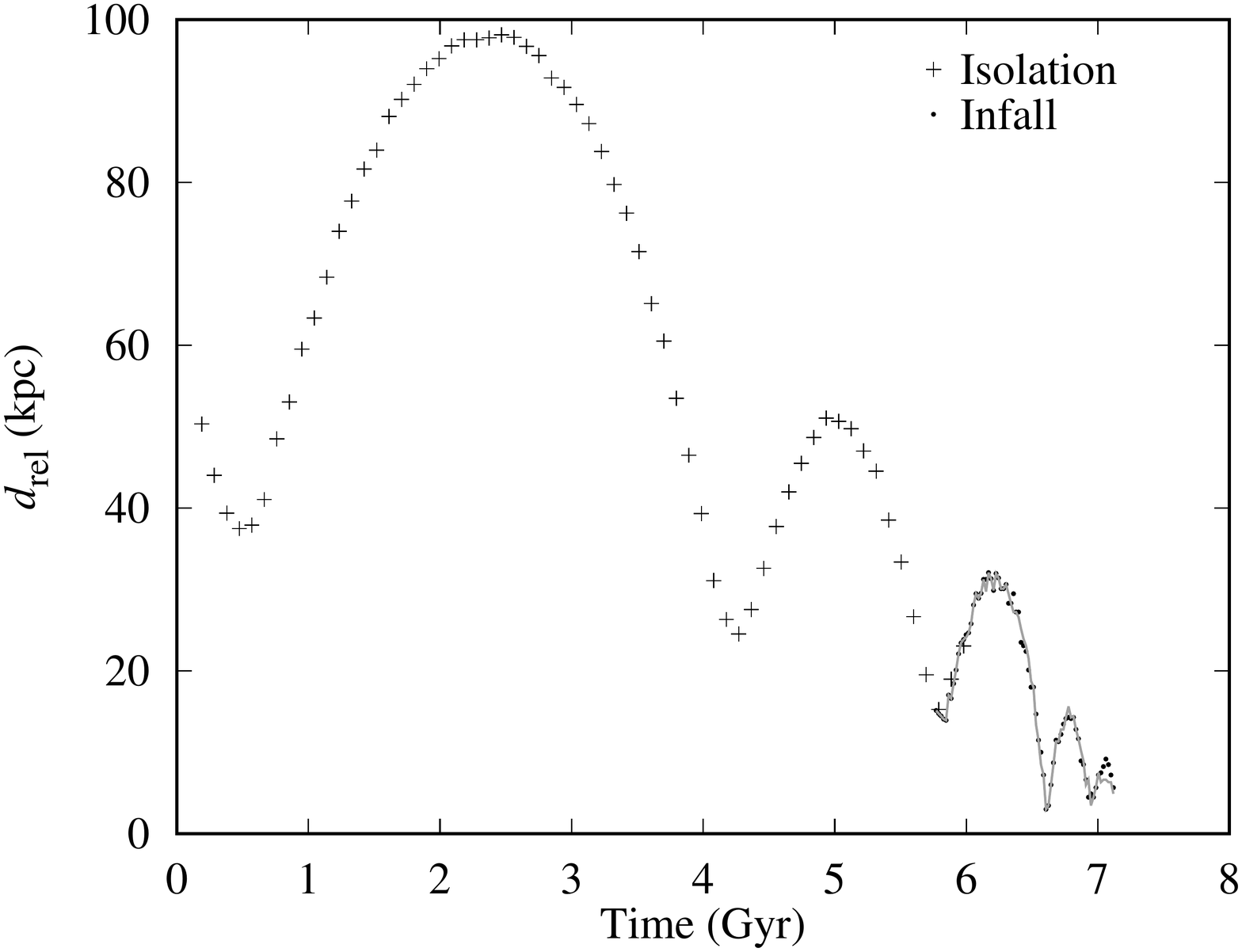}
\includegraphics[width=0.33\textwidth]{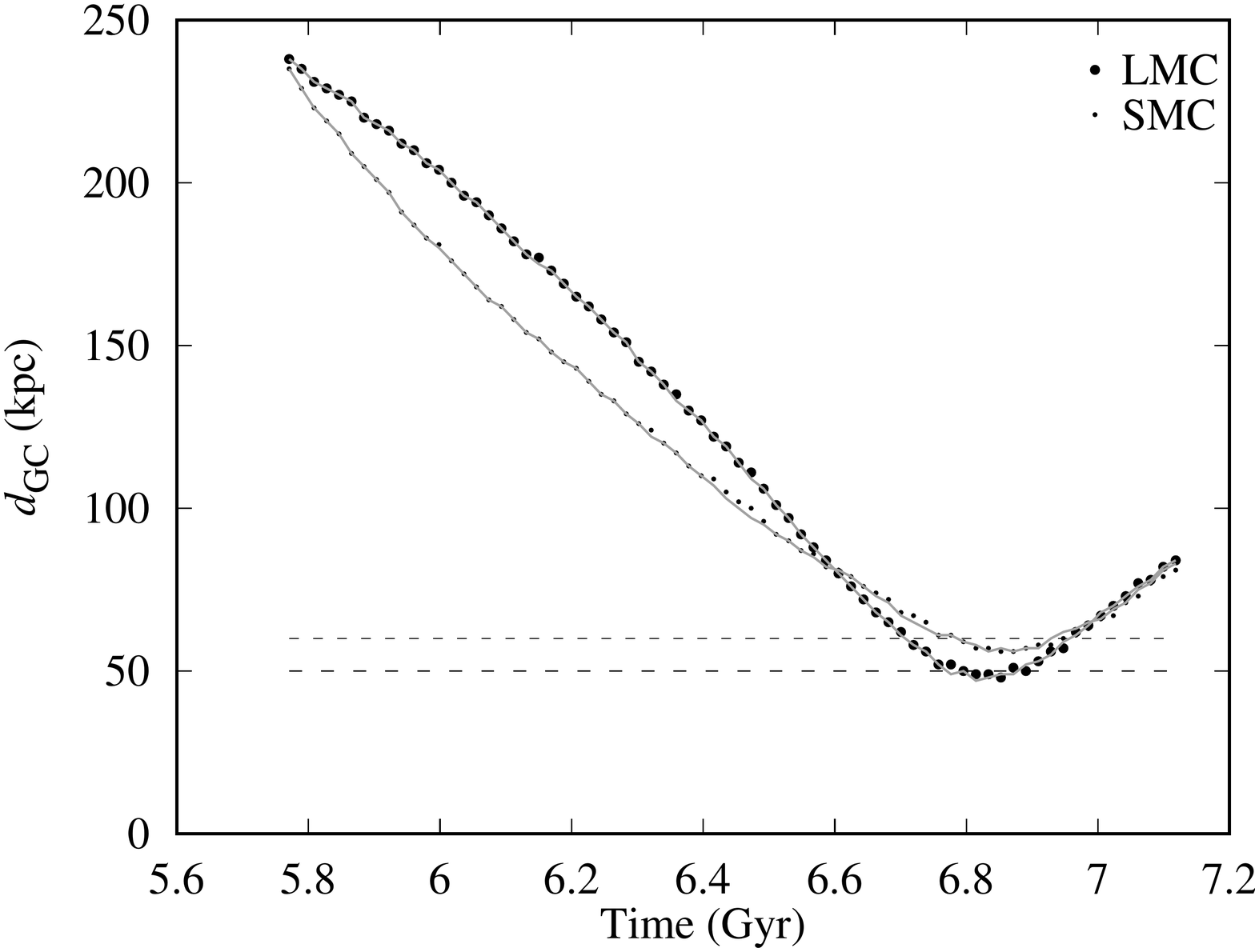}
\includegraphics[width=0.33\textwidth]{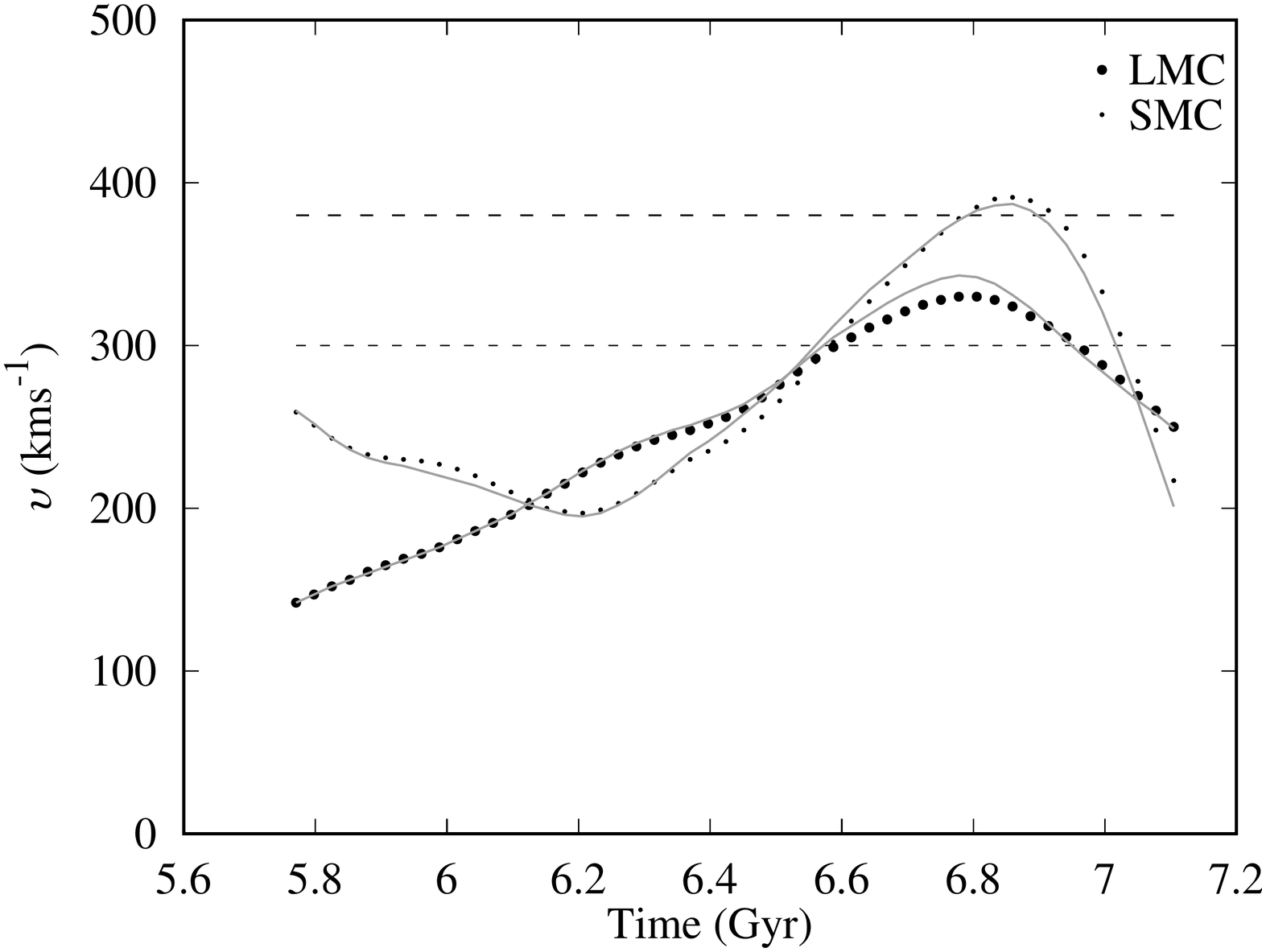}\\
\includegraphics[width=0.33\textwidth]{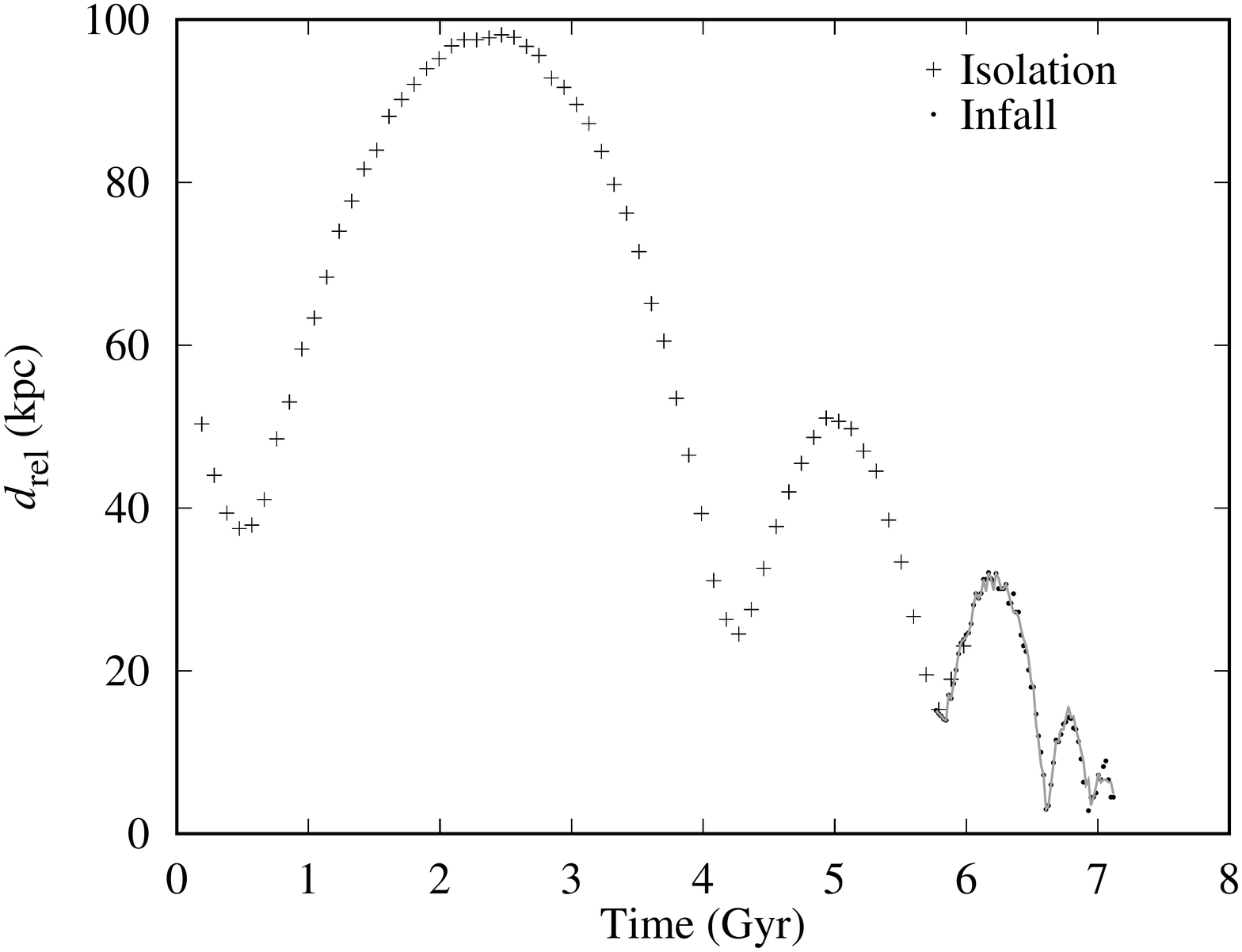}
\includegraphics[width=0.33\textwidth]{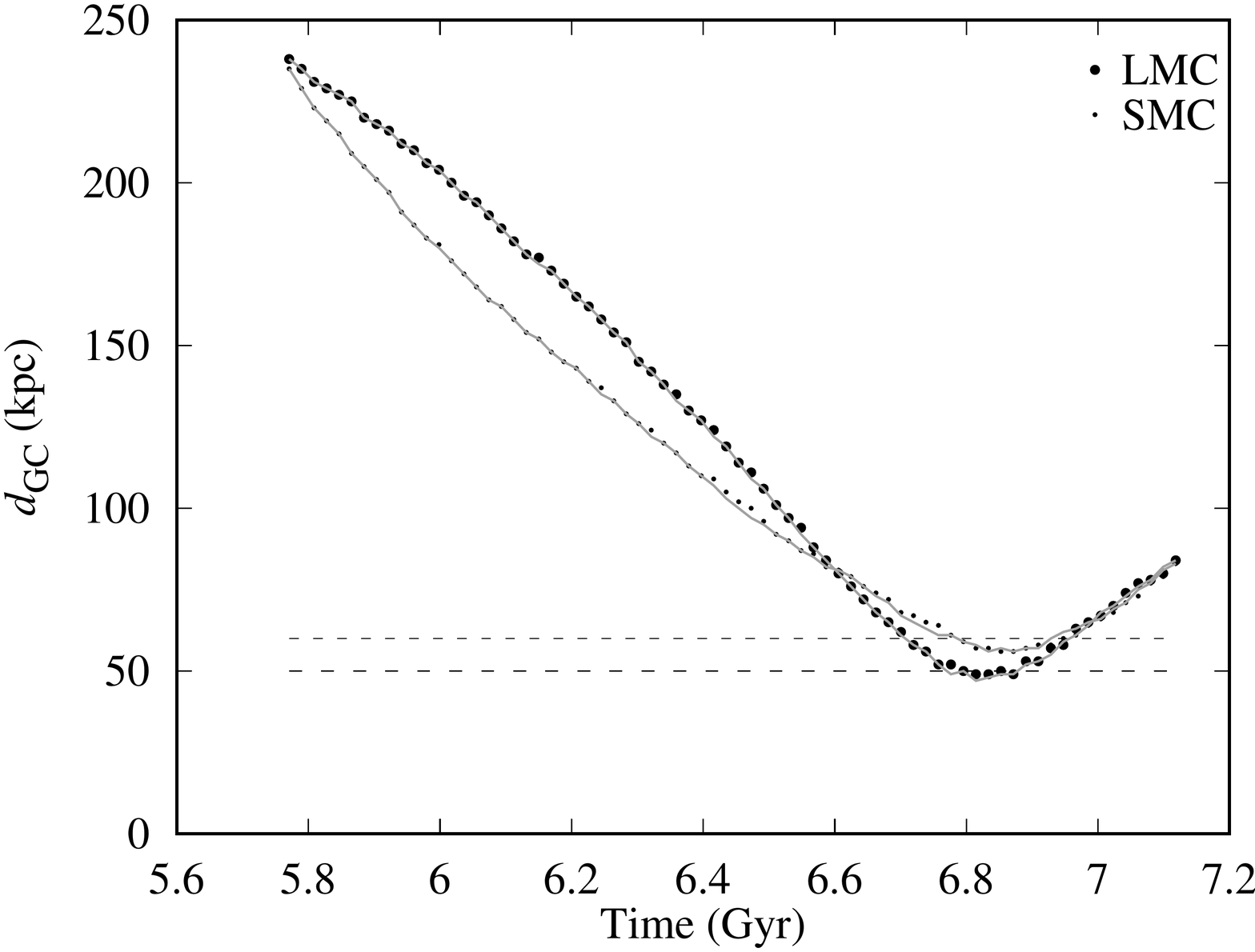}
\includegraphics[width=0.33\textwidth]{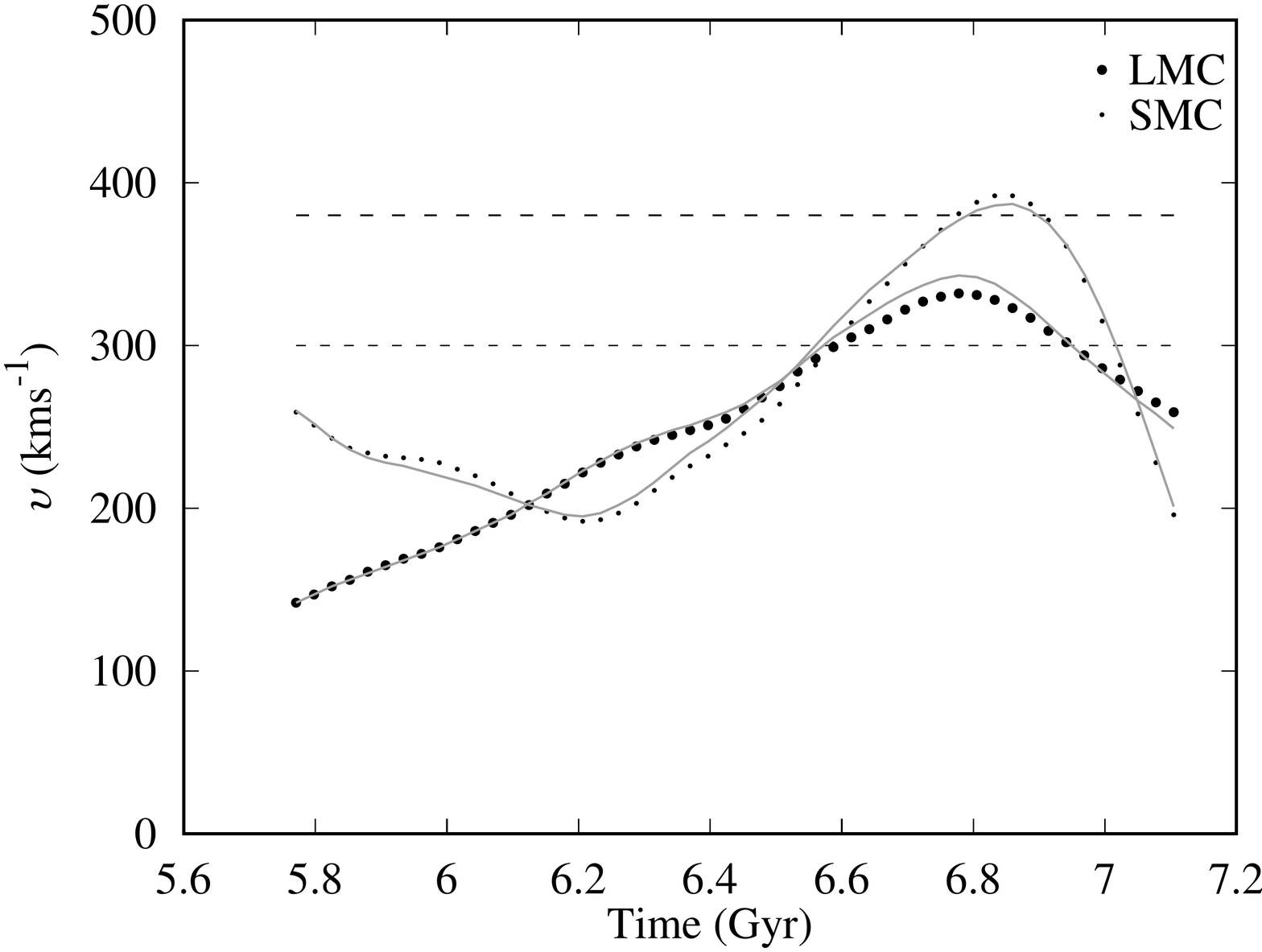}
\caption[  ]{ Same as Figure \ref{fig:ver1}, but for Galactic models that include a hot halo. From top to bottom: Static corona, Slow corona, Fast corona, Magnetised Static corona.  The corresponding results of the `Live DM halo' model shown in the bottom panel of Figure \ref{fig:ver1} are included in each panel (grey thick curves) for comparison. Note that the presence of a (magnetised) corona does not affect the orbital history of the infalling system in any appreciable way compared to the model where the corona is absent.}
\label{fig:ver3}
\end{figure*}

\section{Convergence tests} \label{sec:conv}

We considered in in Sec. \ref{sec:lim} the possibility that our results have been affected by a lack of adequate resolution. This is motivated by recent numerical work which demonstrates that a progressively increasing resolution leads to a less efficient drag, but at the same time decreases a cloud's lifetime. We have run two additional simulations, which are in essence identical to our `Slow corona' model, but at a progressively higher resolution. The model chosen is irrelevant, as any other model behaves roughly in the same way. In one of these simulations we adopt a refinement strategy whereby a cell is refined if the gas mass exceeds $\sim 5 \times 10^5 ~\Msun$ (rather than $\sim 2.5 \times 10^6 ~\Msun$ as we have adopted throughout). In the other simulation we adopt a gas mass threshold of $\sim 2.5 \times 10^5 ~\Msun$. These correspond to an improvement in the gas mass resolution by a factor 5 and 10, respectively, with respect to our initial adopted setting. The {\em spatial} resolution is progressively increased by an additional factor of 2 (per dimension), thus yielding a minimum cell size of 30 pc and 15 pc, respectively.

\begin{figure*}
\centering
\includegraphics[width=0.33\textwidth]{./figures/dice_dice_MW_MCs_q_restart_corona_snpart_lev07_08_xproj_snap_00731_frac-eps-converted-to.pdf}
\includegraphics[width=0.33\textwidth]{./figures/dice_dice_MW_MCs_q_restart_corona_snpart_lev07_08_zproj_snap_00731_frac-eps-converted-to.pdf}
\includegraphics[width=0.33\textwidth]{./figures/gas_snpart_binned_lonlat_dice_MW_MCs_q_restart_corona_lev08_snap00731_hammer-aitoff-eps-converted-to.pdf}\\
\includegraphics[width=0.33\textwidth]{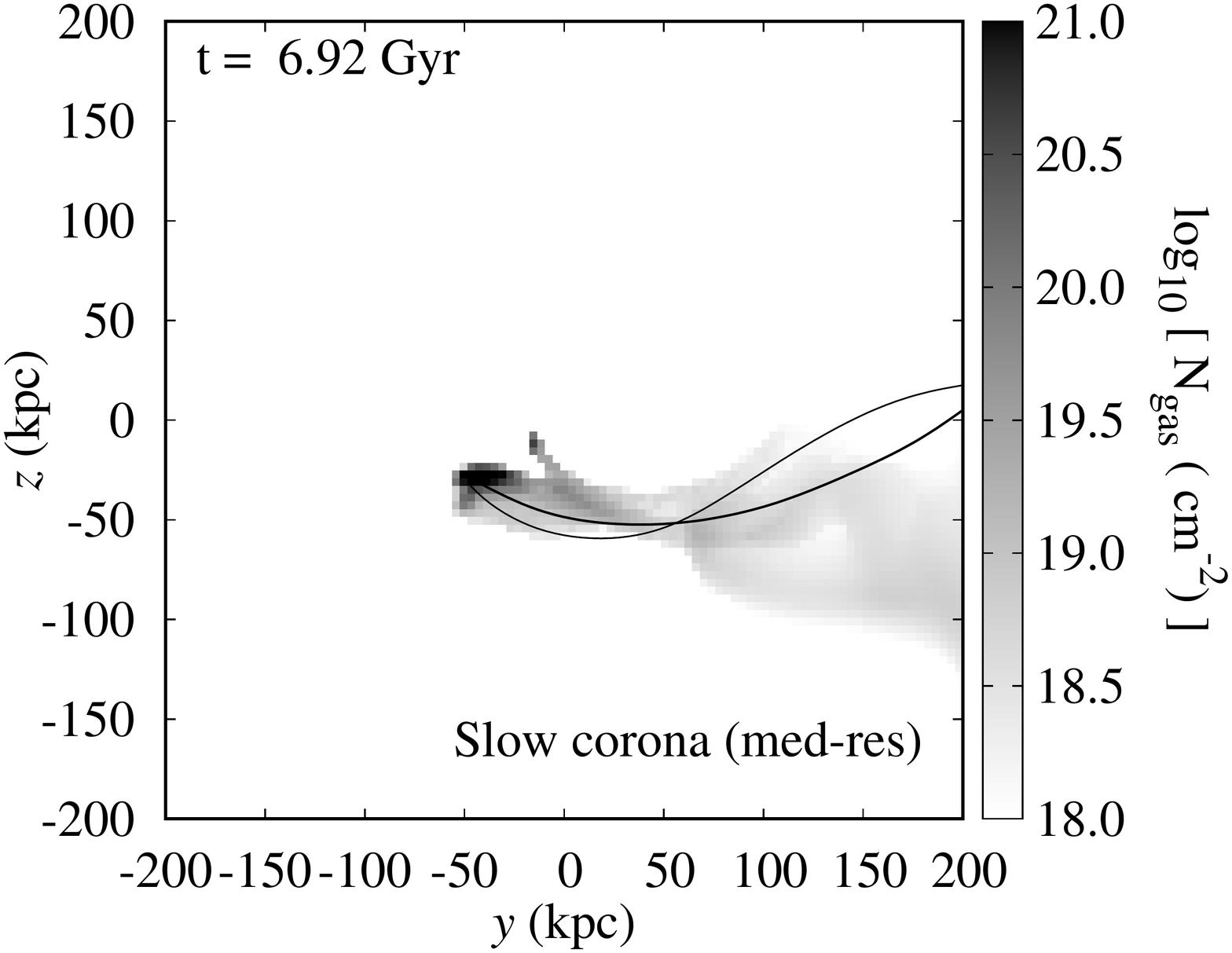}
\includegraphics[width=0.33\textwidth]{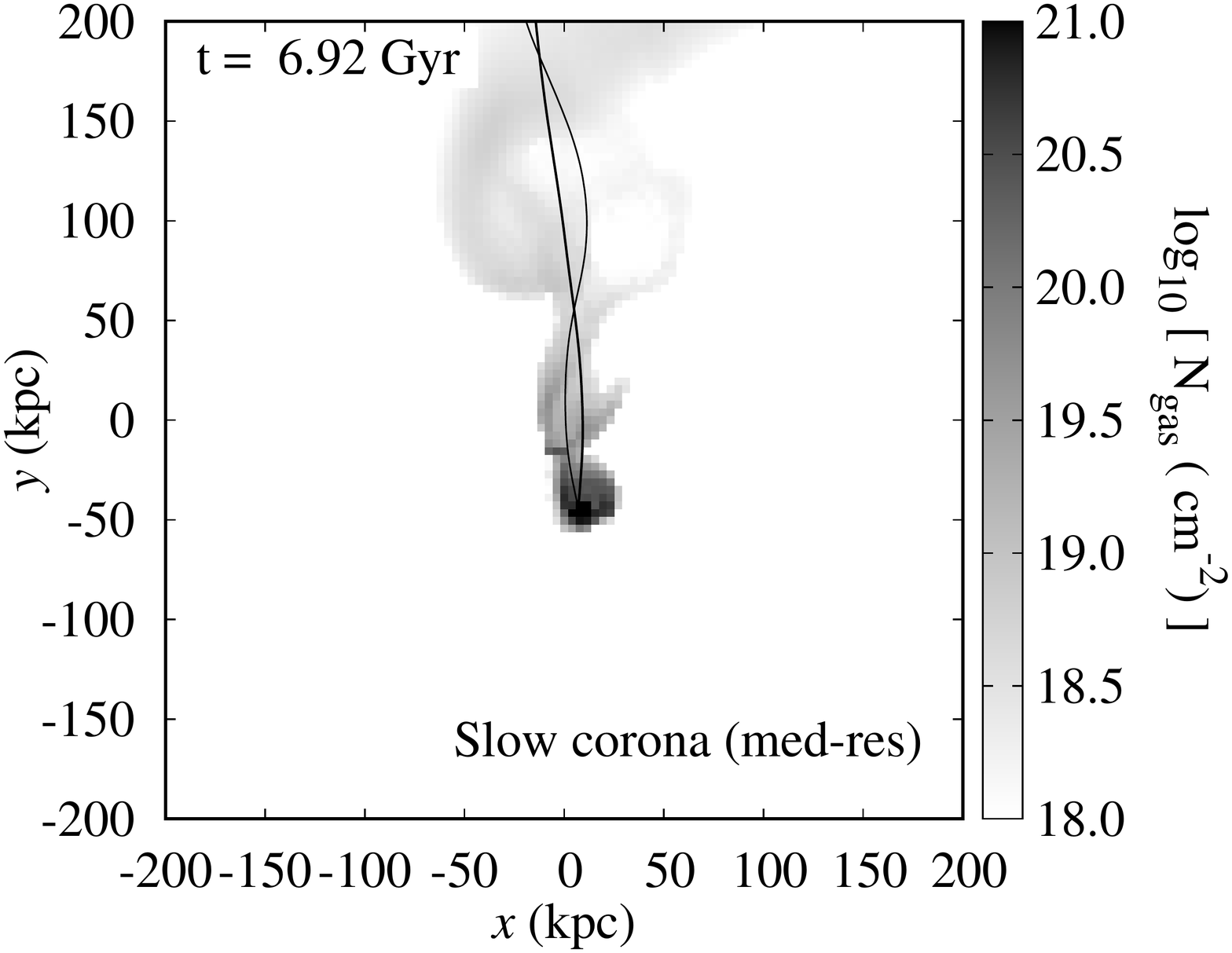}
\includegraphics[width=0.33\textwidth]{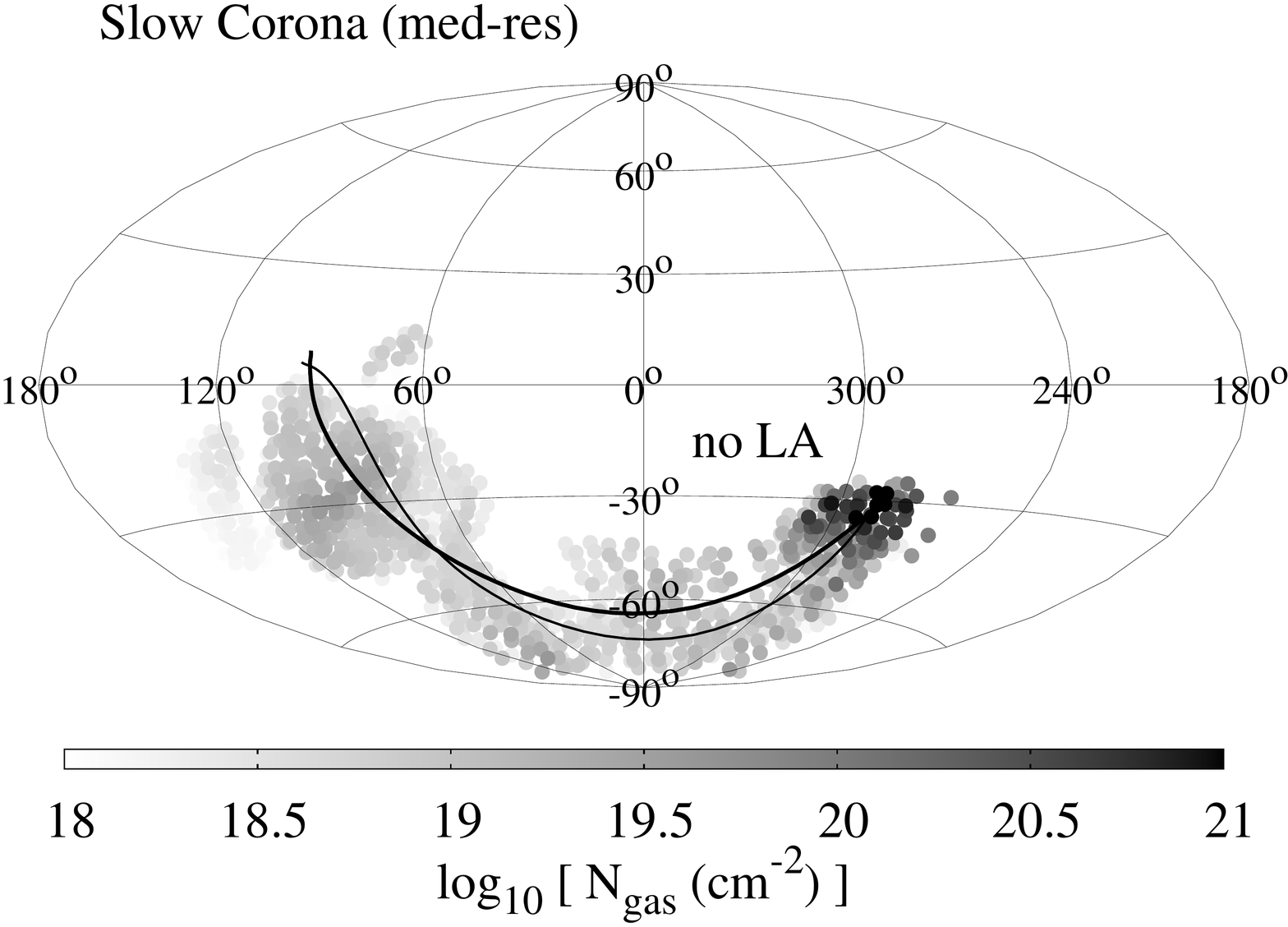}\\
\includegraphics[width=0.33\textwidth]{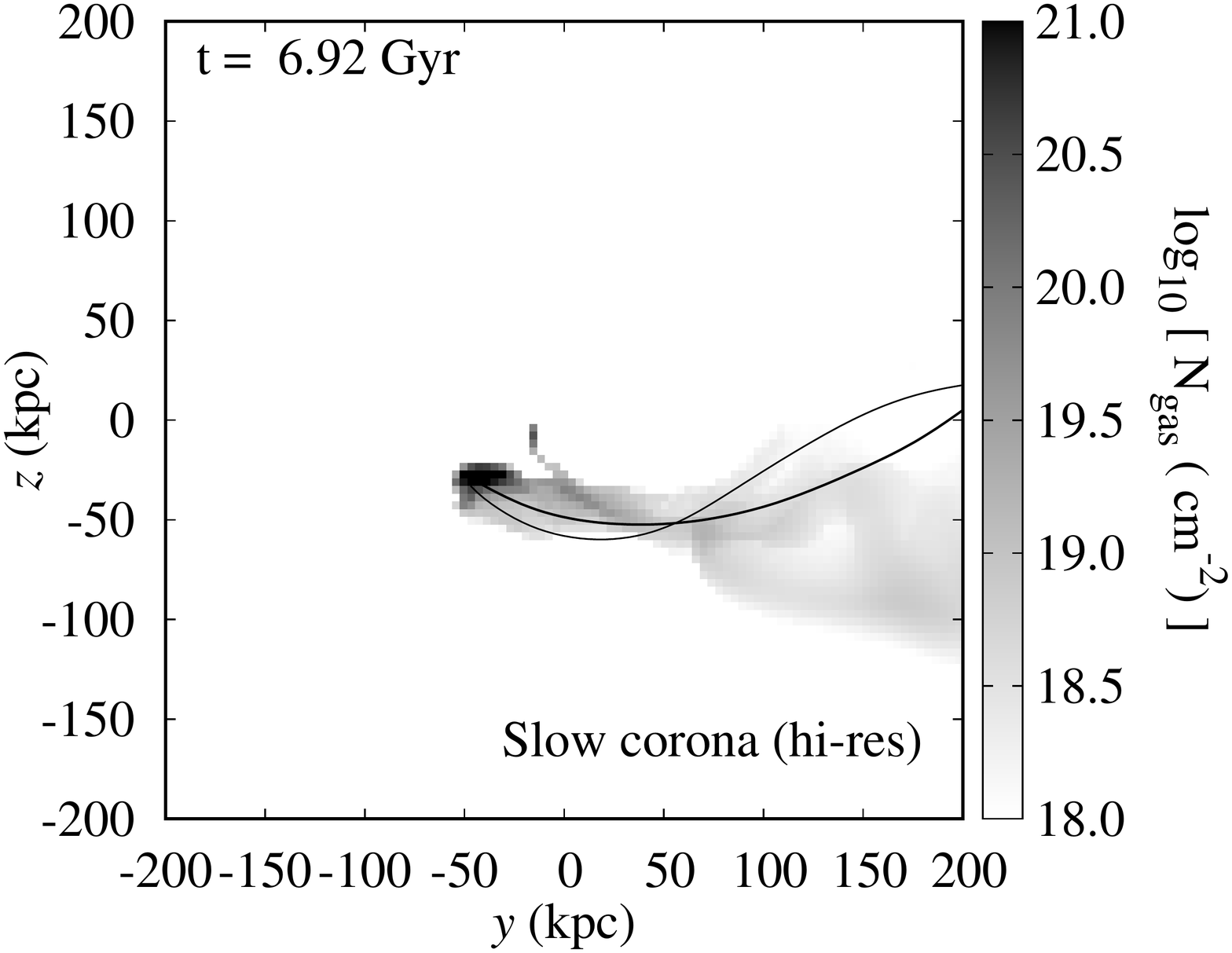}
\includegraphics[width=0.33\textwidth]{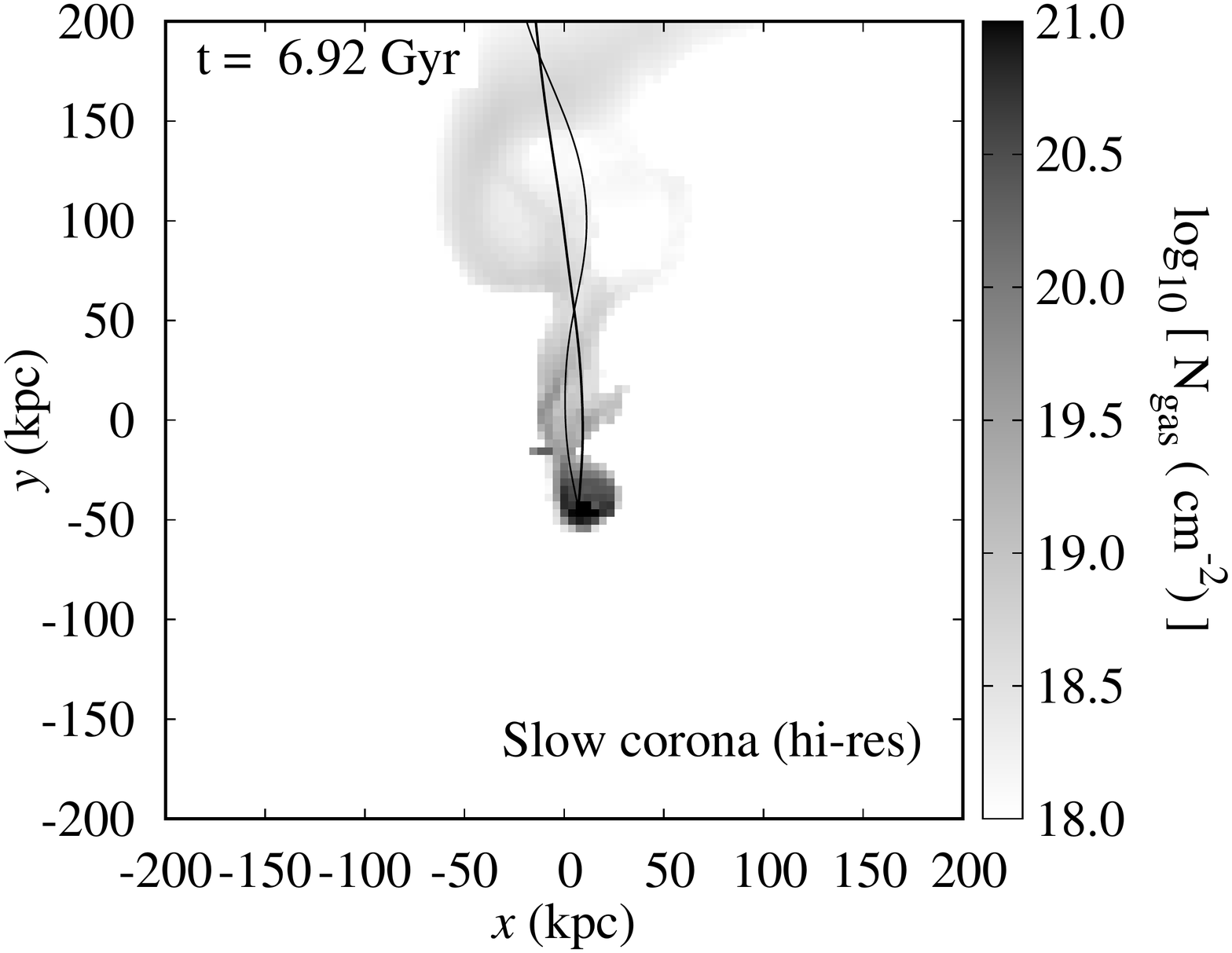}
\includegraphics[width=0.33\textwidth]{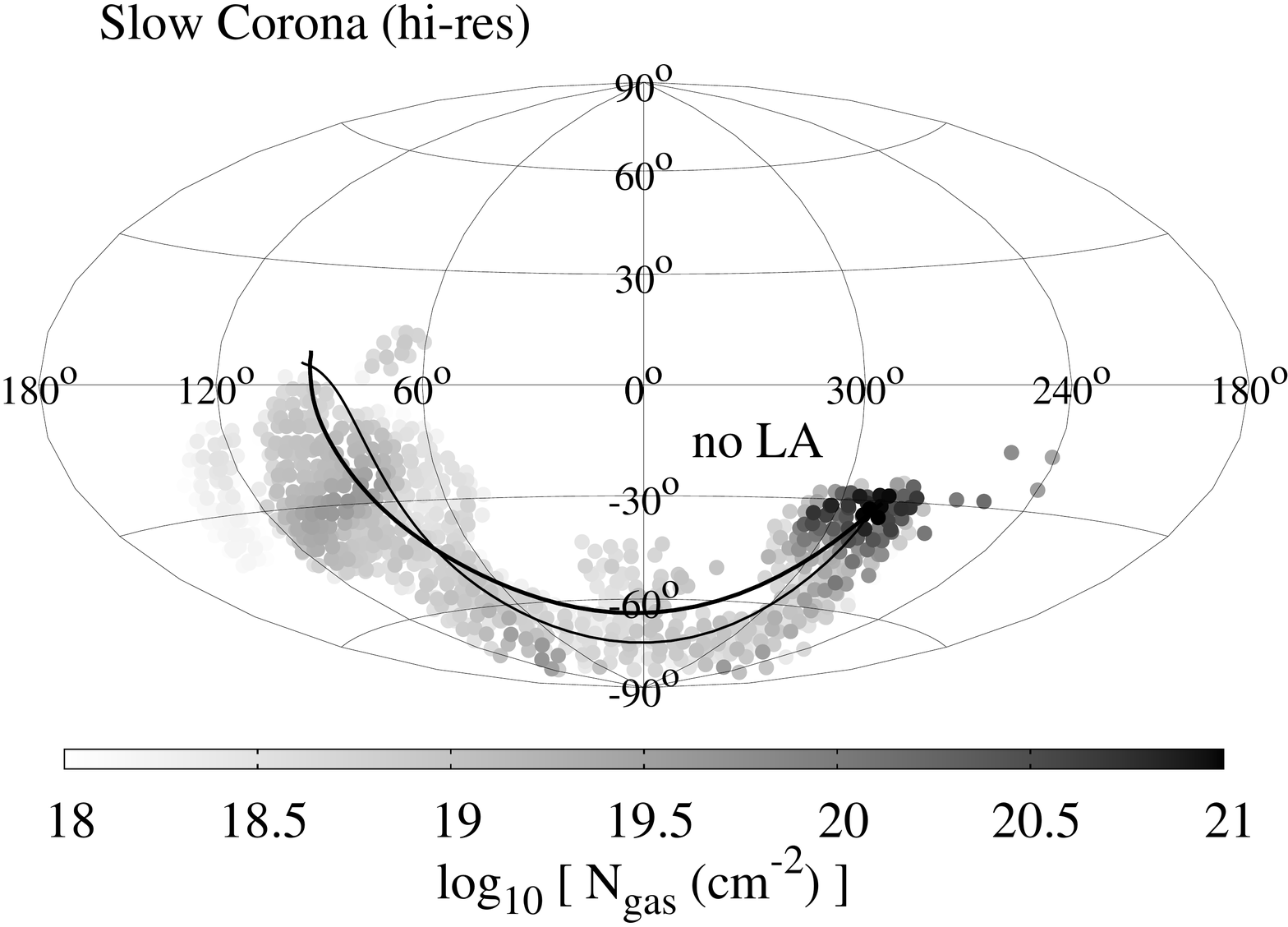}\\
\caption[  ]{ Distribution of gas in the the Magellanic System at the present epoch, i.e. $\sim 1$ Gyr after infall into the Galaxy, in the `Slow corona' model at three different (i.e. progressively increasing) resolutions. The top row displays the results for our standard resolution and each panel is identical to its counterpart in the middle row of Fig. \ref{fig:ms2}. The second and third two row show, respectively, the results for a simulation where the density threshold use to (de-)refine the AMR grid has been decreased by 5$\times$ and 10$\times$, and the limiting spatial resolution improved by 2$\times$ and 4$\times$ (per dimension), with respect to our standard resolution. Note that the gas blobs ahead of the MCs in the high-resolution run apparent in the all-sky projection (third row, right panel) are in fact {\em not} leading the MCs, but falling behind and towards the Galactic centre, as can be clearly seen in the side-on projections in physical space (third row, left panel).}
\label{fig:ms4}
\end{figure*}

To assess the effect of the increase in resolution on our results, we compare the final distribution of gas in each of these simulations, paying especial attention to whether a leading gas stream is able to survive in any of them. The result is shown in Fig. \ref{fig:ms4}. Clearly, the gas distribution is not appreciably different among these three simulations. In particular, none of the simulations features a leading gas stream, despite an order of magnitude increase in the refinement density threshold, and an increase in spatial resolution of 4$\times$ per linear dimension.

\bsp	
\label{lastpage}
\end{document}